%% file: thesis.tex
\documentclass[final,10pt,a5paper]{phdimt}

\def\Title{The World Trade Web: A Multiple-Network Perspective}
\def\Author{Paolo Sgrignoli}
\def\Subject{Economics}
\def\Keywords{world trade web, international trade, migration, international migration network, foreign direct investment, FDI, gravity equation, spatial econometrics, heckman selection process, multiple networks, networks}
\def\Producer{pdfTeX}
\def\myCreationDate{D:20140515193000}	

\title{\Title}
\author{\Author}
\mail{paolo.sgrignoli@imtlucca.it}
\program{Economics, Markets, Institutions}
\coordinator{Prof. Davide Ticchi}
\coordinatorinst{IMT Institute for Advanced Studies Lucca}
\cycle{XXVI}
\year{2014}

\supervisor{Prof. Massimo Riccaboni}
\supervisorinst{IMT Institute for Advanced Studies Lucca}
\tutor{Alexander Petersen}
\tutorinst{IMT Institute for Advanced Studies Lucca}

\firstreviewer{Prof. Giorgio Fagiolo}
\firstreviewerinst{Scuola Superiore Sant'Anna, Pisa}
\secondreviewer{Prof. Javier A. Reyes}
\secondreviewerinst{Walton College of Business, University of Arkansas}

\hypersetup{
	pdfinfo={
		Title={\Title},
		Author={\Author},
		Producer={\Producer},
		CreationDate={\myCreationDate},
		ModDate={\pdfcreationdate},
		Subject={\Subject},
		Keywords={\Keywords}
	}
}

\begin{document}

\def\vec#1{{\bf #1}}

\renewcommand\baselinestretch{1}
\baselineskip=14pt

\def\sym#1{\ifmmode^{#1}\else\(^{#1}\)\fi}

\hyphenation{Ein-ste-in}
\hyphenation{a-ch-ie-ved}
\hyphenation{ap-pro-ach-es}
\hyphenation{mo-del}
\hyphenation{meth-od-ol-o-gy}
\hyphenation{ro-dol-fo}
\hyphenation{o-ver-lap-ping}

\frontmatter

\pagestyle{empty} 
\maketitle
\makereviewerspage
\include{frontmatter/dedication}
\pagestyle{plain} 

\tableofcontents
\listoffigures
\listoftables
\include{frontmatter/acknowledgement}
\include{frontmatter/vita}
\include{frontmatter/abstract}

\mainmatter

\include{mainmatter/1/introduction}
\include{mainmatter/2/chapt2}	
\include{mainmatter/3/chapt3}	
\include{mainmatter/4/chapt4}	

\appendix

{\small
	\bibliographystyle{plainnat}
	\renewcommand{\bibname}{References} 
	\bibliography{backmatter/references}
}

\makecopyright

\pagestyle{empty}
\clearpage
\null\vfill
\begin{center}
This thesis contains at least one error.
\end{center}


\end{document}

%% file: frontmatter/dedication.tex
%

\begin{dedication}
Life is like riding a bicycle.\\To keep your balance you must keep moving.\\
\vspace{0.2cm}
\textit{A. Einstein}
\end{dedication}

%% file: frontmatter/acknowledgement.tex
%

\begin{acknowledgements}
\addcontentsline{toc}{chapter}{Acknowledgements}

\textbf{Chapter 2} is the reproduction of the paper ``The Relation Between Global Migration and Trade Networks'', co-authored with Rodolfo Metulini, Stefano Schiavo and Massimo Riccaboni, currently submitted to \textit{Physica A} (Elsevier);

\textbf{Chapter 3} is the reproduction of the paper ``The Migration Network Effect on International Trade'', co-authored with Rodolfo Metulini, Stefano Schiavo and Massimo Riccaboni, currently submitted to \textit{Papers in Regional Sciences} (Wiley);

\textbf{Chapter 4} is based on a working paper, currently in progress, titled ``FDI \& Trade: A Comparative Network Analysis Approach'' and co-authored with Rodolfo Metulini, Armando Rungi and Massimo Riccaboni.

\vline
\vline

First of all I want to thank my supervisor, Massimo Riccaboni. He has been an endless and valuable source of support, stimuli, suggestions and ideas.

I am grateful to Rodolfo Metulini, Stefano Schiavo and Armando Rungi, for the support, comments and contributions to the different chapters of this thesis.

I also want to acknowledge all the insightful and useful comments from the two referee of this dissertation: Giorgio Fagiolo and Javier A. Reyes.

\newpage
\vline

Last but not least I would like to thank all the beautiful people and good friends I have met at IMT, that made my permanence in Lucca much more interesting and fun:

\vline

(in \textit{random} order)

Farshad Shams, Paula Jimena Matiz L\'opez, Giulia Ghiani,
Andrea Vandin,  \"{O}znur \"{O}zdamar, Samuel Morrison Gallacher,
Benedetto Zaccaria, Stefano Sebastio, Riccardo Di Clemente,
Iulian Romanyshyn, Kristian Gervasi Vidal, Niccolò Fiorini,
Omar Doria, Sahizer Samuk, Nicola De Vivo, Martino Bianchi,
Paola Varotto, Lorena Pullumbi, Rodrigo Lopez Farias, Luca
Anchora, \textit{A}lessandro \textit{Bel}monte, Marco Mori, Yesim Tonga,
Maria Gesualdo, Tiziano Distefano, Alexander Dmitrishin,
Massimo Minervini, Tiziano Antognozzi, Caterina De Vivo,
Olga Chiappinelli, Letizia Montinari, Lorenzo Ferrari, Ana
Ljubojevic, Giovanni Marin (fazzista), Bassma Abou El Fadl,
Lorenzo Cicchi, Konstantinos Karatzias, Maria Romaniello,
Iole Pina Fontana, Panagiotis Patrinos, Nabil Shokri, Michael
Rochlitz, Dario Gianluigi De Maio, Michele Stawowy,, Rafael
Uriarte, Olga Pustovalova, Magdalena Jarosik, Evinc Dogan,
Laura Puglia, Dmytro Karamshuk, Koteswararao Kondepu,
Eleftherios Giovanis, Gabriele Ranco, Alessandro Celestini,
Simona Losito, Nicola Catenacci, Marco Modica, Daria Brasca,
Anvarjon Rahmetov, Francesca Pampaloni, Alberto Guiggiani,
Lorenzo Stella, Isilay Gursu, Steven Daniels, Pavel Belchev,
Lisa Gianmoena, Enrico De Angelis, Luca Bernardi.

\vline

And in general many thanks to all the IMT community for the good time I had in Lucca.

\end{acknowledgements}

%% file: frontmatter/vita.tex
%

\begin{center}
\vspace*{0.5cm}
{\Large \bf  Vita}
\addcontentsline{toc}{chapter}{Vita and Publications}
\end{center}
\begin{table}[h!]
\begin{center}
\renewcommand{\arraystretch}{1.25}
\begin{tabular*}{1\textwidth}{l p{8cm}}

{\bf October 18, 1984} & Born, Reggio nell'Emilia (RE), Italy \\
& \\
{\bf 2006} & Bachelor of Science (B.Sc.) in Physics\\
& Final mark: 108/110\\
& Universit\'a Degli Studi Di Parma, Italy \\
& \\
{\bf 2009} & Master of Science (M.Sc.) in Theoretical Physics\\  
& Final mark: 110/110 cum laude\\
& Universit\'a Degli Studi Di Parma, Italy \\
& \\
{\bf 2011} & Master in Quantitative Finance\\  
& Final mark: 30/30\\
& Universit\'a di Bologna, Italy \\
& \\
{\bf 2013} & Visiting Scholar\\  
& Boston University\\
& Boston, United States \\

\end{tabular*}
\end{center}
\end{table}
\clearpage
\begin{center}
\vspace*{0.5cm}
{\Large \bf  Publications}
\end{center}
\vspace*{0.5cm}
{\small
\begin{enumerate}


\item E.~Agliari, R.~Burioni, P.~Sgrignoli, ``A two-populations Ising model on a diluted random graph,'' in \emph{J. Stat. Mech.}, P07021, 2010.

\item P.~Sgrignoli, R.~Metulini, S.~Schiavo, M.~Riccaboni, ``The Relation Between Global Migration And Trade Networks,'' \emph{Signal-Image Technology \& Internet-Based Systems (SITIS), 2013 International Conference on}, pp. 553 -- 560, 2013.

\end{enumerate}
}
\begin{center}
\vspace*{3.5cm}
{\Large \bf  Presentations}
\end{center}
\vspace*{0.5cm}
{\small
\begin{enumerate}


\item P.~Sgrignoli, ``Two populations Ising model on an Erd\"os-R\`enyi random graph: social and economical applications,'' at \emph{ETH University}, Z\"urich, Switzerland, 2010.

\item P.~Sgrignoli, ``Instability And Network Effects In Innovative Markets,'' at \emph{$7^{th}$ Workshop on ``Dynamic Models in Economics and Finance''}, Urbino, Italy, 2012.

\end{enumerate}
}

%% file: frontmatter/abstract.tex
%

\begin{abstract} 
\addcontentsline{toc}{chapter}{Abstract}

International Trade (IT) plays a fundamental role in today's economy: by connecting world countries production and consumption processes, it radically contributes in shaping their economy and development path.
Although its evolving structure and determinants have been widely analyzed in the literature, much less has been done to understand its interplay with other complex phenomena. The aim of this work is, precisely in this direction, to study the relations of IT with \textit{International Migration} (IM) and \textit{Foreign Direct Investments} (FDI).
In both cases the procedure used is to first approach the problem in a multiple-networks perspective and than deepen the analysis by using ad hoc econometrics techniques.

With respect to IM, a general positive correlation with IT is highlighted and product categories for which this effect is stronger are identified and cross-checked with previous classifications. Next, employing spatial econometric techniques and proposing a new way to define country neighbors based on the most intense IM flows, direct/indirect network effects are studied and a stronger competitive effect of third country migrants is identified for a specific product class.

In the case of FDI, first correlations between the two networks are identified, highlighting how they can be mostly explained by countries economic/demographic size and geographical distance. Then, using the Heckman selection model with a gravity equation, (non-linear) components arising from distance, position in the Global Supply Chain and presence of Regional Trade Agreements are studied. Finally, it is shown how IT and FDI correlation changes with sectors: they are complements in manufacturing, but substitutes in services.

\end{abstract}

%% file: mainmatter/1/introduction.tex
%

\chapter{Introduction}
\label{chapt:introduction}

\graphicspath{{1/figures/PNG/}{1/figures/PDF/}{1/figures/}}

International trade is fundamental phenomenon that in today's globalized world, by linking countries production and consumption structures, plays a key role in shaping their economy and development path.
Globally it has a huge impact and in the last decades it grew at a tremendous rate: measured in gross terms, the dollar value of world merchandise trade increased by more than 7\% per year on average between 1980 and 2011, reaching a peak of 18 trillion USD at the end of that period. Trade in commercial services grew even faster, at roughly 8\% per year on average, amounting to some 4 trillion USD in 2011. Moreover, since 1980, world trade has grown on average nearly twice as fast as world production \citep{WTO2013}.

Since easily accessible large amount of data became available, many empirical studies on international trade appeared. Some of them, basically the first to be published, were based on a gravity-like equation estimations and aimed to understand the determinants of the observed trade regularities: pioneered by \citet{tinbergen1962shaping} and theoretically founded by \citet{anderson1979theoretical}, some of these are for example \citet{bergstrand1985gravity, helpman2008estimating}. Since its first introduction in 1962, the \textit{gravity model of trade} has proved to be incredibly successful in fitting international trade data \citep{van2010gravity,hema13} and gave rise to a huge branch of empirical literature in international trade. Its basic idea is that bilateral trade flows are well explained by a gravity-like equation involving country sizes (depending on the specification this is in general proxied by Gross Domestic Product and population, some combination of them or country fixed effects) and, inversely, geographical distance.

Another popular approach is instead to consider trade as a macroeconomic network, i.e. a graph where nodes are world countries and links represent their possible interaction channels along many economic and social dimensions, and then studying its topological properties \citep{serrano2003topology, garlaschelli2004fitness, Fagiolo2008} and their evolution over time \citep{Garlaschelli2005, Bhattacharya2008, Fagiolo2009}, as well as its community structure \citep{barigozzi2011identifying,piccardi2012existence}. 
Other examples of this broad literature are \citet{Smith1992}, one of the first papers to consider international trade in a network perspective, and \citet{ward2013} that used an extended gravity model to incorporate network dependencies.

In particular, the complex-network perspective lets investigate the properties of the intensive and extensive time evolution of international trade channels, analyzing the intricate and complicated web of relationships between countries over the years \citep{Schweitzer24072009}. Moreover it has recently been argued that knowledge of the topological properties of these networks may be important to understand how economic shocks propagate and how well countries perform over time \citep{Abaix2011, Acemoglu2012, Lee2011, ductor2011social, Chinazzi20131692}.

Furthermore, more recently, a new stream of literature wants to identify countries growth and development path, starting from their exports structure and economies complexity \citep{Hidalgo2007, Hidalgo2009, Tacchella2012, Caldarelli2012}.

Although with different approaches and techniques, a common feature of existing works is the focus on the properties, evolving structure and dynamics of the World Trade Web (WTW) alone, treating the phenomenon as completely independent.
In other words, the topological properties of the WTW has been investigated as if it was a disconnected layer of the directed-weighted multi-graph where nodes are world countries and links represent their macroeconomic interaction channels. Nevertheless, in an increasingly interconnected and globalized world, an important fact to understand is that trade is not an independent or isolated phenomenon, being instead continuously influenced by other aspects of our economy. In particular demography, investment, technology, energy, transportation costs and institutions are among the fundamental factors that shape the overall nature of the World Trade Web and explain why countries trade \citep{WTO2013}.
Still, in this direction much less has been done.

Only in recent years a growing interest among scholars has been observed and this has led to an increasing number of studies focusing on the junction between trade and other phenomena, for example migration \citep{rauch2001business, rauch2002ethnic, Fagiolo2014}, finance \citep{Schiavo2010}, the Internet \citep{riccaboni2013global} or, limited to OECD countries, more than one altogether \citep{Lee2012}; but much more has to be understood.
It is easy to think that a worldwide mechanism like trade is intimately connected with many social and economic aspect of our society and that these interconnections will be mostly complex and non-linear, concealing important information on the fundamental mechanisms driving our economy.

More generally one might build a multi-graph representation of the macroeconomic network, where between any two countries there may exist many links, each representing a different type of between-country interaction (trade, mobility, finance, foreign investment, etc.). This may allow one to explore whether different layers display similar topological properties, and whether such properties are correlated, or causally linked, between layers. This empirical research approach may in fact convey new and interesting insights on the importance of networks structure in shaping aggregate dynamics of the societies and economies where we live.

In this thesis I want to give my contribution to the current literature by studying the interactions and correlations of the World Trade Web with other two worldwide phenomena: in particular Chapters \ref{chapt:migr1} and \ref{chapt:migr2} will be devoted to the interplay of trade with people \textit{International Migration}; while Chapter \ref{chapt:fdi} will study the relations of trade with the Transnational Corporations Control network, that will be interpreted as stock \textit{Foreign Direct Investments}.

\vline

In the remainder of this introductory chapter I will give a brief description of the research ideas and main results related to the subsequent parts of this thesis.

\section*{Research questions}

\begin{itemize}
	\item Chapter \ref{chapt:migr1}
	\begin{itemize}
		\item How are the World Trade Web and people migration networks related? Do they have the same structure or share any property?
		\item Do people migration influence international trade on a global scale? Are there product categories (or countries) for which this effect is stronger?
	\end{itemize}
	
	\item Chapter \ref{chapt:migr2}
	\begin{itemize}
		\item Is there any third-country (network) effect playing in the relation between trade and migration on a global scale?
		\item Are these network effects stronger for any specific category of goods?
	\end{itemize}
	
	\item Chapter \ref{chapt:fdi}
	\begin{itemize}
		\item Are the properties of the World Trade Web and Foreign Direct Investment networks correlated? Do the two networks have similar structures?
		\item Do Foreign Direct Investments influence International Trade on a global scale? What are the factor/conditions that influence this correlation?
	\end{itemize}
\end{itemize}

\section*{Main results and contribution to the literature}

\subsection*{Chapter \ref{chapt:migr1}}
First the World Trade Web and International Migration networks main characteristics are compared, finding results consistent with the previous literature. Then the product categories for which the presence of a community of migrants significantly increases trade intensity are identified by studying the similarities between the two networks and where to assure comparability a filter based on the hypergeometric distribution is applied. Next, proposing a new way to define country neighbors based on the most intense links in the trade network and employing spatial econometrics techniques, the effect of migration on international trade is measured, also controlling for network interdependencies. Overall, migration significantly boosts trade across countries and it is highlighted how this effect is stronger for a new product category we introduce, based on the similarity between trade and migration networks topology.

The main contribution to the literature of this work is the investigation of the effects of migration on international trade in a global perspective, rather than focusing on a single ethnic network as had been done so far in the literature. A new methodology to compare two phenomena and to investigate their correlation and similarities is also proposed, mixing network analysis and Jaccard and Revealed Comparative Advantage indexes. Moreover this methodology led naturally to a new classification of goods, aimed at identifying those that are more correlated with migrant stocks: this proves to be a good alternative to any previously proposed one as it confirms all the results from previous works, with the advantage of being robust to the addition of network effects and better considering subtle product categories (i.e. intermediate goods), as well as having an easier to use, more straightforward definition.

\subsection*{Chapter \ref{chapt:migr2}}
The relationship between international trade and migration is studied, with a specific focus on measuring both direct and indirect network effects, finding that migration significantly affects trade across categories in both ways. The analysis is carried out also differentiating trade in product categories, in order to identify how migration influence trade of different kind of goods. The indirect impact highlights a stronger competitive effect of third country migrants for homogeneous goods.\footnote{For the definition of \textit{homogeneous} and \textit{differentiated} goods see Section \ref{chapt2_sec:data} or \citet{rauch2002ethnic}} Furthermore, from a qualitative point of view we confirm the finding that migration has a larger impact on differentiated products, both at direct and global (network) level. Indeed, the negative effect that third-country migrants have on trade of homogeneous goods (testified by the negative indirect impact found in the estimation results) and that we rationalize as a competition effect, is no longer there when we focus on differentiated goods.

This work also contributes to the literature by proposing a new way of defining the weights matrix for spatial econometric techniques, exploiting topological distances on the network, rather than the usual geographical one.
Some contribution to the literature of spatial economics / econometrics that aims to control for the multilateral resistance terms in the constraint gravity equation for trade is given as well: it is concluded that, accounting for the multilateral resistance terms by means of a Spatial Durbin Model (SDM) specification and using the migration network weight matrix, residuals autocorrelation are filtered out.

\subsection*{Chapter \ref{chapt:fdi}}
International trade and transnational corporations (in the following interpreted as stock Foreign Direct Investment) networks are first compared: as expected considering the role of these two phenomena in the Global Value Chain (GVC), they are strongly correlated and such correlation can be mostly explained by country economic / demographic size and geographical distance. Then, using the Heckman selection model with a gravity equation, this result is confirmed and some factors that (non-linearly) influence this correlation are identified: the industry position in the Global Supply Chain, countries distance and the presence of Regional Trade Agreements. In particular, the more goods are downstream or countries are further away, the more the two phenomena tend to be complements; while the presence of trade agreement lowers the general positive correlation existing between the two. The intuition for all these results resides in the cost-effectiveness of transnational corporations networks and give us important insights about their international strategies. Finally we distinguish the cases of the three main economic macro-sectors, finding that trade and Foreign Direct Investments are \textit{complements} in manufacturing, but \textit{substitutes} in services. In the primary sector instead, their relation depends on the relative direction in which we consider the two: when taken in the same way, i.e. both trade and FDI exports, they are substitutes, while if considered in opposite directions, i.e. trade export and FDI import, no statistically significant relation is observed.

This chapter contributes to the literature by analyzing a newly published dataset for worldwide stock foreign investments, represented by transnational corporations control network. Such dataset make it possible to compare this phenomenon with international trade on a global scale, finding many insights on the interplay between the two. To the best of my knowledge, this is the first work to propose this comparison on such a broad scale.

%% file: mainmatter/2/chapt2.tex
%

\chapter[Trade and Migration (1)]{The Relation Between Global Migration and Trade Networks}
\label{chapt:migr1}

\graphicspath{{2/figures/PNG/}{2/figures/PDF/}{mainmatter/2/figures/}}

\section{Introduction}
An increasing number of interdependent phenomena on a global scale are analyzed as complex networks: international trade, human mobility, communication and transportation infrastructures are just a few examples. Furthermore, researchers are more and more aware that many of these networks are interrelated and cannot be analyzed in isolation. However, only recently the academic literature has started to develop new methodologies to analyze the dynamics of intertwined networks, including cascading failures and the transmission of shocks across multiple and heterogeneous network structures \citep{buldyrev2010catastrophic,PhysRevE.88.050803}. We want to contribute to this emerging field of research on multiple networks by analyzing the relationship between the World Trade Web (WTW) and the International Migration Network (IMN).

In fact, since the mid 1990s a growing body of economic research has investigated the relation between international trade and migrations. 
Whereas standard economic theory suggests that the movement of goods across borders can provide a substitute for the movement of production factors (such as labor), the more recent empirical evidence points toward a complementarity among the two phenomena. In particular many studies find quite robust evidences indicating that bilateral migration affects international trade flows \citep{JOES:JOES696,egger2012migration}.
Moreover, as argued for example in \citet{gould1994immigrant}, trade between any two countries may be enhanced by the stock of immigrants present in either country and coming from the other one.

Since the seminal contributions by Rauch and coauthors \citep[see for instance][]{rauch2002ethnic}, the main argument to rationalize these empirical findings is that formal and informal links among co-ethnic migrants in other countries and at home (the ``network'') facilitate trade by providing potential trading partners with easier access to information. The pro-trade effect thus stems from the reduction of the trade barriers and search costs associated with market transactions. Since these costs are likely to be larger for international trade due to distance, language and cultural differences, legal provisions and the like, networks end up being especially relevant in facilitating cross-border transactions.\footnote{This complementarity appears to hold for different countries \citep[for the US, Canada and Spain respectively, see][]{gould1994immigrant,head1998immigration,peri2010trade} and has recently been confirmed by a meta-analysis covering 48 different studies \citep{gen2011impact}} In fact we find that network effects result to be larger as the differences across economies increase.

As a corollary, the literature finds that the positive effect of migration on trade is larger for ``differentiated goods'', i.e. those items that are not homogeneous and are not traded in organized exchanges therefore rendering that knowledge about counterpart reputation particularly valuable.

Easier access to information via co-ethnic migrants is not only positively correlated with export from the recipient to the home countries, but also facilitates ``chain migration'': it is typical for communities with a significant presence of expatriates to attract more migrants from the same communities. This feature is consistent with the preferential attachment mechanism that accurately describes the evolution of many real-world networks (from airline traffic to the World Wide Web, from social ties to financial networks) and makes migration interesting in terms of complex network analysis. Yet, not much has been written on the subject: \citet{slater2008hubs} studied clustering in US internal migration, whereas \citet{simini2012universal} presented a stochastic radiation model that can be used to predict international migration patterns.
More recently, \citet{fama13pre} studied the topology of the IMN, and its evolution over the period 1960-2000. They find that the network (where links between two countries A and B are given by the stock of migrants originated in country A and living in country B in a given year) is disassortative and highly clustered, and displays a small-world binary pattern. Furthermore, they show that the structural properties of the network are mainly driven by socio-economic, geographical, and political factors.

Given the different characteristics of the available data and of the underlying phenomena, we first make the WTW and IMN comparable by using a hypergeometric benchmark. This allows us to evaluate whether the intensity of each link between any two countries is significantly higher than expected, relative to a purely random network. Next, we use the resulting (filtered) networks as the basis of our study and analyze their topological properties and main features. Similarly, we define a new product classification, that we later compare with Rauch's original distinction between homogeneous and differentiated goods, looking at the overlap between IMN and WTW.

Finally, we run a set of regressions where we control for network interdependencies. Most of the literature referred to above shares a common empirical strategy, based on the estimation of a log-linear gravity-type model where bilateral trade flows are regressed over standard explanatory variables (economic mass and distance), the stock of immigrants from specific partner countries and other controls capturing various trade costs. By means of spatial econometric techniques, we are able to account for network (auto)correlations between trade and migration, using the previously defined network matrices as weights. Hence, we use topological distance rather than the usual geographical space to measure possible spillover effects.

Therefore our approach allow us to define a new category of goods, starting from the similarity between WTW and IMN topologies \citep{Sgrignoli2013}, and then, using spatial econometric regressions, confirm them as the subset of products whose export/import is the most highly related to the presence of migrants communities \citep[see][for a complementary analysis on the WTW and IMN global patterns of correlation and where nodes centrality in the IMN is used to explain bilateral trade]{Fagiolo2014}.

The rest of the chapter is structured as follows. In Section \ref{chapt2_sec:data} we describe migration and trade data, as well as our methodological approach. In Section \ref{chapt2_sec:network} we analyze and compare the IMN and WTW by distinguishing different product types. In Section \ref{chapt2_sec:econ_app} we presents the econometric analysis and discuss the findings. Finally, Section \ref{chapt2_sec:concl} contains discussions of our main results, further research directions and conclusions.

\section{Data \& Methodology}\label{chapt2_sec:data}

\subsection{Migration and trade data}
Data regarding migrants come from the World Bank's Global Bilateral Migration dataset \citep{ozden2011earth}: it is composed of matrices of bilateral migrant stocks spanning 1960-2000 (5 census rounds), disaggregated by gender and based primarily on the foreign-born definition of migrants. It is the first and only comprehensive picture of bilateral global migration over the second half of the 20th century, taking into account a total of 232 countries.
The data reveal that the global migrant stock increased from 92 million in 1960 to 165 million in 2000. Quantitatively, migration between developing countries dominates, constituting half of all international migration in 2000, whereas flows from developing to developed countries represent the fastest growing component of international migration in both absolute and relative terms.

For international trade, we use the NBER-UN dataset described by \citet{NBERw11040}, disaggregated according to the Standardized International Trade Code at the four-digit level (SITC-4 rev. 2). For each country it provides the value (expressed in thousands of US dollars) exported to all other countries, for 775 product classes. In our analysis, we focus on year 2000 (and 1970 for some analysis), although choosing a different year does not qualitatively alter the results of this work.

Looking at the SITC product code of goods traded between each country pair allows us to apply Rauch's classification \citep{rauch2002ethnic} and distinguish among homogeneous and differentiated goods. The former are those that have a reference price, whether it being the result of organized exchanges or simply of price quotations in a specialized journal, while the latter lack it and can be thought of as ``branded'' commodities. An important and typical distinction between these two categories, that ease the understanding of their differences, is that homogeneous products price can be quoted without mentioning the name of the manufacturer, e.g. the prize of a ton of iron.
On the contrary, differentiated products are such that their commodity categories are usually not well defined (e.g. footwear): they need to be disaggregated into various sub-types, a process that leads to the limit where each category contains only one supplier. These products are, in this respect, ``branded'' or differentiated.

From this definition one would expect international trade to be more heavily influenced by migrant networks for this second class of products, as buyers and sellers need to be matched in the product characteristics space. This is indeed the result found by \citet{rauch2002ethnic} and one of the aspects we test in this chapter.

Using the two datasets together, we retain only the countries present in both of them to enhance comparability. From this matching we obtain a set of 146 countries (nodes), that populate both the WTW and IMN.

The controls used for the regressions in Section \ref{chapt2_sec:econometric-results} (e.g. contiguity, common language, distance, etc.) have been retrieved from the CEPII dataset \citep{mayer2011notes,head2010erosion}.

\subsection{Methodology}\label{chapt2_sec:method}
The different nature of the data, with regards to both type and measurement unit, make a direct comparison of the WTW and IMN unreliable. As such what is needed is a way to make the two datasets comparable.

The non-zero flow threshold is probably the simplest way to define edges; but, discriminating only on flows existence, literally throws away a lot of information contained in the links weights (i.e. the flows volume), that could carry important insights on the system.

In the following we want to introduce a different way of defining edges, based on \textit{hypergeometric filtering}: it is a stochastic benchmark for normalization purposes, as recently used in \citet{Sgrignoli2013,riccaboni2013global} and previously introduced in \citet{tumminello2011statistically}. This method has been used to identify statistically significant portions of data in fields ranging from genetics to network theory \citep{tavazoie1999systematic,wuchty2006stable} and to study the relatedness of corporation activities to understand their business coherence \citep{Teece1994a}.

In particular, for two countries, $A$ and $B$, let $N_A$ be the value of goods exported by country $A$ and $N_B$ the value of goods imported by country $B$. The total value of traded goods is $N_k$ and the observed value of goods exported from $A$ to $B$ is $N_{AB}$. Under the null hypothesis of random co-occurrence, i.e. customers in country $B$ are indifferent to the nationality of the exporter, the probability of observing $X$ thousands US dollars of goods traded is given by the hypergeometric distribution
\begin{equation}\label{chapt2_eq:hypergeo}
H(X|N_k,N_A,N_B) = \frac{\binom{N_A}{X} \binom{N_k-N_A}{N_B-X}}{\binom{N_k}{N_B}} \ ,
\end{equation}
and we can associate a \textit{p}-value with the observed $N_{AB}$ as
\begin{equation}\label{chapt2_eq:pvalue}
p(N_{AB}) = 1 - \sum_{X=0}^{N_{AB}-1} H(X|N_k,N_A,N_B) \ .
\end{equation}
Note that the described null hypothesis directly takes into account the heterogeneity of countries with respect to the total value of goods traded. For each pair of countries, we separately evaluate the \textit{p}-value and then use a cutoff ($\widetilde{p}$) to select only those links that represent a significant departure from the hypergeometric benchmark. The resulting matrices are then dichotomized.\footnote{The hypergeometric multi-urn benchmark is equivalent to the Monte Carlo degree-preserving network rewiring procedure \citep{maslov2002specificity}.} 

In other words this approach consist in comparing the observed value of trade flows to the value that would be expected under the hypothesis that trade partners are chosen randomly. The intuition is straightforward: $N_{AB}$ is expected to be larger for stronger trade channels, but it also can be expected to increase with $N_A$ and with $N_B$. Thus if $N_A$ and $N_B$ are large one would expect to see a large trade flow even if the link is not particularly important for any of the two countries involved. Conversely, if $N_A$ or $N_B$ is small one would not expect to see prominent flows of trade even if the link is relevant for one of the two economies. Hence the information in $N_{AB}$ about link importance can be extracted by comparing it to the amount of trade flows that would be observed for a given $N_A$, $N_B$ and $N_k$ if trade partners were randomly assigned.

We are therefore assuming that \textit{important trade partnerships are characterized by particularly large trade flows}. The only (exogenous) parameter to be set with this technique is the cutoff threshold $\widetilde{p}$: we use $\widetilde{p}=0.01$ as the \textit{p}-values resulting from our data are polarized at the two extremes, 0 and 1, with very few sparse values in between. The chosen value let us select the entire block at the low boundary, representing approximately the 13\% of the total links.

Thanks to our stochastic approach we can treat a weighted network as a non-weighted one, in which links represent a sensibly high connection with respect to randomly chosen connections. We apply this filter to both the WTW and IMN.

In the next Section we study the structural properties of these filtered WTW and IMN and then in Section \ref{chapt2_sec:econ_app} use them in place of distance matrices in a spatial gravity model of trade flows.

\section{Network analysis}\label{chapt2_sec:network}
In this section we first describe and compare some basic topological properties of the two networks separately (Section \ref{chapt2_sec:top_prop}); then analyze the interconnections and correlations among their links (Section \ref{chapt2_sec:links_corr}).

\subsection{Topological properties}\label{chapt2_sec:top_prop}
\begin{figure}[tb]
	\centering
	\caption[Cumulative degree distribution and ANND for the WTW and IMN]{Main: cumulative degree distribution $P_c(k)$ for the WTW and IMN. The dashed line is the degree distribution for a random graph with the same average degree. The solid line is a power law fit of the form $P_c(k) \sim k^{\gamma-1}$, with $\gamma\approx3.0$. Inset: Average Nearest Neighbors Degree (ANND) as a function of the total vertex degree. Year 2000.}
	\vspace{-20pt}
	\includegraphics[width=\myFiguresSize]{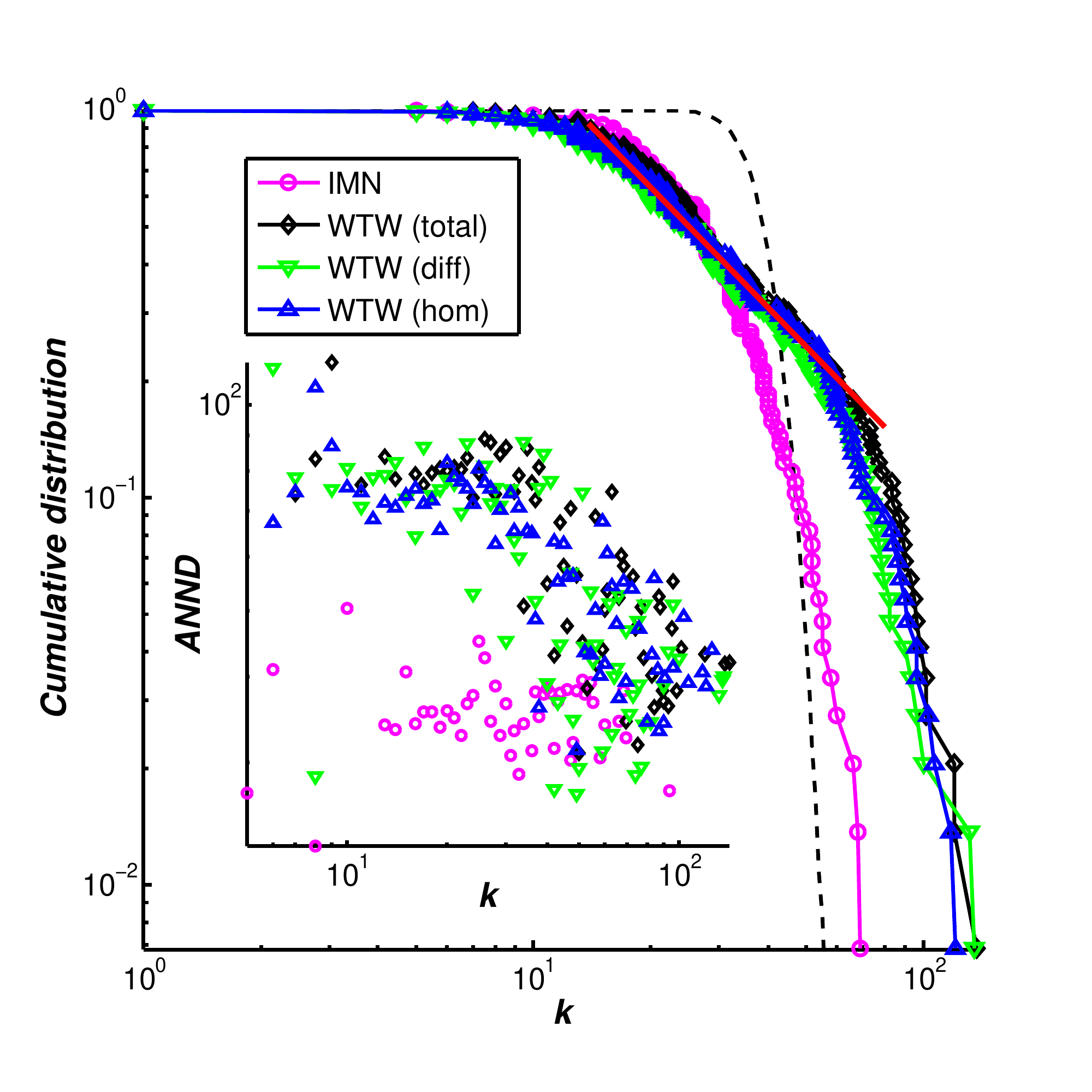}
	\vspace{-10pt}
	\label{chapt2_fig:annd}
\end{figure}

One of the basic and yet most important global topological properties of a network is its degree distribution, $P(k)$: it represent the probability of a randomly chosen vertex to have $k$ neighbors. Figure \ref{chapt2_fig:annd} (main) shows the undirected cumulative distribution of the filtered WTW and IMN, defined as $P_c(k)\equiv \sum_{k*>k}P(k*)$. Trade is also considered split by products categories according to Rauch's classification. In all cases the cumulative distributions show a flat approach to the origin, indicating the presence of a maximum at $k\sim10$ and, in this respect, similar to the Erd\"{o}s-R\'{e}nyi (ER) network; However, for $k>10$, the cumulative distributions follow a power law decay $P_c(k) \sim k^{\gamma-1}$, with $\gamma\approx3.0$, showing a strong deviation from the exponential tail predicted by the classical random graph theory. The exponent $\gamma$ is found to be within the range defined by many other complex networks \citep{serrano2003topology,Dorogovtsev2002}. In fact the distributions present a power law behavior just in a very short range, especially the migration one, strongly and rapidly deviating from it. This is probably due to the hypergeometric filtering that suppress part of the links and therefore a portion of the networks heterogeneity.

Table \ref{chapt2_tab:top} presents a selection of well-known statistics for these networks. We can observe that the three trade networks (total trade plus the two subsets) share very similar properties; while they are slightly different for the migration case. In particular, while the IMN has a lower connectivity, as indicated by the nodes' average degree and density, its links are more reciprocal, meaning that a bilateral bond is more common than in commercial trades. These results are consistent with the previous literature separately analyzing the two networks \citep[see for example][]{serrano2003topology,fama13pre}.\footnote{Some of the statistics are quantitatively different from what found in the literature, due to the reduced connectivity resulting from the application of the hypergeometric filter.}

Another important aspect is the hierarchical structure of the networks, which is usually analyzed by means of the clustering coefficient and degree-degree correlation. The \textit{clustering coefficient} of vertex $i$ is defined as $C \equiv 2n_i/k_i(k_i-1)$, where $n_i$ is the number of neighbors of $i$ that are interconnected. 

\begin{table}[tb]
	\centering
	\caption[WTW and IMN network properties]{WTW and IMN network properties, year 2000. D = Differentiated, H = Homogeneous.}
	\begin{tabular}{lcccc}
		\hline\hline
		\rule{0pt}{2ex} & IMN & WTW & WTW (D) & WTW (H)\\
		\hline
		\rule{0pt}{3ex}Average degree & 29.45 & 38.47 & 34.30 & 36.19\\
		Density (\%) & 15.1 & 20.4 & 18.5 & 19.3\\
		Corr. coeff. & 0.45 & 0.79 & 0.70 & 0.78\\
		Reciprocity (\%)& 34.5 & 30.2 & 28.2 & 29.1\\
		Average cluster coeff. & 0.099 & 0.088 & 0.073 & 0.089\\
		Assortativity & -0.037 & -0.399 & -0.356 & -0.391\\
		\hline\hline
	\end{tabular}
	\label{chapt2_tab:top}
\end{table}

The resulting values of the clustering coefficient for our networks are all $C \simeq 0.1$. In previous works, \citet{serrano2003topology} found the clustering coefficient for a binary version of the WTW to be $C=0.65$; however in their analysis they had limited data about only the forty most exchanged merchandises, thus truncating the network connectivity and biasing upward their result. In \citet{Fagiolo2008} $C$ is found to be varying from $\sim 0.001$ to $\sim 0.8$, depending if one considered the binary or weighted version of the network respectively. For the IMN, \citet{fama13pre} found $C=0.13 \sim 0.15$. Our results, shown in Table \ref{chapt2_tab:top}, seems to be compatible with these previous works, once accounted for the different strategies adopted.\footnote{Still, we have to recall that here we applied the hypergeometric filtering, hence probably biasing downward our results: as noted in \citet{riccaboni2013global}, clustering coefficient value decreases controlling for the hypergeometric benchmark, indicating that some of the triangles observed in the network show weak ties which do not stand up to the hypergeometric test.
Nevertheless, the loss of a small part of the clustering structure does not spoil the current analysis, which aim is not to study the detailed structures of the two networks \citep[for this see][]{Bhattacharya2008,Fagiolo2008,Fagiolo2009} but instead to determine the interplay between trade and migration, taking advantage of their main topological properties. In this sense, the hypergeometric filtering let us obtain a binary and more tractable version of the two networks, making them more easily comparable, yet exploiting all the information available in the data.}

Hierarchy is also reflected in the degree-degree correlation through the conditional probability $P(k|k')$, i.e. the probability that a vertex of degree $k'$ is linked to a vertex of degree $k$. This function is difficult to measure, due to statistical fluctuations, and it is usually substituted by the \textit{Average Nearest Neighbors Degree} (\textit{ANND}), defined as $\left\langle k_{nn}(k)\right\rangle =\sum_{k'}k'P(k'|k)$ \citep{pastor2001dynamical}. For independent networks this quantity would result independent of $k$. Figure \ref{chapt2_fig:annd} (inset) reports the ANND for the WTW and IMN, showing a dependency on the vertex's degree and indicating that in all the networks highly connected vertexes tend to connect to poorly connected vertexes, i.e. they show a \textit{disassortative} behavior. It is clear how this phenomenon is much more pronounced in trade than in migration, as can also be seen by the assortativity coefficients\footnote{Also known as Pearson's degree correlation.} in Table \ref{chapt2_tab:top}.

\subsection{The Relation between migration and trade}\label{chapt2_sec:links_corr}
A first approach to analyzing the interplay between international trade and human migration is to observe whether, for a pair of countries, a strong bond on one side corresponds to a strong bond on the other. To see this we make use of the Jaccard index \citep{jacc1901} that, given two sets of events $A$ and $B$, is defined as $J=\left| A\cap B\right| /\left| A\cup B\right|$, therefore representing the ratio between the number of events shared by the two sets, over the number of events in at least one of them. In our case the events will be represented by significant\footnote{Significant in the sense of Section \ref{chapt2_sec:method}, so to say with the hypergeometric filter \textit{p}-value below the $1\%$ threshold.} links between two countries, both in the WTW and IMN.

\begin{figure}[tb]
	\centering
	\caption[Jaccard index for product overlapping and number of overlaps in individual countries]{Main figure: Jaccard index relative to the overlapping of single products with the IMN for years 1970 and 2000 vs. product rank. Red lines are power law fits with exponents $\chi=-0.15$ and $\chi=-0.11$ respectively. Inset: Number of overlaps between the WTW and IMN for individual countries vs. country rank. All cases show a Zipf law behavior.}
	\vspace{-20pt}
	\includegraphics[width=\myFiguresSize]{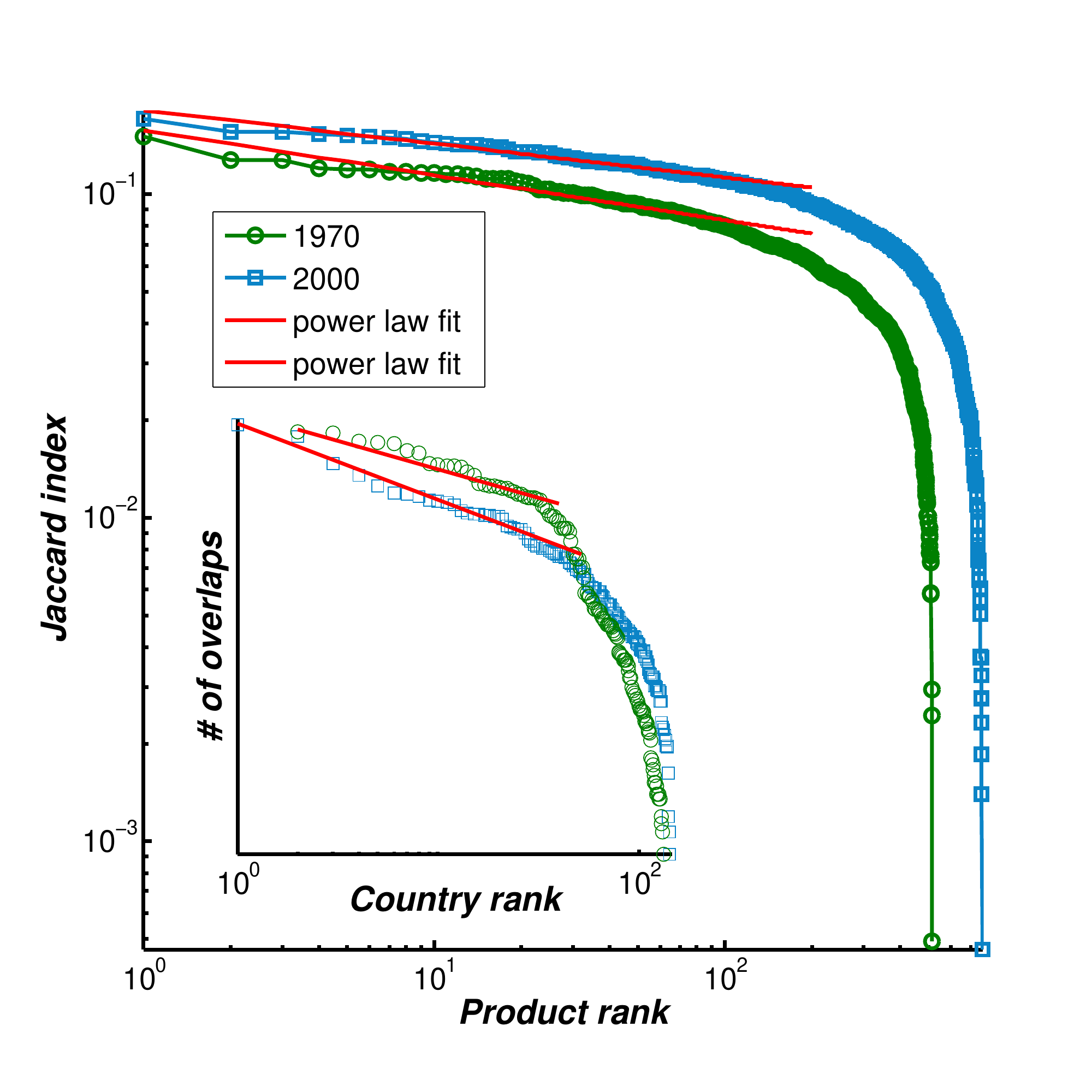}
	\vspace{-10pt}
	\label{chapt2_fig:total}
\end{figure}

In Figure \ref{chapt2_fig:total} (main) we show the relation of the Jaccard index ($J$) with product rank\footnote{We intend rank in the sense of our overlaps analysis, see Tables \ref{chapt2_tab:products_list} and \ref{chapt2_tab:countries_list}.} for all the individual products in the years 1970 and 2000, along with power law fits with exponents $\chi=-0.15\pm0.02$ and $\chi=-0.11\pm0.02$ respectively. For the year 2000, the overlap with migration is generally higher. In the inset we plot the total number of overlaps in the networks for individual countries vs. the country rank, again with power-law fits with exponents $\chi=-0.50\pm0.02$ and $\chi=-0.62\pm0.03$. All the cases show a Zipf law behavior, i.e. Jaccard index (number of overlaps) is proportional to product (country) $rank^{-\chi}$.

\begin{table}[tb]
	\centering
	\caption[List of the countries with the most and least overlap between the WTW and IMN]{List of the countries with the most and least overlap between the WTW and IMN, based on the Jaccard index. Year 2000.}
	\begin{tabular}{cl}
		\hline\hline
		\rule{0pt}{2ex}\textbf{\#} & \textbf{Country}\\
		\hline
		\rule{0pt}{3ex}1 & Germany\\
		2 & United Kingdom\\
		3 & Italy\\
		4 & Spain\\
		5 & Netherlands\\
		6 & Belgium\\
		7 & United States\\
		$\cdots$ & $\cdots$\\
		11 & China\\
		$\cdots$ & $\cdots$\\
		142 & Sierra Leone\\
		143 & Somalia\\
		144 & Saint Pierre and Miquelon\\
		145 & Tanzania, United Republic of\\
		146 & Uganda\\
		\hline\hline
	\end{tabular}
	\label{chapt2_tab:countries_list}
\end{table}

\begin{table}[tb]
	\centering
	\caption[List of the products with the most and least overlap with the IMN]{List of the products (SITC-4) with the most and least overlap with the IMN, based on the Jaccard index. Year 2000. D = Differentiated, H = Homogeneous.}
	\begin{adjustbox}{width=\textwidth,keepaspectratio}
		\begin{tabular}{clc}
			\hline\hline
			\rule{0pt}{2ex}\textbf{\#} & \textbf{Product} & \textbf{Type}\\
			\hline
			\rule{0pt}{3ex}1 	& Miscellaneous articles of plastic	& D \\
			2 	& Insulated electric wire. cable. bars. etc	& D \\
			3 	& Plastic packing containers. lids. stoppers and other closures	& D \\
			4 	& Switches. relays. fuses...; switchboards and control panels... & D \\
			5 	& Edible products and preparations. nes	& D \\
			6 	& Chemical products and preparations. nes & D \\
			7 	& Other furniture and parts thereof. nes & D \\
			8 	& Machinery for specialized industries and parts thereof. nes & D \\
			9 	& Other polymerization and copolymerization products & D \\
			$\cdots$ & $\cdots$ & $\cdots$\\
			762 & Copra & H \\
			763 & Palm nuts and kernels & H \\
			764 & Manila hemp. raw or processed but not spun. ... & H\\
			765 & Uranium depleted in U235. thorium. and alloys. nes; ... & H \\
			766 & Ores and concentrates of uranium and thorium & H \\
			767 & Castor oil seeds & D \\
			768 & Coal gas. water gas and similar gases & H \\
			769 & Wood-based panels. nes & D \\
			770 & Knitted or crocheted fabrics. elastic or rubberized & D \\
			\hline\hline
		\end{tabular}
	\end{adjustbox}
	\label{chapt2_tab:products_list}
\end{table}

In Tables \ref{chapt2_tab:countries_list} and \ref{chapt2_tab:products_list} we also list the top and bottom countries and products resulting from our overlapping ranking.

\begin{table}[!tb]
	\centering
	\caption[List of the top overlapping products for selected countries]{List of the top overlapping products for the selected set of countries exports.}
	\begin{adjustbox}{width=\textwidth,totalheight=\textheight,keepaspectratio}
		\begin{tabular}{llccc}
			\hline\hline
			\rule{0pt}{2ex}\textbf{Country} & \textbf{Product description} & \textbf{SITC rev2} & \textbf{RCA\_J} & \textbf{Trade vol. (\$ x 1000)} \\
			\hline
			\rule{0pt}{3ex} & Rye. unmilled & 0451 & 7,40 & 151.630 \\
			& Mechanically propelled railway. tramway. trolleys. etc & 7913 & 5,10 & 268.669 \\
			& Domestic dishwashing machines & 7753 & 3,26 & 547.025 \\
			Germany & Steam power units (mobile engines but not steam tractors. etc) & 7126 & 2,97 & 163.593 \\
			& Other fixed vegetable oils. soft & 4239 & 2,91 & 190.492 \\
			& Rape and colza seeds & 2226 & 2,77 & 164.289 \\
			& Other wheat and meslin. unmilled & 0412 & 2,67 & 562.651 \\
			& & & &\\
			& Bacon. ham. other dried. salted or smoked meat of domestic swine & 0121 & 3,46 & 158.457 \\
			& Machines for extruding man-made textile; other textile machinery & 7244 & 2,47 & 501.709 \\
			& Olive oil & 4235 & 2,43 & 712.817 \\
			Italy & Fabrics. woven. of silk. of noil or other waste silk & 6541 & 2,30 & 233.199 \\
			& Pins. needles. etc. of iron. steel; metal fittings for clothing & 6993 & 2,14 & 128.604 \\
			& Motor vehicles piston engines. headings: 722; 78; 74411 and 95101 & 7132 & 2,11 & 477.544 \\
			& Gas turbines. nes & 7148 & 2,10 & 211.731 \\
			& & & &\\
			& Raw silk (not thrown) & 2613 & 10,72 & 209.056 \\
			& Natural calcium phosphates. natural aluminium. etc & 2713 & 5,29 & 123.881 \\
			& Fine animal hair. not carded or combed & 2683 & 5,00 & 398.841 \\
			China & Sesame seeds & 2225 & 4,70 & 81.881 \\
			& Anthracite. not agglomerated & 3221 & 4,65 & 187.453 \\
			& Synthetic or reconstructed precious or semi-precious stones & 6674 & 4,27 & 74.667 \\
			& Railway. tramway passenger coaches. etc. not mechanically propelled & 7914 & 3,91 & 74.377 \\
			& & & &\\
			& Aircraft of an unladen weight exceeding 15000 kg & 7924 & 5,64 & 21.418.960 \\
			& Chemical wood pulp. dissolving grades & 2516 & 4,96 & 299.909 \\
			& Nuclear reactors. and parts thereof. nes & 7187 & 4,83 & 555.022 \\
			United States & Other rail locomotives; tenders & 7912 & 4,38 & 462.136 \\
			& Cellulose acetates & 5843 & 3,87 & 198.565 \\
			& Durum wheat. unmilled & 0411 & 3,61 & 539.424 \\
			& Other wheat and meslin. unmilled & 0412 & 3,16 & 2.141.316 \\
			& & & &\\
			& Mate & 0742 & 7,65 & 28.138 \\
			& Waxes of animal or vegetable origin & 4314 & 5,98 & 36.684 \\
			& Iron ore agglomerates & 2816 & 5,21 & 1.263.631 \\
			Brazil & Raw silk (not thrown) & 2613 & 4,81 & 26.401 \\
			& Aircraft of an unladen weight from 2000 kg to 15000 kg & 7923 & 4,40 & 2.480.803 \\
			& Tobacco refuse & 1213 & 4,20 & 37.114 \\
			& Armoured fighting vehicles. war firearms. ammunition. parts. nes & 9510 & 4,04 & 34.365 \\
			& & & &\\
			& Cork. natural. raw and waste & 2440 & 50,01 & 25.948 \\
			& Zinc ores and concentrates & 2875 & 19,06 & 44.717 \\
			& Cork manufactures & 6330 & 15,40 & 9.169 \\
			Morocco & Molasses & 0615 & 9,02 & 7.941 \\
			& Vegetable products roots and tubers. nes. fresh. dried & 0548 & 8,77 & 5.387 \\
			& Lead. and lead alloys. unwrought & 6851 & 8,25 & 33.139 \\
			& Fuel oils. nes & 3344 & 7,62 & 102.610 \\
			\hline\hline
		\end{tabular}
	\end{adjustbox}
	\label{chapt2_tab:RCA_J_products}
\end{table}

In order to better investigate whether the connection between trade and migrants is comparable among countries or whether it exist for the same product in different places, we calculated a Jaccard index between WTW and IMN, disaggregated by product and country, obtaining a \textit{country} $\times$ \textit{products} matrix of Jaccard indexes. To compare them we apply a measure similar to Balassa's Revealed Comparative Advantage \citep{bala65} to Jaccard indexes themselves, in order to identify for each country the specific exports that are more related to migration stocks. Define $RCA_J \equiv RCA(J_{p,c}) \equiv \frac{J_{p,c}/\sum_{p}J_{p,c}}{\sum_{c}J_{p,c}/\sum_{p,c}J_{p,c}}$, where $J_{p,c}$ is the Jaccard index specific to country $c$ and product $p$. In Table \ref{chapt2_tab:RCA_J_products} we list the top products resulting by this new index for a selection of countries: by the very low superposition among these lists, our method highlights how migrants with different home countries are correlated with trade in different product categories, letting us identify each country with different characteristic trade footprints.

\begin{figure}[tb]
	\centering
	\caption[Jaccard index separately for differentiated and homogeneous products and total, for different years]{Main figure: Jaccard index relative to the overlapping of single products with the IMN vs. product rank, shown separately for \textit{differentiated} and \textit{homogeneous} products, for years 1970 and 2000. Inset: Jaccard index relative to the overlap of the entire WTW and IMN for different years. A distinction is made between all the products and just differentiated or homogeneous ones.}
	\vspace{-20pt}
	\includegraphics[width=\myFiguresSize]{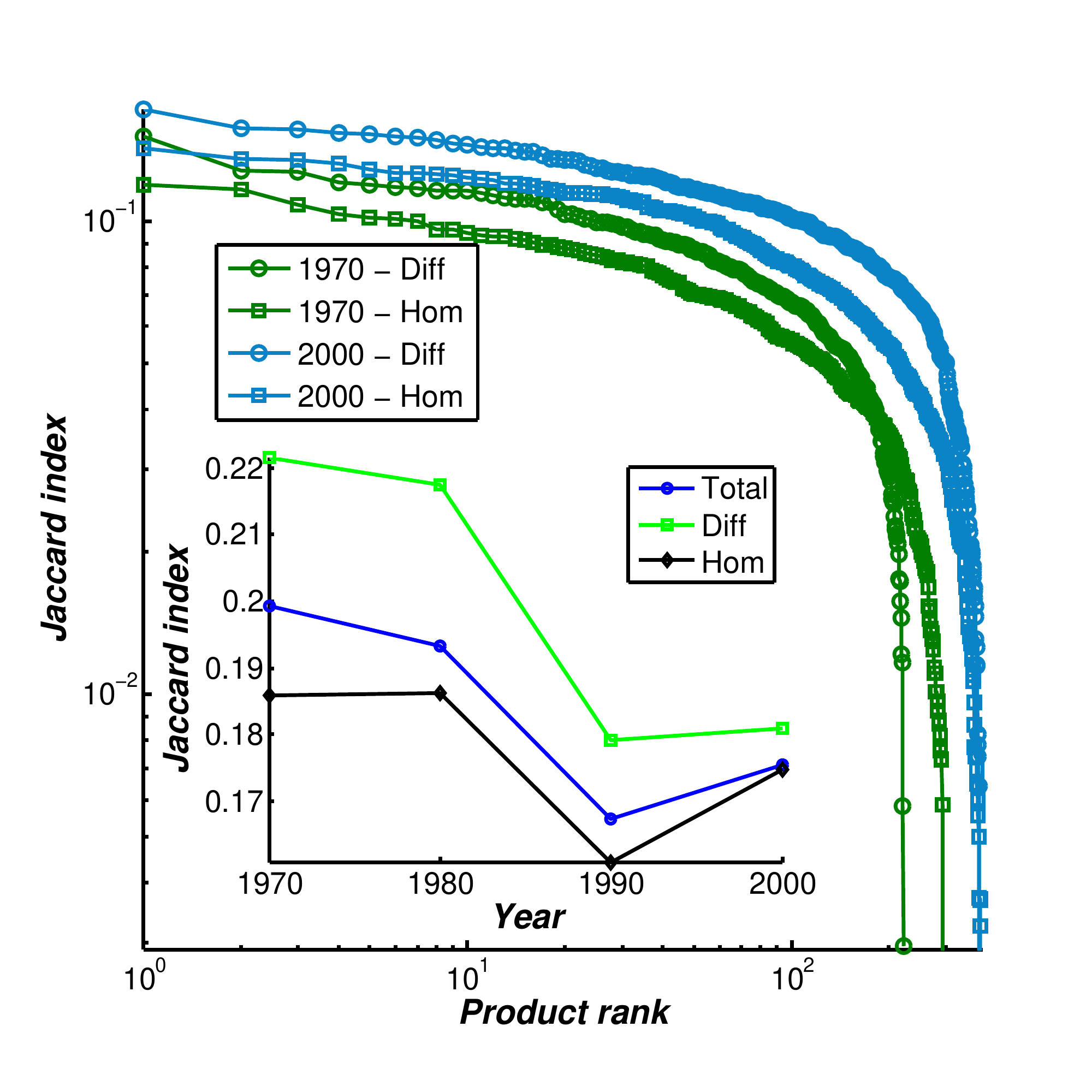}
	\vspace{-10pt}
	\label{chapt2_fig:diff}
\end{figure}

To verify the validity of the Rauch classification \citep{rauch2002ethnic} we repeated the previous analysis separating \textit{differentiated} and \textit{homogeneous} goods: in the main panel of Figure \ref{chapt2_fig:diff} we plot the Jaccard index for individual products vs. the product rank; while in the inset we show the Jaccard index for the entire networks in the years 1970-2000. As one can observe, in this first analysis the Rauch classification seems to be verified by our results, with values for the differentiated products generally higher for both individual SITC codes and the whole networks.

\subsection{Differentiated vs. overlapping products}
We want now to introduce a new product classification based on the level of superposition between WTW and IMN: we expect that the products for which a higher number of intense links co-exist in the two networks should also be those for which migration and trade are more related. To make our classification more comparable with Rauch's one, we keep the same number of products present there in the two categories: being N the number of differentiated goods in Rauch's classification, we take the \emph{top-N} products from our Jaccard-ranked list, i.e. Table \ref{chapt2_tab:products_list}, and compare them with Rauch's differentiated ones, doing the same for the bottom ones with Rauch's homogeneous commodities.
We call these two new categories \emph{overlapping} and \emph{non-overlapping}.

\begin{figure}[tb]
	\centering
	\caption[Distribution of the Jaccard index for all countries and all products]{Distribution of the Jaccard index for all countries and all products relative to the two years analyzed, 1970 and 2000. The red vertical line indicates the threshold we used to define our classification of overlapping and non-overlapping products in order to have in the two categories the same number goods as there were in Rauch's classification.}
	\vspace{-10pt}
	\includegraphics[width=\myFiguresSize]{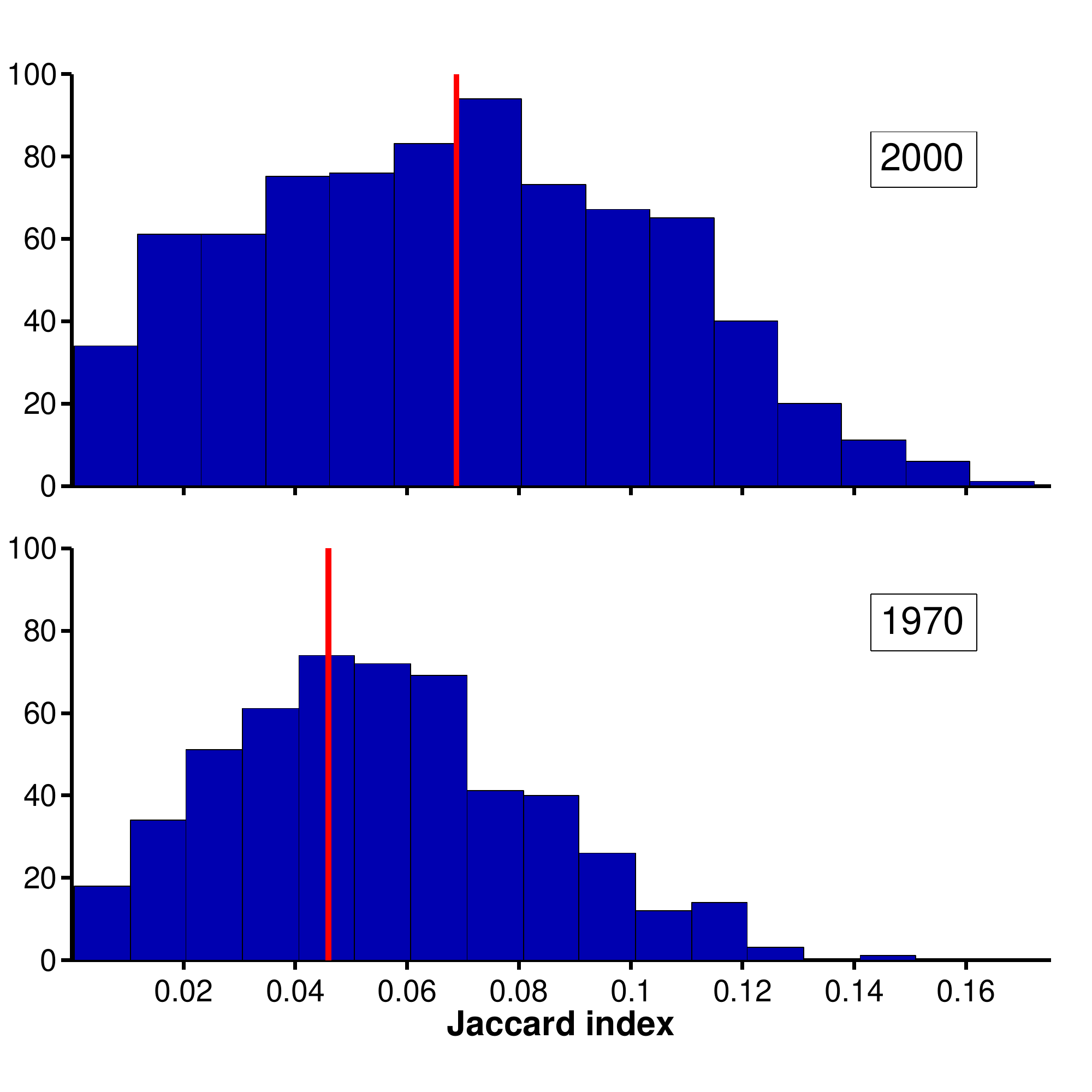}
	\vspace{-10pt}
	\label{chapt2_fig:jaccard}
\end{figure}

To check the validity of taking the number of Rauch's differentiated products as the dimension of our overlapping class, we plot in Figure \ref{chapt2_fig:jaccard} the Jaccard index distribution for the WTW, with the red line indicating the value for the N-th product. In other words, on the right of the red line we have the N products for which the superposition of WTW and IMN is the highest and that we will call \textit{overlapping} products; while on the left lay those for which the Jaccard index between WTW and IMN is low and therefore we will call \textit{non-overlapping} products. In both cases, for years 1970 and 2000, our threshold is reasonably close to the mean of the distribution, which justify our criteria to separate the two product classes.

\begin{table}[!tb]
	\centering
	\caption[List of Rauch's homogeneous products appearing in the top-50 overlapping products]{List of Rauch's homogeneous products appearing in the top-50 overlapping products.}
	\begin{adjustbox}{width=\textwidth,keepaspectratio}
		\begin{tabular}{cclc}
			\hline\hline
			\rule{0pt}{2ex}\textbf{\#} & \textbf{SITC rev2} & \textbf{Product} & \textbf{Type}\\
			\hline
			13 & 6842 & Aluminium and aluminium alloys. worked & H \\
			21 & 6421 & Packing containers. box files. etc. of paper. used in offices & H \\
			24 & 7788 & Other electrical machinery and equipment. nes & H \\
			26 & 5831 & Polyethylene & H \\
			31 & 5922 & Albuminoid substances; glues & H \\
			35 & 5832 & Polypropylene & H \\
			37 & 6822 & Copper and copper alloys. worked & H \\
			39 & 7781 & Batteries and electric accumulators. and parts thereof. nes & H \\
			43 & 6996 & Miscellaneous articles of base metal & H \\
			48 & 6911 & Structures and parts of. of iron. steel; plates. rods. and the like & H \\
			50 & 6924 & Cask. drums. etc. of iron. steel. aluminium. for packing goods & H \\
			\hline\hline
		\end{tabular}
	\end{adjustbox}
	\label{chapt2_tab:homog_in_our_diff}
\end{table}

\begin{figure}[tb]
	\centering
	\caption[Log of real GDP per capita vs log of the number of overlaps between WTW and IMN]{Log of real GDP per capita vs log of the number of overlaps between WTW and IMN. Marker size proportional to node centrality in the WTW, while color (from blue to red) proportional to the log of the country population. Year 2000.}
	\vspace{-20pt}
	\includegraphics[width=\myFiguresSize]{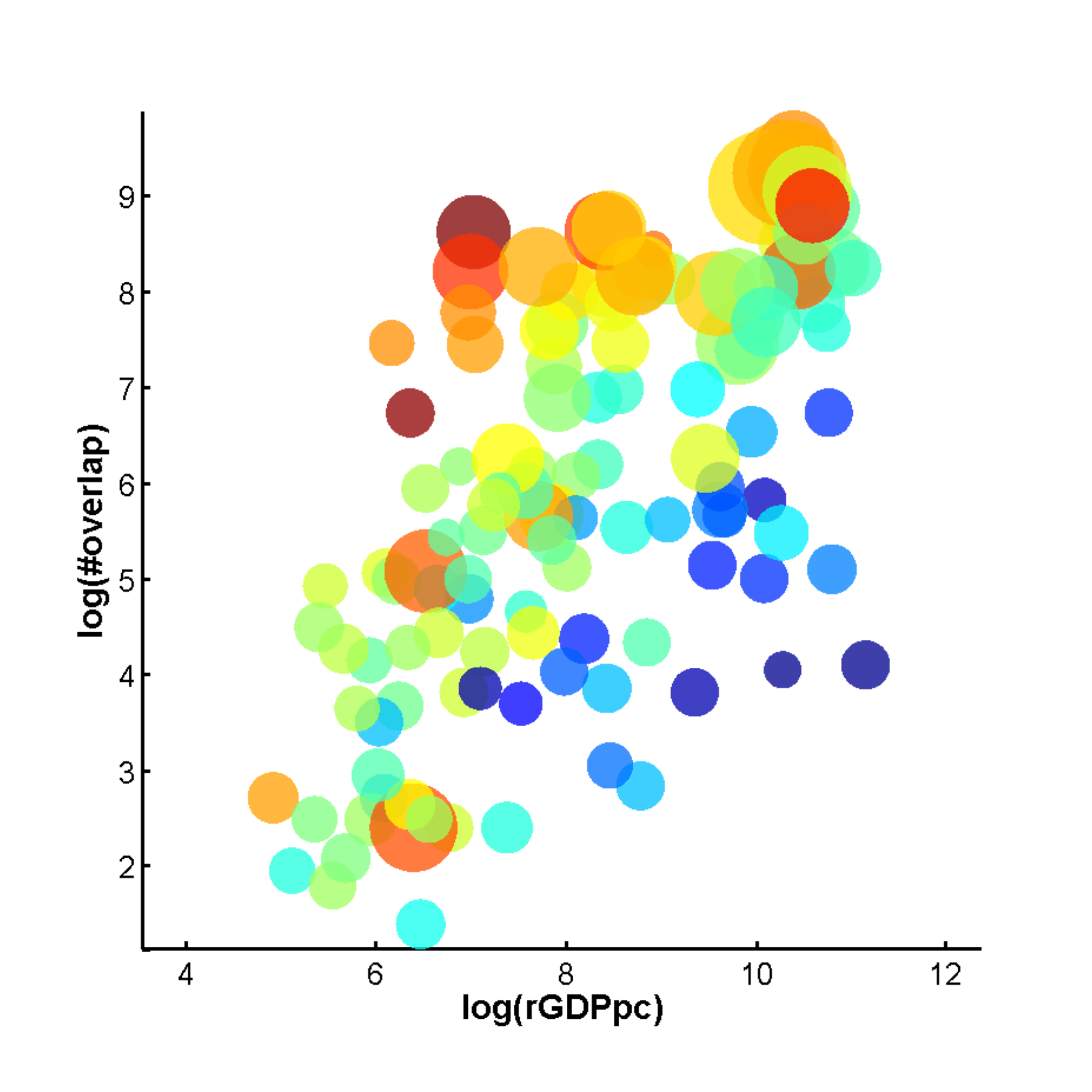}
	\vspace{-10pt}
	\label{chapt2_fig:bubble2}
\end{figure}

In our new classification, for the year 1970 the overlapping (non-overlapping) products are composed by $54\%-46\%$ ($27\%-73\%$) of Rauch's differentiated --- homogeneous products; while for the year 2000 the figure changes to $62\%-38\%$ ($37\%-63\%$). Even if the agreement between the two classification is in general quite good, they are not always the same: given their different origin it is common to find differentiated products among our non-overlapping ones and vice-versa. While evidences of the former occurrence can already be seen in Table \ref{chapt2_tab:products_list}, we want to further analyze the case where products that were classified as homogeneous by Rauch (i.e. with a reference price, therefore whose trade in his reasoning, should be less stimulated by migration) appear in our overlapping class of products.

In Table \ref{chapt2_tab:homog_in_our_diff} we list such products found among the first 50 of our ranking. It is important to point out that all of them are \emph{intermediate products}. This result about intermediate products was already known in the literature \citep[see][]{mundra2005immigration} and while Rauch's classification fails on this point, our method let us identify their behavior correctly.

To further investigate this point, in Section \ref{chapt2_sec:econ_app} we compare the results for the classification proposed by Rauch and the one that emerges from the overlapping of the WTW and IMN (see Table \ref{chapt2_tab:products_list}).

Moreover, the number of links that result to be strong in both networks, i.e. overlaps, is related to other country indicators: in Figure \ref{chapt2_fig:bubble2} the log of GDP is plotted versus the log overlaps number. Note how highly populated countries (red colored), at fixed GDP per capita, tend to show more overlapping of WTW and IMN. This can be thought as larger populations having more emigrants, thus, independently from their economic power, more ties on which they can influence trade.
Also, countries that are more central in the trade network, have in general a higher number of overlaps.
The figure indicate that some relation among these indicators exist; however further analysis is required to better understand their intensity, implications and scope.
In the next section we perform a comprehensive econometric analysis to quantitatively investigate the causal links among trade, migration, using the networks introduced in this section to define the spatial relations among countries.
In addition, we will also compare the results for our and Rauch's classification, to check which one is more explanatory with respect to the relation between international trade and migration.

\section{Econometric analysis}\label{chapt2_sec:econ_app}
In the econometric literature, the standard way to investigate the empirical effect of migration on bilateral trade flows is the estimation of a gravity model augmented with the stock of migrants. A large trunk of the literature covers both the theoretical foundations \citep{anderson1979theoretical,bergstrand1985gravity,anderson2003gravity,ande11}, as well as the proper econometric specification \citep[see for instance][]{egger2000note,bata06,hema13} of the gravity model, with its baseline describing trade links intensity as a function of the distance plus variables pertaining to country-specific characteristics.

In the standard gravity model all trade links are treated as independent observations, that is to say no network interdependence is allowed. Nevertheless it has been recently fully recognized that a spatial interaction effect, not included in the geographical distance, exists \citep{porojan2001trade}, essentially due to the spatial spillover and third country effects. Such idea is exploited here to estimate spatial interdependence: by defining a squared matrix ($W$) that identifies neighborhood structures we allow for network interactions. More specifically, we borrow from spatial econometric techniques and, using the weight matrix describing the WTW ($W^{(T)}$), we take into account the neighborhood structure as defined by the trade flows intensity among nodes. 

Therefore, in analyzing the trade link between country $i$ and $j$, we also consider neighbors interactions. That is to say, we assume trade between nodes $i$ and $j$ to depend on link-specific characteristics, as well as on characteristics of the set of network neighbors of $i$ ($N(i)$) and $j$ ($N(j)$). In this way we can identify and separately analyze the direct and indirect impact of regressors on trade. By doing this we explicitly take into account network interdependence in our regression model framework: when we investigate direct trade between $i$ and $j$ ($i \rightarrow j$), we also control for characteristics of the links from the neighbors of $i$ to $j$ ($N(i) \rightarrow j$) and  for characteristics of the links from the neighbors of $j$ to $i$ ($N(j) \rightarrow i$).

There is no unique view on the exact specification of the weight matrix to use: it has recently been argued that it can be either spatial\footnote{\citet[See ][ for a list of many possible definitions.]{anselin1988spatial}} or \textit{non}-spatial. In the latter case many proposals have been advanced in the literature: \citet{case1993budget} use the regional differences in per-capita income, \citet{behrens2012dual} employ the relative size of regions as reflected by population shares, or \citet{lesage2011pitfalls, elhorst2012model} even discuss about jointly modeling spatial and non-spatial dependencies through a double autoregressive component that make use of two different weight matrix specifications.

We instead propose the use the trade binary matrix defined in \ref{chapt2_sec:method} to define the neighborhood structure, in order to exploit the largest amount of information available from our data and the WTW network topology, as opposed to all previously proposed specifications and the more commonly used spatial distance.
Therefore the specification of our $W^{(T)}$ matrix becomes:
$$
W^{(T)}:=
\left\{ 
\begin{array}{rl}
w_{i,j} = 1, & \mbox{ if \textit{i} significantly exports to \textit{j}}\\
w_{i,j} = 0, & \mbox{ otherwise.}
\end{array}
\right.
$$

Moreover, to identify \textit{origin} and \textit{destination} neighborhood effects as discussed above, we need to redefine our proximity matrix as the sum of the Kronecker product of $W^{(T)}$ with the identity matrix $I$ and the Kronecker product of $I$ with $W^{(T)}$, i.e.
$$
W^{(T)} := W^{(T)} \oplus W^{(T)} = W^{(T)} \otimes I + I \otimes W^{(T)} =: W^{(T)}_o + W^{(T)}_d \ ,
$$
thus ending up with $n^2 \times n^2$ matrix \citep{lesage2008spatial}.

\subsection{Model specification}\label{chapt2_sec:mod_spec}
The analysis of the spatial (network) \textit{origin-destination} trade flows can be based on two classes of models \citep{lesage2008spatial}: \emph{Spatial autoregressive models} (SAR) and \emph{Spatial Durbin / Spatial error models} (SDM/SEM). The former consists in the inclusion either of a spatially lagged dependent variable or of a spatial autoregressive process in the residual term, motivated by significant spatial autocorrelation in the dependent variable; the latter is best described as a proxy for missing variables that follows a meaningful spatial pattern.\footnote{Intended in a broad sense which may include non-geographical distance.} The Durbin model can also take into account both the spatially lagged dependent variable and the spatial autoregressive process in the residuals: this is the augmented version of the SDM model (also called \textit{Manski}) that accounts for all possible spatial dependency, and can be written as:
\begin{equation}\label{chapt2_eq:sdm2}
y = \rho W y + X \beta + W X \gamma + \lambda W\epsilon + \epsilon\ ,
\end{equation}
where $y$ is the vector of the dependent variable, $X$ is the matrix of the explanatory variables, $\epsilon$ represents the vector of the stochastic residuals, $W$ is the weight matrix\footnote{As discussed above, as weight matrix $W$ we will use our specification $W^{(T)}$.} and $\beta, \gamma, \lambda $ and $\rho$ are the coefficients to be estimated. From now on, with the term SDM or Durbin we will refer to this last specification of SDM model.

To compute this formulation of the spatial gravity model, the Concentrated Maximum Likelihood (CML) estimator, as proposed by \citet{anselin1988spatial} and revised by \citet{lesage2008spatial}, is the most commonly used because it overcomes the issue of intrinsic endogeneity emerging from dependent variable's spatial (network) lag inclusion among the regressors (which make the OLS estimator not correct anymore). Moreover it permits to calculate the direct and indirect impacts for each explanatory variable \citep{pace2009introduction}, where the former is related to the dyad, while the latter express to network spillovers. In fact, the change of an explanatory variable in a single dyad value, affects the dyad itself but potentially all other dyads too. This rich set of information increases the difficulty of interpreting the results and therefore make necessary the calculation of direct and indirect impacts, as proposed by \citet{pace2009introduction}.

The CML method is based on the log-determinant, which can be computed with new approximation algorithm \citep{pace2004chebyshev, barry1999monte, smirnov2001fast}; nonetheless IV / GMM estimation techniques \citep{kelejian1998generalized, kelejian1999generalized} have also been proposed as an alternative to CML, being them less computation demanding. This class of estimators does not need for residuals to be normal, however they are only asymptotically correct.

Therefore we choose to employ the SDM model with teh CML estimator, as these let us control for both dyad and lagged explanatory variables in a consistent way.

As additional controls, GDP per capita is generally used as a proxy to control for purchasing power of importing and exporting countries, or for their endowment ratio \citep{sohn2005does, bergstrand1985gravity}, while population serves as a control for country size. These variables can be used either by including both the origin and the destination value, or by using a pair-specific measure, e.g. the sum of GDP or populations. Following \citet{baltagi2007estimating} we make use of this second formulation, as it allows a better interpretation of our variables of interest. Besides GDP and population, we also use the stock of migrants and a number of standard controls like countries contiguity, language commonality, presence of free trade agreements, etc.
Our gravity specification uses GDP per capita to control for purchasing power and population to control for country size. The log-in-log model variables are constructed as follows: $gdpcapsum_{ij} = log(gdpcap_i + gdpcap_j)$ and $popsum_{ij} = log(pop_i + pop_j)$ indicate the bilateral purchase power and size; $gdpcapsim_{ij} = (1-(\frac{gdpcap_i}{gdpcap_i + gdpcap_j}) ^2 - (\frac{gdpcap_j}{gdpcap_i + gdpcap_j}) ^2)$ and $popsim_{ij} = (1-(\frac{pop_i}{pop_i + pop_j}) ^2 - (\frac{pop_j}{pop_i + pop_j}) ^2)$ are GDP and population similarity.\footnote{This specification follows that of \citet{helpman1985market, Helpman198762, baltagi2007estimating}.} We also include controls for contiguity ($contig$), common language ($comlang$), common currency ($comcur$), colony relationship ($colony$) and regional trade agreements ($rta$). Moreover distance between countries (\textit{distw}) is taken from GeoDist in the specification weighted with the geographic distribution of population: the basic idea is to calculate distance between two countries based on bilateral distances between the biggest cities of those two countries, those inter-city distances being weighted by the share of the city in the overall country's population.\footnote{Nevertheless our results are robust to the use of different distance definitions.} Migration data (\textit{migrant}) represent the stock of migrants originating in country $i$ and present in destination $j$, where migrant status is consistently defined in terms of country of birth.

Notice that since we use logs, in particular of migration stocks and trade flows, we automatically fit only strictly-positive links, i.e. we only consider non-zero weights. In fact fitting a CML estimator on a log-log gravity model generally disregard the presence of zero trade flows and, to the best of our knowledge, no extension of this approach exists that combines it with spatial autoregressive models

All the regressions will be conducted on cross-section data relative to year 2000,\footnote{We did not employ panel data regressions as the years for which the trade and migration datasets overlap are too few for practical purposes. Instead we prefer to concentrate our analysis on the year 2000, the most recent available. This strategy let us isolate our results from the past decades and, concentrating on the latest mechanisms of interplay between the two phenomena, make them more usable for actual policy enforcement. On the other hand, in Chapter \ref{chapt:migr2} we will use pooled panel regressions.} on bilateral trade flows from 146 world countries. As dependent variables we employ five different product categories export trade flows: $(i)$ total; $(ii)$ differentiated goods; $(iii)$ non-differentiated goods; $(iv)$ overlapping goods and $(v)$ non-overlapping goods.

\subsection{Results}\label{chapt2_sec:econometric-results}
We first present a preliminary baseline model, estimated by ordinary least square (OLS) method:
\begin{equation}\label{chapt2_eq:ols}
T = \sum_{k=1}^{K} X_k \beta_k + \epsilon \ ,
\end{equation}
where $T$ represents the $n^2 \times 1$ vector of trade flows between each possible country-pair, $\beta_k$ are the $k \times 1$ vectors of coefficients related to the $n^2 \times k$ matrix of the explanatory variables $X_k$ and $\epsilon$ is the $n^2 \times 1$ vector of residuals.

\begin{table}[tb]
	\centering
	\caption[OLS regression results]{Regression results for the OLS. Year 2000.}
	\begin{adjustbox}{width=\textwidth,keepaspectratio}
		\begin{tabular}{l|ccccc}
			\hline\hline
			& Total & Differentiated & Homogeneous & Overlapping & Non-Overlapping \\
			\hline
			migrant & 0.072*** & 0.054*** & 0.078*** & 0.085*** & 0.059*** \\
			distw & -0.720*** & -0.652*** & -0.697*** & -0.612*** & -0.686*** \\
			rta & 0.224* & 0.340*** & 0.161* & 0.089 & 0.290** \\
			gdpcapsum & 1.806*** & 1.797*** & 1.590*** & 1.415*** & 1.846*** \\
			gdpcapsim & 0.914*** & 0.869*** & 0.817*** & 0.717*** & 0.919*** \\
			popsum & 3.545*** & 3.445*** & 3.194*** & 2.911*** & 3.540*** \\
			popsim & 1.691*** & 1.652*** & 1.571*** & 1.437*** & 1.697*** \\
			contig & 0.238*** & 0.223*** & 0.266*** & 0.321*** & 0.266*** \\
			comlang & 0.203*** & 0.163*** & 0.233*** & 0.213*** & 0.187*** \\
			colony & 0.288*** & 0.239*** & 0.216*** & 0.114*** & 0.273*** \\
			comcur & 0.164*** & 0.306*** & 0.161*** & 0.126*** & 0.219*** \\
			\hline
			$R^2$-adjusted & 0.669 & 0.625 & 0.611 & 0.547 & 0.658 \\
			\hline
			observations & 6829 & 5845 & 6270 & 5447 & 6404 \\
			\hline\hline
			\multicolumn{6}{l}{\footnotesize \sym{\cdot} \(p<0.1\), \sym{*} \(p<0.05\), \sym{**} \(p<0.01\), \sym{***} \(p<0.001\)}\\
		\end{tabular}
	\end{adjustbox}
	\label{chapt2_tab:baseline_cros}
\end{table}

Results are reported in table \ref{chapt2_tab:baseline_cros}. We note that the coefficients of the baseline gravity specification are in line with those in the literature: distance is $\cong -0.7$, sums and differences of population and GDPpc are all positive and significant. Migration coefficients are slightly lower than expected, but still coherent with the analysis by \citet{gen2011impact}. In particular, in this formulation, the migration coefficient is higher for Rauch's homogeneous goods than differentiated ones: this is not the case of our classification, where overlapping goods are more correlated with migration than non-overlapping goods. This also confirms the results from other recent works \citep[see for example][]{felb2012iza}.

Appending the spatial autoregressive components to the model seems to be necessary in order to grasp the potential contributions of the network effects of migration on trade. In addition, the Moran I test on the residuals of the baseline gravity model is significant (above $0.9$) for all the product categories, confirming the presence of omitted network correlation and motivating the use of SDM/SEM models.

\begin{table}[tb]
	\centering
	\caption[Durbin model regression results]{Regression results for the cross section Durbin (SDM2) model. Year 2000.} 
	\begin{adjustbox}{width=\textwidth,keepaspectratio}
		\begin{tabular}{l|ccccc}
			\hline\hline
			& Total & Differentiated & Homogeneous & Overlapping & Non-Overlapping \\
			\hline
			migrant & 0.082*** & 0.063*** & 0.097*** & 0.073*** & 0.098*** \\
			distw & -0.729*** & -0.670*** & -0.686*** & -0.689*** & -0.622***  \\
			rta & 0.212*** & 0.339*** & 0.142* & 0.278*** & 0.075$\cdot$ \\
			gdpcapsum & 1.816*** & 1.849*** & 1.530*** & 1.874*** & 1.351***  \\
			gdpcapsim & 0.966*** & 0.901*** & 0.938*** & 0.991*** & 0.807***  \\
			popsum & 1.736*** & 1.670***& 1.557*** & 1.692*** & 1.441***  \\
			popsim & 0.799*** & 0.720*** & 0.799*** & 0.766*** & 0.742***  \\
			contig & 0.225*** & 0.201*** & 0.234*** & 0.239*** & 0.285*** \\
			comlang & 0.185** & 0.149*** & 0.208*** & 0.165*** & 0.199*** \\
			colony & 0.300*** & 0.235*** & 0.245*** & 0.279*** & 0.119 \\
			comcur & 0.148*** & 0.301*** & 0.137*** & 0.207*** & 0.101** \\
			& & & & & \\
			$W^{(T)}$.migrant & 0.051* & 0.037* & 0.079** & 0.032& -0.142**  \\
			$W^{(T)}$.distw & 0.002 & -0.089** & -0.246*** & -0.096** & 0.557***  \\
			$W^{(T)}$.rta & 0.048 & 0.110* & -0.069 & 0.131* & 0.048 \\
			$W^{(T)}$.gdpcapsum & 0.071** & 0.076 & 0.214*** & 0.079** & -0.286**  \\
			$W^{(T)}$.gdpcapsim & 0.046** & 0.118 & 0.122 & 0.099** & -0.465***  \\
			$W^{(T)}$.popsum & 0.173*** & 0.210*** & 0.324*** & 0.258*** & -0.409***  \\
			$W^{(T)}$.popsim & 0.059** & 0.103** & 0.136** & 0.095** & -0.239***  \\
			$W^{(T)}$.contig & 0.315*** & 0.214*** & 0.119** & 0.142*** & -0.024  \\
			$W^{(T)}$.comlang & -0.0273& 0.068 & -0.010 & 0.055 & -0.141**  \\
			$W^{(T)}$.colony & -0.050 & -0.086 & 0.049 & -0.134* & 0.064*  \\
			$W^{(T)}$.comcur & 0.489*** & 0.314*** & 0.618*** & 0.348*** & 0.230***  \\
			\hline
			$\rho$ & -0.275*** & -0.339*** & -0.428*** & -0.310*** & -0.421***  \\
			$\lambda$ & 0.564*** & 0.539*** & 0.680*** & 0.580*** & 0.578***  \\
			\hline\hline
			\multicolumn{6}{l}{\footnotesize \sym{\cdot} \(p<0.1\), \sym{*} \(p<0.05\), \sym{**} \(p<0.01\), \sym{***} \(p<0.001\)}\\
		\end{tabular}
	\end{adjustbox}
	\label{chapt2_tab:durbin_cros}
\end{table}
\begin{table}[!ht]
	\centering
	\caption[Impacts for the Durbin model regression]{Impacts for the cross section Durbin (SDM2) model. D = direct, I = indirect. Year 2000.} 
	\begin{adjustbox}{width=\textwidth,keepaspectratio}
		\begin{tabular}{l|ccc|ccc|ccc|ccc|ccc}
			\hline\hline
			& \multicolumn{3}{c|}{Total} & \multicolumn{3}{c|}{Differentiated} & \multicolumn{3}{c|}{Homogeneous} & \multicolumn{3}{c|}{Overlapping} & \multicolumn{3}{c}{Non-Overlapping} \\
			\hline
			& D & I & Tot & D & I & Tot & D & I & Tot & D & I & Tot & D & I & Tot \\
			\hline
			migrant & .081 & .024 & .104 & .062 & .013 & .075 & .095 & .028 & .123 & .073 & .008 & .080 & .090 & -.166 & -.076 \\
			distw & -0.736 & 0.166 & -0.570 & -0.677 & 0.110 & -0.567 & -0.689 & 0.036 & -0.652 & -0.694 & 0.094 & -0.600 & -0.595 & 0.485 & -0.110 \\
			rta & 0.213 & -0.008 & 0.205 & 0.339 & -0.003 & 0.336 & 0.150 & -0.098 & 0.051 & 0.277 & 0.036 & 0.313 & 0.082 & 0.131 & 0.213 \\
			gdpcapsum & 1.832 & -.352 & 1.480 & 1.877 & -.439 & 1.438 & 1.555 & -.335 & 1.221 & 1.895 & -.404 & 1.491 & 1.377 & .465 & 1.841 \\
			gdpcapsim & .974 & -.180 & .794 & .910 & -.149 & .761 & .954 & -.212 & .742 & 1.000 & -.168 & .832 & .796 & -.206 & 0.591 \\
			popsum & 1.748 & -.250 & 1.498 & 1.688 & -.283 & 1.404 & 1.577 & -.260 & 1.317 & 1.704 & -.215 & 1.489 & 1.459 & .324 & 1.784 \\
			popsim & .806 & -.132 & .674 & .727 & -.113 & .614 & .810 & -.156 & .655 & .773 & -.115 & .658 & .749 & .120 & .870 \\
			contig & 0.216 & 0.208 & 0.425 & 0.193 & 0.117 & 0.310 & 0.233 & 0.014 & 0.247 & 0.236 & 0.055 & 0.291 & 0.294 & 0.157 & 0.451 \\
			comlang & 0.188 & -0.064 & 0.124 & 0.148 & 0.014 & 0.162 & 0.213 & -0.075 & 0.138 & 0.166 & 0.004 & 0.169 & 0.195 & -0.091 & 0.104 \\
			colony & 0.306 & -0.109 & 0.197 & 0.243 & -0.132 & 0.111 & 0.248 & -0.043 & 0.205 & 0.289 & -0.178 & 0.111 & 0.129 & 0.188 & 0.317 \\
			comcur & 0.133 & 0.367 & 0.500 & 0.290 & 0.169 & 0.459 & 0.104 & 0.424 & 0.528 & 0.196 & 0.229 & 0.424 & 0.125 & 0.448 & 0.574 \\
			\hline\hline
		\end{tabular}
	\end{adjustbox}
	\label{chapt2_tab:durbin_cros_imp}
\end{table}

To choose the model specification that best fit to our case we perform a series of Likelihood Ratio (LR) and Common Factor Wald tests: for all the product categories, the results indicate the SDM2 model to be the best fit. We therefore employ it with the CML estimator, using our filtered total WTW for year 2000 as weights.
The resulting model can be written as follow:
\begin{equation}\label{chapt2_eq:sdm2bis}
T = \rho W^{(T)}_t T + \sum_{k=1}^{K} X_k \beta_k + \sum_{k=1}^{K} W^{(T)}_t X_k \gamma_k + \lambda W^{(T)}_t \epsilon + \epsilon\ ,
\end{equation}
where $\rho$ and $\lambda$ are the scalar coefficient of the lagged trade and residuals; $\gamma$ is the $k \times 1$ vector of coefficients relative to the lagged explanatory variables,\footnote{All the variables (except dummies) are taken in $\log_{10}$.} i.e. $k=\{$\textit{lag.distw}, \textit{lag.gdpcapsum}, \textit{lag.popsum}, \textit{lag.gdpcapsim}, \textit{lag.popsim}, \textit{lag.migrat}, \textit{lag.contig}, \textit{lag.comcur}, \textit{lag.comlang}, \textit{lag.colony}, \textit{lag.rta}$\}$; $W^{(T)}$ is the $n^2 \times n^2$ network weight matrix relative to the total trade ties.\footnote{For a better comparison, all the reported results are computed using the total trade matrix specification.}
Results are presented in Table \ref{chapt2_tab:durbin_cros}. Direct and indirect impacts are shown in Table \ref{chapt2_tab:durbin_cros_imp}.

\noindent The main results can be summarized as follows:
\begin{itemize}
	\item $\rho$ is always negative and significant, meaning that an increase in the trade of neighbor countries decreases trade in the dyad. This can be interpreted as a competition effect, indicating that larger trade flows between country $i$ (or $j$) and a third one, reduces the tie of the country pair itself. $\lambda$ instead have positive value, which account for the autocorrelation in the residual terms.
	
	\item Looking at indirect impacts, lagged GDP and population (both as sum and similarity) have a negative impact on trade: if neighbors are big in terms of GDP and population, they decrease the intensity of trade in the link.
	
	\item Analyzing the total migration impact (direct + indirect), we find that it is bigger for overlapping goods with respect to non-overlap\-ping ones, while this is not the case for Rauch's classification where homogeneous goods are subject to a bigger influence than differentiated ones. Our classification is robust to the addition of network interdependencies.
	
	\item Negative indirect impact in homogeneous and non-overlapping categories tells us that migration to/from neighboring countries have a \emph{substitution effect} on trade.
\end{itemize} 

In conclusion network interdependencies, which are not included in the baseline specification of Eq. \ref{chapt2_eq:ols}, are always significant and considerably influence the results of our regressions. In fact, comparing Tables \ref{chapt2_tab:baseline_cros}, \ref{chapt2_tab:durbin_cros} and \ref{chapt2_tab:durbin_cros_imp} it is clear how the direct part of SDM2 estimations is consistent with the OLS one; but, considering also network effects, OLS estimation result to be biased, in particular as expected underestimating the effect of migration on trade. These considerations make us conclude that indirect effects play a relevant role in shaping the WTW and are important to fully understand its relations with other phenomena. Furthermore, trade in differentiated goods (both \'a la Rauch and with our definition) are, as expected, more strongly related with people migration than other product categories.

\FloatBarrier

\section{Conclusions}\label{chapt2_sec:concl}
Sets of interdependent phenomena on a global scale are increasingly analyzed as complex networks. International trade, human mobility, communication and transportation infrastructures being just a few examples. Only recently new methodologies have been developed to analyze the dynamics of intertwined networks, including cascading failures and the transmission of shocks across multiple and heterogeneous network structures. In this chapter we contribute to this emerging field of research by analyzing the relationship between the International Migration Network and the World Trade Web.

Increased data availability both at national and international levels has triggered a host of research on the relationship between trade and migration. We contribute to this line of research by using network analysis to compare the two phenomena and the Jaccard and RCA indexes to investigate their correlation and similarities.

We then propose a new methodology to classify trade goods, aimed at identifying which of them are more connected by migrant stocks and compare it with Rauch's one. Analyzing the two we conclude that our classification is a good alternative to any previously proposed one as it confirms the same main results, with the advantage of being robust to the addition of network effects and better considering subtle good categories (i.e. intermediate goods), as well as having an easier to use, more straightforward definition.

Moreover we make use of spatial econometric techniques exploiting the network structure, rather than standard geographic distances, in order to analyze direct and indirect impacts of migration on trade. We are then able to investigate the network impacts suggested by Rauch's seminal paper from a global perspective, rather than focusing on a single ethnic network as had been done in the literature so far.
When we test a gravity model of trade controlling for network interdependencies, we find that our classification behave correctly, with overlapping goods subjected to higher overall impact than non-overlapping goods. Negative indirect impacts in non-overlapping goods also suggest that migration to/from neighboring countries have a \emph{substitution effect} on trade.

In conclusion our work, as other numerous statistical and case studies, provide evidence that transnational business and social networks promote international trade by alleviating problems of contract enforcement and providing information about trading opportunities.
All these findings, and the general prominent part that migration has in shaping international trade, open up space for greater consideration of the role of personal contacts and relationship-building in determining the geographic distribution of economic activity.

%% file: mainmatter/3/chapt3.tex
%

\chapter[Trade and Migration (2)]{The Migration Network Effect on International Trade}
\label{chapt:migr2}

\graphicspath{{3/figures/PNG/}{3/figures/PDF/}{mainmatter/3/figures/}}

\section{Introduction}
Since the mid Nineties a growing body of research has investigated the relation between human migration and international trade. 
Whereas the standard Heckscher-Ohlin model suggests that the movement of goods across borders can provide a substitute for the movement of production factors, the empirical bottom line of these more recent works is that the two actually complement each other. 
This appears to hold for different countries \citep[the US, Canada, Spain, Italy and France, to name just a few, see respectively][]{gould1994immigrant, head1998immigration, peri2010trade, debeneC2012, briant2014twe} and has recently been confirmed by a meta-analysis covering 48 different studies \citep{gen2011impact}.
As it has often happened in the international trade literature, empirical findings have percolated to economic theory, with recent models being able to accommodate the complementarity between migration and trade \citep{felb2012iza}.

We contribute to this growing field of research with a novel methodological approach that combines network analysis and spatial econometric techniques.
On the one hand, this allows us to assess both the direct and the indirect effect of migration on trade without focusing on a single ethnic community at a time, as customarily done in the existing literature.
On the other hand, spatial econometrics allows us to effectively account for the interdependences among trade flows that would otherwise lead to inconsistent (or even biased) estimates. 

Most of the empirical literature we refer to shares a common strategy, based on the estimation of a log-linear gravity model where bilateral trade flows are regressed over standard explanatory variables (economic mass and distance), the stock of migrants from specific partner countries and other controls aiming at capturing various types of trade costs (common language, colonial relationships and the like).
The two main strands of research that have emerged investigate the direct relation between trade and migration (i.e. the impact of migration from A to B on import/export flows between the same countries) and the existence of indirect or ``network'' effects (migration from A to both B and C not only affects trade from A to B and from A to C, but also establishes a connection between B and C due to the presence of a community of expatriates with the same background in both countries).
The core of the argument \citep[see for instance the seminal contribution by][]{rauch2002ethnic} is that formal and informal links among co-ethnic migrants in other countries and at home facilitate trade by providing potential trading partners with easier access to valuable, i.e. qualified, information. The pro-trade effect thus stems from the reduction of the trade barriers and search costs associated with market transactions. Since these costs are likely to be larger for international trade due to distance, language and cultural differences, legal provisions and the like, ethnic networks end up being especially relevant in facilitating cross-border transactions. 

Indeed, one of the central results in the literature is that the positive effect of migration on trade is larger for ``differentiated goods", i.e. those items that are not homogeneous and are not traded in organized exchanges therefore rendering that knowledge about counterpart reputation particularly valuable \citep{rauch2002ethnic}.\footnote{Although subsequent work has shown that the actual magnitude of this pro-trade effect is smaller than originally estimated \citep[see][]{felb2010}, its existence and its specific importance for differentiated goods is confirmed.}
Similar results have been replicated by a number of subsequent works using a variety of datasets and techniques. 
\citet{peri2010trade}, for instance, analyze the Spanish case and find that doubling the number of immigrants from a given country increases export to the same destination by 10 percent. This effect is higher for firms selling differentiated products and for more distant countries (geographically or culturally). All of these elements are consistent with the notion that networks (in this case the presence of a large community of expatriates and their connections with co-nationals at home and abroad) lower the hurdle in terms of economic interactions, providing better access to information and trade opportunities and reducing the fixed costs associated with entry into a foreign market. 
\citet{aleksynska2013isolating} focus on the share of migrants involved in business activities rather than on the total migrant population, and find a significant effect, even after controlling for the overall bilateral stock of migrants. 
Using trade data on Italian provinces, \citet{debeneC2012} find that the presence of migrants boosts both import from and export to their home countries, with the former effect being much larger. In the literature, this difference is interpreted as signaling a second channel through which migration affects trade, namely a home-country bias in demand by ethnic communities. 
\citet{briant2014twe} also use a fine geographical disaggregation based on French departments to investigate the effect of migration on trade in goods with different degrees of complexity, as well as across countries with various levels of institutional quality. Migration is more relevant for complex goods, regardless the quality of institutions in the partner country, whereas it matters also for simple products only when the institutional quality of the source country is low. 
A similar substitution effect between migrants and institutions is found in \citet{ehrhart2014}, who focus on African countries.

In parallel to these developments in the trade-migration literature, the past decade has witnessed important advances in both the theoretical foundations of the gravity model and its estimation methods \citep{anderson2003gravity,deardorff1998determinants}. 
The literature has suggested that special care has to be applied in the empirical analysis to account for the interdependencies between trade flows that are inherent to the estimation of a general equilibrium model.
In fact, \citet{anderson2003gravity} show that bilateral export does not only depend on bilateral trade costs, the size of the trading economies and other dyad-specific characteristics, but also on \emph{Multilateral Trade Resistance} (MTR) i.e. the overall set of trade barriers that exporter and importer countries face. 
Several ways to account for MTR have been proposed: these involve the use of country-specific effects \citep{feenstra2003advanced}, export- and import-specific dummies \citep{anderson2004trade}, measures of geographic remoteness \citep{hell98}, as well as more sophisticated methods \citep[see][for an excellent survey]{hema13}. 
\citet{behrens2012dual} tackle the issue borrowing from the spatial econometrics literature \citep[see also][for an earlier contribution along the same lines, with an application to bilateral migration flows]{lesage2008spatial}: they suggest using a spatial autoregressive moving average specification as a proxy for MTR, which results in a consistent estimation of the gravity equation.\footnote{The need to account for spatial autocorrelation in trade flows had been already recognized in \citet{porojan2001trade}, although that paper suffers from serious methodological limitations pointed out by \citet{johnC03}.} 

We build on both the aforementioned streams of literature to estimate the effect of migration on trade using spatial econometrics to adequately account for interdependences in trade flows.\footnote{See \citet{Fagiolo2014} for a complementary analysis with a focus on the role of third-party (indirect) common and non-overlapping inward migration channels.}
In fact, the key innovation proposed in this chapter rests on the fact that spatial autocorrelation matrix is based on topological rather than geographical distance.
More precisely, we build the International Migration Network (IMN) connecting countries and use distance in the network to define proximity.\footnote{As opposed to Chapter \ref{chapt:migr1}, where we used for the same purpose a similar matrix built from International Trade data.} 
Hence, we proxy MTR introducing the IMN into the model, assuming that migration network filters out the heterogeneity on the relative trade costs faced by exporting and importing countries.
Our tests confirm that controlling for the IMN eliminates the spatial autocorrelation, thus supporting our intuition.

The rest of the chapter is structured as follows. 
Our empirical strategy is laid down in Section \ref{chapt3_sec:empirics}, which illustrates the rationale for our approach, the model specification and the data used.
Section \ref{chapt3_sec:results} discusses our main results, while some concluding remarks are elaborated in Section \ref{chapt3_sec:concl}.

\section{Empirical strategy}\label{chapt3_sec:empirics}
\begin{figure}[thb]
	\centering
	\caption[Exemplifying representation of the direct and indirect migration channels]{Exemplifying representation of the direct and indirect migration channels (origin-side).}
	\vspace{-10pt}
	\includegraphics[width=\myFiguresSize]{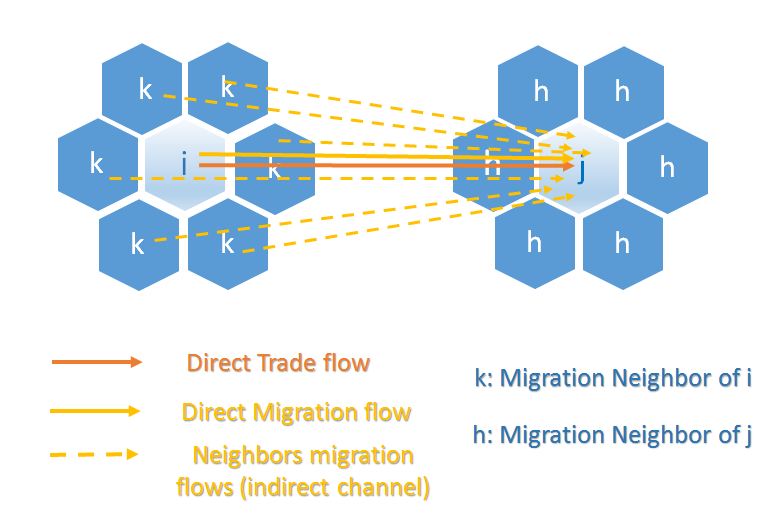}
	\vspace{-10pt}
	\label{chapt3_fig:indirect}
\end{figure}

The combination of network analysis and spatial econometrics we propose in this chapter is summarized in Figure \ref{chapt3_fig:indirect}.
We assume that trade between $i$ and $j$ depends both on variables specific to the country-pair (e.g distance, stock of bilateral migrants), but is also affected by third-country effects. 
In particular, we focus our attention on the potential impact that migrants from third countries (say $k$) may have on bilateral trade between $i$ and $j$. 
Let $k$ labels neighbors of the origin country $i$ in the IMN: this means that there is a significant number of people born in $i$ and resident in $k$.\footnote{What represents a \emph{significant} number of migrants is explained in Section \ref{chapt3_sec:econ_app} below.}
Migration from $k$ to $j$ represents the third-country (indirect) effect we take into account in the empirical analysis. 
In other words, we investigate whether migration from $k$ to $j$ affects export from $i$ to $j$, given the existence of a strong migration link from $i$ to $k$. 
Similarly, we could let $h$ be a migration neighbor of the destination country $j$. 
In this case migrants from $i$ to $h$ should represents the indirect channel affecting trade from $i$ to $j$. However, we have no theoretical and empirical reason to model this second type of network dependence. 

\subsection{Gravity models and spatial interaction}\label{chapt3_sec:econ_app}
As mentioned above, the standard approach used in the empirical literature on migration and trade entails the estimation of a gravity model augmented with the stock of migrants. 
We follow a similar strategy and model bilateral trade in terms of per capita GDP to control for purchasing power and population to control for size. 
Following \citet{baltagi2007estimating} we construct pair-specific measures of both GDP and population (POP) rather than separately including information on both the origin and the destination countries, as this allows us to better interpret of our variables of interest.
The control variables are defined as $GDPpc\_sum_{ij} = log(GDPpc_i + GDPpc_j)$ and $population\_sum_{ij} = log(POP_i + POP_j)$.
Moreover, we also introduce similarities indexes defined as as $GDPpc\_sim_{ij} = (1-(\frac{GDPpc_i}{GDPpc_i + GDPpc_j}) ^2 - (\frac{GDPpc_j}{GDPpc_i + GDPpc_j}) ^2)$ and $POP\_sim_{ij} = (1-(\frac{POP_i}{POP_i + POP_j}) ^2 - (\frac{POP_j}{POP_i + POP_j}) ^2)$. 
Last, the model includes the stock of migrants and a number of standard controls such as geographic contiguity ($contig$), common language ($comlang$), common currency ($comcur$), colonial ties ($colony$) and participation into regional trade agreements ($rta$).

Since the seminal contribution by \citet{anderson2003gravity} recent empirical works recognize the importance of adequately account for MTR, i.e. to consider interdependencies among trade flows, that stem from the estimation of a model resulting from a general equilibrium framework. 
A number of alternative methods have been proposed to deal with this issue, most of which are very effectively summarized by \citet{hema13}.
Here we concentrate on two: the first entails augmenting the gravity model with exporter- and importer-specific dummies; the second models MTR in a way similar to spatial autocorrelation. 
In particular, \citet{behrens2012dual} suggest a spatial autoregressive moving average specification for the gravity model, which results in consistent estimates of the parameters. 
They argue that the baseline fixed effects specification does not fully succeed in capturing the MTR dependencies in the error structure, and indeed find that the residuals still display a significant amount of autocorrelation. 
\citet{anselin2013spatial} demonstrated by means of a series of simulation experiments that fixed effects correctly remove autocorrelation only in some specific cases.
In the empirical analysis we use the Moran I test to check for the presence of autocorrelation in the standard gravity model and the ability of our specification accounting for spatial contiguity in the migration matrix to adequately proxy for MTR, and therefore remove this autocorrelation in the residuals.

To model the spatial autoregressive component one generally uses an $n\times n$ weight matrix ($W$) that defines the set of neighbors: most frequently $W$ is based on spatial contiguity, so that $[w_{ij}]=1$ if $i$ and $j$ share geographical borders, and 0 otherwise.\footnote{Other formulations are based on inverse distance.}

It was recently argued that the matrix can be both spatial or non-spatial. Accordingly, several proposals have been made in the literature, such as using the technological similarities or the transport costs instead of spatial metrics. One of the newest suggestions, however, is to analyze the effect of network-propagation, viewed both as an alternative and a complement to the spatial effect. 
\citet{lesage2011pitfalls} discuss the possibility of jointly modeling spatial and non-spatial dependence through a double autoregressive component that make use of two different weight matrix specifications \citep{elhorst2012model}. 
In general, network theory and spatial econometrics are intimately connected. \citet{leenders2002modeling} proposes using Spatial Autoregressive models employing an ad-hoc $W$ matrix based on network relations (in terms of social influences and communication); \citet{farber2010topology} analyze the relationship between the topology property of networks and the properties of spatial models, performing several simulation tests. 
\citet{manski1993identification} gives a seminal contribution, as it lays the foundation for analyzing the exogenous, endogenous and correlated effects that researchers encounter both in network and econometric theory. 
\citet{lee2010specification}, following \citeauthor{manski1993identification}'s work, propose a specification for estimating network models in presence of exogenous, endogenous and correlated effects. Furthermore, the correct specification for the estimation of network models has become a popular object of study as of late \citep{bramoulle2009identification, chandrasekhar2011econometrics}.

To control for autocorrelation we use the matrix describing the IMN ($W^{(M)}$), so that topological distance in the network replaces the more usual geographical one. 
In order to identify the significant links, we use a stochastic benchmark based on the hypergeometric distribution, as recently done in \citet{riccaboni2013global}.
The procedure starts from the null hypothesis that treat all links are randomly assigned following an hypergeometric probability distribution. For each pair of countries, we can thus compute the probability that the observed link weight comes from the same distribution, which takes as parameters the out- and in-strength of the nodes, plus the total amount of migrants observed in the network. 
Hence the procedures takes into account the heterogeneity of countries with respect to the total number of migrants and allows us to retain only those links that represent a significant departure from the hypergeometric benchmark.\footnote{We set the cutoff at $1\%$.}
The specification of the $W^{(M)}$ matrix then becomes:
$$
W^{(M)}:=
\left\{ 
\begin{array}{rl}
w_{i,j}=1, &\mbox{ if \textit{i} has a significant migration}\\
&\mbox{relationship with \textit{j}}\\
w_{i,j}=0, &\mbox{ otherwise,}
\end{array}
\right.
$$
and where a specific Kronecker transformation is applied so that the set of neighbors for each country-pair includes neighbors of the exporter country.\footnote{The resulting transformed matrix ($W^{(M)}_K$) has dimension $n^2 \times n^2$ and is generally constructed as the Kronecker product of $W$ with the identity matrix $I$ \citep[as proposed in][]{lesage2008spatial}: $$W^{(M)}_K = W^{(M)} \otimes I\ .$$ In a panel framework one needs to account for the time index so that the matrix has to be pre-multiplied by a diagonal matrix of dimension $t$: $W^{(M)}_{K,t} = I_t \otimes W^{(M)}_K. $}

\subsection{Model specification and estimators}\label{chapt3_sec:model}
Using spatial econometrics, the measure of the spatial (network) association in the origin-destination trade flow specification can be based on two classes of models \citep{lesage2008spatial}: \emph{Spatial autoregressive models} (SAR) and \emph{Spatial Durbin / Spatial error models} (SDM/SEM). The former consists in the inclusion of either a spatially lagged dependent variable or of a spatial autoregressive process in the residual term, motivated by significant spatial autocorrelation in the dependent variable. This model can be augmented with the inclusion of the spatial lagged residuals, and it is called \emph{Spatial autoregressive error model} (SARAR). The latter can be motivated by a statistical nuisance and it is best described as a proxy for missing variables that follow a meaningful spatial pattern.
The econometric representation of the models can be illustrated as:
\begin{align}
	\mbox{SAR} & \qquad y = \rho W y + X \beta + \epsilon \\
	\mbox{SARAR} & \qquad y = \rho W y + X \beta + \lambda W \epsilon + \epsilon \\
	\mbox{SDM} & \qquad y = \rho W y + X \beta + W X \gamma + \epsilon\ ,
\end{align}
which becomes the SEM model in the event that included and excluded variables are not correlated \citep[common factor tests can be performed, see][]{lesage2008spatial}
\begin{equation}
\mbox{SEM} \qquad y = X \beta + \lambda W\epsilon + \epsilon\ .
\end{equation}
The Durbin model can also take into account both the spatially lagged dependent variable and the spatial autoregressive process in the residuals: this augmented version of the SDM model (also called \emph{Manski}) that fully accounts for all possible spatial dependency takes the form:
\begin{equation}
\mbox{Manski} \qquad y = \rho W y + X \beta + W X \gamma + \lambda W\epsilon + \epsilon\ ,
\end{equation}
where $y$ is the dependent variable; $X$ is the matrix of the explanatory variables; $\epsilon$ represents the residuals; 
$W$ is the (spatial) weight matrix; while $\beta$, $\gamma$, $\lambda$ and $\rho$ are the coefficients to be estimated.
However, \citet{elhorst2010applied} argues that the SDM is the only model that provides unbiased parameter estimates and correct standard errors, even if the true data-generation process is any of the other spatial regression models mentioned above. 

In spatial models, the presence of intrinsic endogeneity due to the inclusion of a spatial lag of the dependent variable among the controls renders OLS estimation inconsistent. The standard alternative in this literature is the concentrated maximum likelihood (CML) estimator proposed by \citet{anselin1988spatial} and revised by \citet{lesage2008spatial}, that is known to overcome this problem.\footnote{Another source of possible endogeneity is the origin of the spatial weight matrix: ``in the standard estimation and testing approaches, the weights matrix W is taken to be exogenous'' \citep[p.~244]{anselin1998spatial}, meaning that, in the case it is not, it may lead to biased estimates. It is not clear if our choice of the weight matrix introduced additional bias in the estimates, and further tests are necessary to quantify and possibly control for it. This observation in fact applies also to the analysis in Chapter \ref{chapt:migr1}. Only very recently, \citet{kelejian2014estimation} proposed a model specification and estimation that is unbiased even when using an endogenous lag matrix.}
Fitting a CML estimator on a log-log gravity model disregards the presence of zero trade flows, which represent around 20 percent of our sample. The standard literature has addressed this by considering trade flows as count processes and fitting Poisson or negative binomial models. However, to the best of our knowledge no extension of this approach exists that combines it with spatial autoregressive models. The alternative to fit a zero inflated Poisson model in which the spatial effect is captured by spatial-filtering eigenvectors \citep[see][]{lionetti2009trading} would however prevent us from distinguishing between direct and indirect spatial effects. Therefore in the present work we fit only strictly positive migration stocks and trade flows, limiting our analysis to non-zero weights.
Last, we are aware of the fact that a log-log model implies non-realistic assumptions about homoscedasticity in the residuals, and will explicitly test for this in the empirical analysis. 

\subsection{Data}\label{chapt3_sec:data}
Data regarding migrants come from the World Bank's Global Bilateral Migration dataset \citep{ozden2011earth}: it is composed of matrices of bilateral migrant stocks spanning five decades from 1960 to 2000 (5 census rounds), based primarily on the foreign-born definition of migrants. It is the first and only comprehensive picture of bilateral global migration over the second half of the 20th century, taking into account a total of 232 countries.
The data reveal that the global migrant stock increased from 92 million in 1960 to 165 million in 2000. Quantitatively, migration between developing countries dominates, constituting half of all international migration in 2000, whereas flows from developing to developed countries represent the fastest growing component of international migration in both absolute and relative terms.

For international trade, we use the NBER-UN dataset described by \citet{NBERw11040}, disaggregated according to the Standardized International Trade Code at the four-digit level (SITC-4). For each country it provides the value (expressed in thousands of US dollars) exported to all other countries, for 775 product classes. In our analysis, we focus on the years 1970, 1980, 1990 and 2000.

Looking at the SITC product code of goods traded between each country pair allows us to apply \citeauthor{rauch2002ethnic}'s (2002) classification and distinguish between homogeneous and differentiated goods. 
Trade in the latter type of products are more heavily influenced by the presence of migrant networks, as buyers and sellers need to look for relevant information that is not easily embedded in prices. 

We only consider countries present in both datasets: this results in a final sample of 146 countries (nodes) that have active interactions in both trade and migration. All the other controls used in the regressions (e.g. contiguity, common language, etc.) have been retrieved from the CEPII dataset documented in \citet{mayer2011notes}.

\section{Results}\label{chapt3_sec:results}
We conduct a panel regression estimation using pooled data from the years 1970, 1980, 1990 and 2000.\footnote{A cross sectional analysis was also performed for the years 1970 and 2000 as a robustness check. Results are available upon request.}
We employ three different dependent variables: $(i)$ total exports; $(ii)$ export of differentiated goods; and $(iii)$ export of homogeneous goods.

We start by estimating a baseline gravity model for total trade without migration using pooled OLS; results, presented in the first column of Table \ref{chapt3_tab:baseline_panel}, are in line with the literature. In column 2 of the table we add the stock of migrants to the model, where we note that the migration coefficient (0.129) is in line with the meta-analysis by \citet{gen2011impact}, who report coefficients that vary between 0.13--0.15. 
Moreover, we find that adding migration to the explanatory variables lowers the impact of distance. 
This is in good agreement with the literature \citep[see for instance][]{felb2012iza} and suggests that distance picks up a host of formal and informal informational barriers.

\begin{table}[tb]
	\centering
	\caption[Gravity results with OLS and FE models]{Gravity results with OLS and FE models, with and without instrumenting migration for reverse causality.}
	\begin{adjustbox}{width=\textwidth,totalheight=\textheight,keepaspectratio}
		\small
		\begin{tabular}{lrrrrrrr}
			\toprule
			\textbf{Non instrumented} & \multicolumn{1}{c}{base} & \multicolumn{2}{c}{total trade} & \multicolumn{2}{c}{diff. goods} & \multicolumn{2}{c}{homog. goods} \\
			\cmidrule(lr){2-2} \cmidrule(lr){3-4} \cmidrule(lr){5-6} \cmidrule(lr){7-8} 
			& \multicolumn{1}{c}{ols} & \multicolumn{1}{c}{ols}& \multicolumn{1}{c}{fe} & \multicolumn{1}{c}{ols} & \multicolumn{1}{c}{fe} & \multicolumn{1}{c}{ols}& \multicolumn{1}{c}{fe} \\
			\midrule
			distance 	& -.858*** & -.706*** & -1.002*** & -.669*** & -.055*** & -.728*** & -1.011*** \\
			GDPpc\_sum & 1.746*** & 1.654*** & - & 1.732*** & - & 1.497*** & - \\
			GDPpc\_sim & .933*** & .888*** & - & .851*** & - & .868*** & - \\
			POP\_sum & 1.622*** & 1.476*** & - & 1.446*** & - & 1.438*** & - \\
			POP\_sim & .774*** & .703*** & - & .617*** & - & .742*** & - \\
			contig 	& .268*** & .168*** & .79** & .184*** & .144*** & .118*** & .017 \\
			comlang & .188*** & .082*** & .129*** & .118*** & .244*** & .108*** & .093*** \\
			colony 	& .604*** & .471*** & .455*** & .375*** & .313*** & .401*** & .443*** \\
			comcur 	& .360*** & .289*** & .298*** & .345*** & .270*** & .248*** & .300*** \\
			rta 	& .187*** & .148*** & .005 & .324*** & .009 & .074** & 0.041 \\
			migration 	& & .129*** & .128*** & .133*** & .140*** & .109*** & .113*** \\
			\midrule
			$R^2$ adj 	& .639 & .639 & .752 & .629 & .820 & .604 & .716 \\
			obs 		& 29784 & 24105 & 27217 & 20908 & 23467 & 22256 & 24813 \\
			
			\addlinespace[8pt]
			\textbf{Instrumented} & & \multicolumn{2}{c}{total trade} & \multicolumn{2}{c}{diff. goods} & \multicolumn{2}{c}{homog. goods} \\
			\cmidrule(lr){3-4} \cmidrule(lr){5-6} \cmidrule(lr){7-8} 
			& &\multicolumn{1}{c}{ols} & \multicolumn{1}{c}{fe} & \multicolumn{1}{c}{ols} & \multicolumn{1}{c}{fe} & \multicolumn{1}{c}{ols}& \multicolumn{1}{c}{fe} \\
			\midrule
			distance & & -.776*** & -1.064*** & -0.680*** & -.075*** & -0.805*** & -1.086*** \\
			GDPpc\_sum & & 1.896*** & - & 1.944*** & - & 1.687*** & - \\
			GDPpc\_sim & & .955*** & - & .899*** & - & .915*** & - \\
			POP\_sum & & 1.659*** & - & 1.594*** & - & 1.590*** & - \\
			POP\_sim & & .783*** & - & .676*** & - & .815*** & - \\
			contig & & .229*** & .074* & .228*** & .144*** & .201*** & .025 \\
			comlang & & .149*** & .114*** & .152*** & .239*** & .172*** & .071*** \\
			colony & & .384*** & .429*** & .283*** & .288*** & .339*** & .429*** \\
			comcur & & .175** & .088 & .201** & .067*** & .206** & .128* \\
			rta & & .093*** & -.035 & .280*** & -.029 & .027 & -.007 \\
			migration & & .088*** & .121*** & .109*** & .135*** & .070*** & .105*** \\
			\midrule
			$R^2$ adj & & .636 & .746 & .608 & .806 & .589 & .707 \\
			obs & & 17448 & 18551 & 15261 & 16124 & 16211 & 17039 \\
			\bottomrule
		\end{tabular}
	\end{adjustbox}
	\label{chapt3_tab:baseline_panel}
\end{table}

\begin{table}[tb]
	\centering
	\caption[Tests for migration endogeneity and instruments validity]{Tests for migration endogeneity and instruments validity.}
	\begin{adjustbox}{width=\textwidth,totalheight=\textheight,keepaspectratio}
		\begin{tabular}{lrrr}
			\cmidrule[\heavyrulewidth]{1-4}
			& \multicolumn{1}{c}{total } & \multicolumn{1}{c}{diff. } & \multicolumn{1}{c}{homog. } \\
			& \multicolumn{1}{c}{trade} & \multicolumn{1}{c}{goods} & \multicolumn{1}{c}{ goods} \\
			\cmidrule{1-4}
			Correlation between $Trade_t$ and $Migration_t$ & 0.35 & 0.37 & 0.29 \\ 
			Correlation between $Trade_t$ and $Migration_{t-1}$ & 0.28 & 0.29 & 0.22 \\
			First stage test for the validity of the instrument & $>$37.75 & $>$37.75 & $>$37.75 \\
			Durbin-Wu-Hausman for the endogeneity in the model & 14.16 & 4.70 & 12.77 \\
			\bottomrule
		\end{tabular}
	\end{adjustbox}
	\label{chapt3_tab:ivtests}
\end{table}

A specification that includes origin- and destination-specific fixed effect has been widely applied in estimating the gravity equation for international trade, to accounts for MTR. 
Here we opt for importer and exporter time-varying fixed effects (FE) as suggested by the most recent literature \citep{felb2012iza,hema13} and find a migration coefficient of the same magnitude as before (0.129 with OLS, 0.128 with FE).
Columns 4--7 of Table \ref{chapt3_tab:baseline_panel} report OLS and FE results for export of differentiated and homogeneous goods: the migration coefficient is higher in the former case, in line with expectations.

An important issue that has recently moved to center stage is potential endogeneity biases. 
Since the causal relationship between trade and migration can hold both ways, to disentangle the effect of migration on trade one needs to adopt an instrumental variable strategy. 
We follow the literature \citep{felb2012iza,briant2014twe} and use data from the previous decade ($migration_{t-1}$) as an instrument for contemporaneous migration. 
Results for an F-test on the validity of instruments and a Durbin-Wu-Hausmann test for endogeneity are reported in Table \ref{chapt3_tab:ivtests}: for all the three dependent variables they confirm the presence of endogeneity and the necessity to use instruments, as well as the validity of the IV strategy adopted. 
The migration coefficients using the IV model (columns 8--13 of Table \ref{chapt3_tab:baseline_panel}) are lower than in the standard OLS, but the positive effect of migration on trade persists and remains larger in the case of trade in differentiated goods.

\begin{table}[tb]
	\centering
	\caption[Moran I test for autocorrelation on the residuals of the gravity model]{Moran I test for autocorrelation on the residuals of the gravity model estimated by OLS (colums ii. and iii.), FE (columns iv. and v.) and SDM (columns vi. and vii.).}
	\begin{adjustbox}{width=\textwidth,totalheight=\textheight,keepaspectratio}
		\begin{tabular}{lcccccc}
			\toprule
			& \multicolumn{2}{c}{OLS} & \multicolumn{2}{c}{FE} & \multicolumn{2}{c}{SDM} \\
			& {15 near.neigh. } & {Migration } &{15 near.neigh. } & {Migration } & {15 near.neigh. } & {Migration } \\
			matrix & {contiguity} & {network} & {contiguity} & {network} & {contiguity} & {network} \\
			\midrule
			\textbf{total} & 0.078 & 0.077 & -0.011 & - 0.008 & -0.000 & 0.001 \\
			z-score (p-val) & 29.21(0.000) & 28.01 (0.000) & -4.40 (0.000) & -3.49 (0.000) & -1.05 (0.144) & 0.139 (0.444) \\
			\addlinespace[6pt]
			\textbf{differentiated} & 0.087 & 0.081 & -0.010 & -0.009 & 0.001 & -0.012 \\
			z-score (p-val) & 28.14 (0.000) & 25.44 (0.000) & -3.83 (0.000) & -3.43 (0.000) & 1.152 (0.123) & -2.297 (0.011) \\
			\addlinespace[6pt]	
			\textbf{homogeneous} & 0.075 & 0.081 & -0.012 & -0.011 & -0.001 & 0.001 \\	
			z-score (p-val) & 26.07 (0.000) & 27.31 (0.000) & -4.87 (0.000) & -4.16 (0.000) & -1.209 (0.116)& 0.299 (0.381) \\
			\bottomrule
		\end{tabular}
	\end{adjustbox}
	\label{chapt3_tab:moranI}
\end{table}

The Moran I test on the residuals of the unconstrained gravity model confirms the presence of residual autocorrelation. Here, our unconstrained gravity model corresponds to the baseline OLS. 
As we can see in the columns 2-3 of Table \ref{chapt3_tab:moranI}, the OLS residuals still display some positive autocorrelation, measured with both the spatial weight matrix (column 2) and with our IMN matrix (column 3).\footnote{The \emph{spatial weight matrix} is constructed using the k-nearest neighbors method. To make the network and spatial weight matrices comparable in terms of concentration, we choose $k=15$, resulting in a spatial weight matrix having a mean number of 18.38 neighbors based on geographic proximity.} 
The autocorrelation is significant for all the classifications (all trade, differentiated and homogeneous goods).
The FE model that incorporates origin- and destination-specific effects to account for the MTR does not properly capture all the residual autocorrelation: the Moran I tests (columns 4 and 5 of Table \ref{chapt3_tab:moranI}), still finds a significant (negative) autocorrelation.
This motivates the use of the SDM/SEM model in the rest of the analysis, since we were able to empirically confirm the findings of \citet{behrens2012dual} regarding the lack of the FE formulation to fully filter out all of the residual autocorrelation. 

Adding the spatial autoregressive components to the gravity model seems therefore fundamental in order to grasp the potential contributions of the IMN and to test whether this network structure can capture the residual autocorrelation stemming from MTR.
In order to do so, we make use of the previously computed $146\times 146$ matrices for the $2000$ time period, representing the network of country to country migrations.
We perform both the SAR/SARAR and the SDM/SEM models with the CML estimator using network matrices as weights.\footnote{We also compute a CML estimator, separately, using the spatial matrix based on geographic proximity. Results are available upon request. On this issue, \citet{lesage2011pitfalls} discuss the conjoint use of two or more weight matrices in the same model (one spatial and the other non-spatial), but some pitfalls emerge. We may analyze this in future developments.} 
To choose from different specifications of the model we perform a likelihood ratio test, starting from the most general case (SDM) as suggesed by \citet{elhorst2010applied}.

The first three columns of Table \ref{chapt3_tab:durbin_panel} report the results obtained from the estimation of the following final equation: 
\begin{equation}
T = \rho W^{(M)}_t T + \sum_{k=1}^{K} X_k \beta_k + \sum_{k=1}^{K} W^{(M)}_t X_k \gamma_k + \epsilon\ ,
\end{equation}
where $T$ is the dependent variable, $\rho$ is the scalar coefficient of the lagged trade term to be estimated, $\beta$ and $\gamma$ are the $k \times 1$ vectors of coefficients to be estimated for, respectively, the explanatories and the lagged explanatories $X_k$, where the regressors $k$ are the following: $distance$, $POP\_sum$, $GDPpc\_sum$, $POP\_sim$, $GDPpc\_sim$, $migration$, $contig$, $comcur$, $colony$, $comlang$ and $rta$.\footnote{All the variables, except the dummies, are taken in $\log_{10}$.} 
Finally, $W^{(M)}_t$ is the $(n^2\cdot t) \times (n^2\cdot t)$ network weight matrix relative to migration.

We have performed the common factor test for SDM versus SEM. Results point toward the SDM specification, which accounts for the lagged dependent variable and lagged explanatory variables.
Likelihood ratio tests for the choice between SAR/SARAR models and SDM were also performed, leading to favour the SDM. The SDM, in fact, as confirmed in the literature \citep{elhorst2010applied} is able to correct for the parameters mispecification due to autocorrelated omitted variables, even when the true model is not a SDM. However, in order to let our work comparable with \citet{behrens2012dual}, we also have estimated the SAR and the SARAR specifications. As we can see in table \ref{chapt3_tab:durbin_panel}, the estimated $\rho$ parameter for the lagged dependent variable is positive, while \citet{behrens2012dual} found this parameter to be negative in the SARAR specification. We also found a negative $\rho$ when performing SARAR model.\footnote{SAR and SARAR regression results are available upon request}

\begin{table}[tb]
	\centering
	\caption[Durbin model regression with instrumented migration and controls for import strength]{Results from pooled panel SDM model with instrumented migration. Without (i) and (ii) with controls for import strength.}
	\begin{adjustbox}{width=\textwidth,totalheight=\textheight,keepaspectratio}
	\begin{tabular}{lrrrrrr}
		\toprule
		& \multicolumn{3}{c}{(i) {\bf Baseline}} & \multicolumn{3}{c}{(ii) {\bf Import strength}} \\
		\cmidrule(lr){2-4} \cmidrule(lr){5-7}
		& \multicolumn{1}{c}{total } & \multicolumn{1}{c}{diff. } & \multicolumn{1}{c}{homog. } & \multicolumn{1}{c}{total } & \multicolumn{1}{c}{diff. } & \multicolumn{1}{c}{homog. } \\
		& \multicolumn{1}{c}{trade} & \multicolumn{1}{c}{goods} & \multicolumn{1}{c}{ goods} & \multicolumn{1}{c}{trade} & \multicolumn{1}{c}{goods} & \multicolumn{1}{c}{ goods}\\
		\midrule
		distance & -0.784*** & -0.689*** & -0.810*** & -0.785*** & -0.681*** & -0.816*** \\
		GDPpc\_sum & 1.924*** & 1.989*** & 1.701*** & 1.919*** & 2.229*** & 1.589*** \\
		GDPpc\_sim & 0.967*** & 0.939*** & 0.915*** & 0.964*** & 1.060*** & 0.867*** \\
		POP\_sum & 1.662*** & 1.596*** & 1.588*** & 1.659*** & 1.786*** & 1.497*** \\
		POP\_sim & 0.787*** & 0.679*** & 0.817*** & 0.785*** & 0.799*** & 0.759*** \\
		contig & 0.217*** & 0.215*** & 0.188*** & 0.217*** & 0.180** & 0.207*** \\
		comlang & 0.140*** & 0.137*** & 0.166*** & 0.140*** & 0.145*** & 0.162*** \\
		colony & 0.385*** & 0.279*** & 0.341*** & 0.384*** & 0.253*** & 0.356*** \\
		comcur & 0.166*** & 0.182*** & 0.209*** & 0.167*** & 0.201*** & 0.202*** \\
		rta & 0.115* & 0.297*** & 0.045 & 0.113* & 0.297*** & 0.046 \\
		migration & 0.092*** & 0.115*** & 0.074*** & 0.092*** & 0.141*** & 0.063*** \\
		im\_strength\_net & - & - & - & 0.003 & -0.245*** & 0.125** \\
		
		\addlinespace[6pt]
		$W^{(M)}$.distance & 0.177*** & 0.286*** & 0.110** & 0.174*** & 0.251** & 0.100*** \\
		$W^{(M)}$.GDPpc\_sum & -0.222*** & -0.275*** & -0.159** & -0.234*** & -0.173** & -0.197*** \\
		$W^{(M)}$.GDPpc\_sim & -0.051 & -0.195*** & 0.015* & -0.059** & -0.077* & -0.031 \\
		$W^{(M)}$.POP\_sum & 0.008 & -0.059 & 0.017 & 0.005 & -0.057 & 0.035 \\
		$W^{(M)}$.POP\_sim & -0.060 & -0.050 & -0.071 & -0.059 & -0.063 & -0.058 \\
		$W^{(M)}$.contig & 0.026 & 0.111*** & 0.048 & 0.028 & 0.098*** & -0.043 \\
		$W^{(M)}$.comlang & -0.006 & 0.137*** & -0.002 & -0.005 & -0.006 & -0.002 \\
		$W^{(M)}$.colony & 0.008 & 0.043* & -0.068 & 0.009 & 0.024 & -0.047 \\
		$W^{(M)}$.comcur & 0.090*** & 0.239*** & 0.067** & 0.093** & 0.173*** & 0.103*** \\
		$W^{(M)}$.rta & -0.140*** & -0.080* & -0.179** & -0.143*** & 0.052 & -0.205*** \\
		$W^{(M)}$.migration & -0.023** & -0.014* & -0.044* & -0.026** & -0.004 & -0.043** \\
		$W^{(M)}$.im\_strength\_net & - & - & - & 0.011 & -0.011 & 0.003 \\
		$\rho$ & 0.035 & 0.051 & 0.051 & 0.033 & 0.042 & 0.039 \\
		\bottomrule
	\end{tabular}
	\end{adjustbox}
	\label{chapt3_tab:durbin_panel}
\end{table}

\begin{table}[tb]
	\centering
	\caption[Impacts for Durbin model regression with instrumented migration and controls for import strength]{Impacts from pooled panel SDM model with instrumented migration. Without (i) and with (ii) controlling for import strength.}
	\begin{adjustbox}{width=\textwidth,totalheight=\textheight,keepaspectratio}
	\begin{tabular}{lrrrcrrrcrrr}
		\toprule 
		& \multicolumn{11}{c}{\bf Baseline specification} \\ 
		& \multicolumn{3}{c}{total trade} & & \multicolumn{3}{c}{different. goods} & & \multicolumn{3}{c}{homogen. goods} \\ 
		\cmidrule(lr){2-4} \cmidrule(lr){6-8} \cmidrule(lr){10-12}
		& direct & indirect & total & & direct & indirect & total & & direct & indirect & total \\ 
		\midrule
		
		distance & -0.782 & 0.154 & -0.628 & & -0.685 & 0.261 & -0.425 & & -0.810 & 0.071 & -0.738 \\
		GDPpc\_sum & 1.923 & -0.160 & 1.762 & & 1.987 & -0.180 & 1.807 & & 1.700 & -0.075 & 1.625 \\
		GDPpc\_sim & 0.967 & -0.018 & 0.949 & & 0.937 & -0.153 & 0.783 & & 0.915 & 0.064 & 0.979 \\
		POP\_sum & 1.663 & 0.067 & 1.729 & & 1.596 & 0.024 & 1.620 & & 1.589 & 0.102 & 1.692 \\
		POP\_sim & 0.786 & -0.034 & 0.753 & & 0.679 & -0.015 & 0.663 & & 0.817 & -0.031 & 0.786 \\
		contig & 0.217 & 0.035 & 0.252 & & 0.217 & 0.127 & 0.344 & & 0.189 & 0.060 & 0.248 \\
		comlang & 0.140 & -0.001 & 0.139 & & 0.138 & 0.085 & 0.223 & & 0.166 & -0.035 & 0.130 \\
		colony & 0.385 & 0.022 & 0.407 & & 0.280 & 0.060 & 0.340 & & 0.340 & -0.053 & 0.287 \\
		comcur & 0.167 & 0.098 & 0.265 & & 0.185 & 0.259 & 0.444 & & 0.210 & 0.081 & 0.291 \\
		rta & 0.114 & -0.139 & -0.025 & & 0.296 & -0.068 & 0.228 & & 0.046 & -0.184 & -0.138 \\
		migration & 0.092 & -0.021 & 0.071 & & 0.115 & -0.009 & 0.106 & & 0.073 & -0.042 & 0.031 \\
		\addlinespace[6pt]
		& \multicolumn{11}{c}{\bf Controlling for import strength} \\ 
		& \multicolumn{3}{c}{total trade} & & \multicolumn{3}{c}{different. goods} & & \multicolumn{3}{c}{homogen. goods} \\ 
		\cmidrule(lr){2-4} \cmidrule(lr){6-8} \cmidrule(lr){10-12}
		& direct & indirect & total & & direct & indirect & total & & direct & indirect & total \\ 
		\midrule
		distance & -0.783 & 0.152 & -0.631 & & -0.679 & 0.229 & -0.449 & & -0.815 & 0.070 & -0.745 \\
		GDPpc\_sum & 1.918 & -0.174 & 1.743 & & 2.228 & -0.081 & 2.146 & & 1.588 & -0.140 & 1.448 \\
		GDPpc\_sim & 0.964 & -0.028 & 0.936 & & 1.059 & -0.034 & 1.025 & & 0.867 & 0.003 & 0.870 \\
		POP\_sum & 1.660 & 0.062 & 1.722 & & 1.787 & 0.020 & 1.806 & & 1.498 & 0.096 & 1.593 \\
		POP\_sim & 0.785 & -0.034 & 0.751 & & 0.798 & -0.031 & 0.767 & & 0.759 & -0.030 & 0.729 \\
		contig & 0.217 & 0.036 & 0.253 & & 0.181 & 0.110 & 0.290 & & 0.207 & 0.052 & 0.259 \\
		comlang & 0.140 & 0.002 & 0.142 & & 0.145 & -0.000 & 0.144 & & 0.162 & 0.004 & 0.158 \\
		colony & 0.385 & 0.023 & 0.407 & & 0.253 & 0.036 & 0.289 & & 0.355 & -0.035 & 0.321 \\
		comcur & 0.168 & 0.101 & 0.268 & & 0.203 & 0.187 & 0.390 & & 0.203 & 0.114 & 0.317 \\
		rta & 0.112 & -0.143 & -0.030 & & 0.296 & -0.041 & 0.255 & & 0.044 & -0.210 & -0.165 \\
		migration & 0.092 & -0.023 & 0.068 & & 0.141 & 0.002 & 0.143 & & 0.062 & -0.042 & 0.021 \\
		\bottomrule
	\end{tabular}
	\end{adjustbox}
	\label{chapt3_tab:durbin_panel_imp}
\end{table}

The SDM model controls both for the dyad and for the IMN lagged explanatory variables, in order to allow changes in a given explanatory variable associated with a single country-pair to affect the pair itself and to potentially reverberate across all other dyads indirectly. 
This rich set of information increases the difficulty of interpreting the regression results.
For the sake of clarity, we therefore calculate the direct and indirect impacts as suggested by \citet{pace2009introduction} and discussed by \citet{lesage2014interpreting} for exogenous and endogenous flow models.

We present the figures in Table \ref{chapt3_tab:durbin_panel_imp}.\footnote{We compute these models in R with the spdep package. The models have been fitted using Monte Carlo simulations with 1000 replications using traces of powers of the network weight matrix, which considerably reduces computation time.}
Comparing the first three columns of Table \ref{chapt3_tab:durbin_panel} with the upper panel of Table \ref{chapt3_tab:durbin_panel_imp}, we see that the direct effect of migration is in line with OLS and FE results displayed above (see Table \ref{chapt3_tab:baseline_panel}). 

Analyzing the total effects, we note a negative indirect coefficient for differentiated goods, which significantly lowers the total impact of migration on trade. 
One possible interpretation of this negative indirect effect is that migrants also bring knowledge, competences and business contacts that are particularly relevant for producing and exporting differentiated goods. 
As a result, migration from $i$ to $h$ may erode $i$'s ability to export specific goods to other markets (e.g. to country $j$), making $h$ a better competitor. So, we decided to estimate and report both the SDM regression results without (first three columns of table \ref{chapt3_tab:durbin_panel} and upper panel of table \ref{chapt3_tab:durbin_panel_imp}) and with (last three columns of table \ref{chapt3_tab:durbin_panel} and bottom panel of table \ref{chapt3_tab:durbin_panel_imp}) controlling for this phenomena.
To control for this effect we include total import by $j$ net of imports from $i$ among the controls:
\begin{equation}
im.strength.net_j = \sum_{k\neq i} T_{kj} - T_{ij}\ .
\end{equation} 

Results that account for import strength appear in columns 4--6 of Table \ref{chapt3_tab:durbin_panel} and in the bottom part of Table \ref{chapt3_tab:durbin_panel_imp}. 
The additional control turns out highly significant and negative for differentiated goods, suggesting that export of such products from $i$ to $j$ is substituted by trade from other sources. 
Moreover, migration coefficients change considerably: accounting for import strength of the destination country, the total effect of migration for the differentiated goods is now significantly higher than for homogeneous goods (0.143 versus 0.021).

All in all, looking at the results controlling for import strength, we note that the GDP coefficients slightly decrease for the effect of the inclusion of the lagged GDP terms ($W^{(M)}$.GDPpc\_sum), that highlights a negative indirect impact. 
The distance coefficient also decreases when we introduce lagged terms. In particular, distance matters more for trade of homogeneous goods compared with differentiated goods, to which corresponds a total impact of -0.449, significantly smaller than the traditional distance coefficient for total trade, which vary from -0.7 to -1 in the literature. Interestingly, we find a negative indirect effect for the RTA dummy: this can be easily rationalized if we think that a trade agreement between a country's export partners is likely to have negative ``indirect'' effect on that country's ability to export because of trade diversion effects. 

The impact of migration on trade is significantly higher for differentiated goods compared to homogeneous ones: the gap in the effect becomes even larger when we consider the total impact rather than only the direct one.
In fact, we find a negative indirect impact of migration on total and homogeneous goods trade, while the counterpart for differentiated is close to zero. The negative indirect impact can be interpreted as a competition effect: having more third country migrants in the importer country reduces trade between the country pair. This is true for total trade and homogeneous goods trade, but not for differentiated goods and it is likely to depend on the fact that the latter are more difficult to substitute for, so that they suffer less competition from third countries.

We next control for the residual autocorrelation in the SDM model using the Moran I test. 
Looking to columns 6-7 of Table \ref{chapt3_tab:moranI}, we obtain encouraging results: using both the spatial matrix and the IMN matrix, the autocorrelation that was present both in the OLS and the FE residuals is no longer significant. 
This provides further support to our statement: the SDM model associated with a weight matrix based on IMN successfully captures MTR. 
We also check for the normality assumption of the CML residuals in the selected model: they are normally distributed, confirming that the model is reasonably well-specified.

All in all, the controls for network interdependencies are always significant in our analysis. This means that the baseline gravity model does not account for network effects, which play a relevant role in shaping the World Trade Web. Furthermore, trade in differentiated goods is more strongly affected by migration, as predicted by Rauch.

\section{Conclusions}\label{chapt3_sec:concl}
Increased data availability both at national and international levels has triggered a host of research on the relationship between trade and migration. We contribute to this line of research by applying spatial econometric techniques exploiting the topological distance of the migration network,\footnote{As opposed to the analysis in Chapter \ref{chapt:migr1} where we used instead the trade network.} rather than the usual geographical standard geographic space, in order to look at direct and indirect effects of migration on trade.
In this way we can investigate the network effects suggested by Rauch's seminal papers from a global perspective, rather than focusing on a single ethnic network as done in the literature so far. 

Moreover, using pooled panel data, the results of this chapter are of a broader scope than those obtained in Chapter \ref{chapt:migr1}, where we analyzed data only for the year 2000. In this sense the conclusions of this chapter are more general, covering data from three decades and are less useful for policy implementation as they do not highlight the most recent mechanisms of interplay between trade and migration, but rather the more universal, time-independent ones.

Our work also contributes the literature of spatial economics / econometrics that aims to control for the multilateral resistance terms in the constraint gravity equation for trade. 
We can draw several conclusions. First, accounting the multilateral resistance terms by means of a SDM specification using a International Migration Network (IMN) weight matrix, we filter out the residual autocorrelation. 
Furthermore, from a qualitative point of view we confirm the finding that migration has a larger impact on differentiated products, both at direct and global (network) level.
Indeed, the negative effect that third-country migrants have on trade of homogeneous goods (testified by the negative indirect impact found in the estimation results) and that we rationalize as a competition effect, is no longer there when we focus on differentiated goods.

%% file: mainmatter/4/chapt4.tex
%

\chapter[Trade and Foreign Investments]{Trade \& Foreign Investments: A Comparative Network Analysis Approach}
\label{chapt:fdi}

\graphicspath{{4/figures/PNG/}{4/figures/PDF/}{mainmatter/4/figures/}}

\section{Introduction}
International trade and Foreign Direct Investments (FDI) are two phenomena in continuous evolution, that play a fundamental role in shaping countries economies and development.
In today's world economy these two phenomena are also intimately related through the processes of the Global Value Chain (GVC): transaction costs and technological capability had led in the last decades to a fragmentation of the production process and the building of proper \textit{productive networks} \citep{Gereffi2005,Ntras2013}. Usually coordinated by transnational corporations these networks have the purpose of optimizing the production process by placing different production stages in countries where they are more cost-effective and, by doing so, link many world countries through various FDIs and trade flows. Globally this is a huge phenomenon: in 2010 for example, these international production networks accounted for approximately two thirds of both exports and imports of goods for the United States and France \citep{UNCTAD2013}.

If much has been done to model both trade and FDI separately \citep{linnemann1966econometric, serrano2003topology, baltagi2007estimating}, less is known about the interplay between the two, both with network approaches and with econometric models, since the traditional specifications (e.g. the gravity model) for trade does not account for FDI and vice-versa. This is exactly the gap we want to close in this chapter. For trade we will use BACI \citep{CEPII:2010-23}; while for FDI a dataset on transnational corporations (TNC) control networks, recently published in \citet{Altomonte2013}:\footnote{For a network analysis on similar data see \citet{vitali2011network}.} in particular we will assume TNC affiliates as stocks of FDI, thus not having data on investments flows but rather a snapshot of the network of global corporate control as it was in 2010.\footnote{This choice is due to the scarce availability of data about FDI flows on a worldwide scale. In fact, in this sense, the main dataset available is that provided by OECD, thus limited to 34 countries.}

With this setting, we initially analyze the two phenomena in a network perspective and then employ a gravity equation, where we account for zero flows using the Heckman 2-step (H2S) selection model \citep{Heckman1979}, to study the correlations between them and with other factors.

In particular, to start addressing how these phenomena are related we look at them as layers of a macroeconomic network, thus considering countries as nodes and trade and FDI channels as two different types of directed weighted edges. A similar approach has recently been used also for trade and migration \citep{Fagiolo2014,Sgrignoli2013}, for trade and financial integration \citep{Schiavo2010} or for trade and the Internet \citep{riccaboni2013global}.
The reason why this approach is important is that it lets investigate the influences of topological patterns on various phenomena, understand the extent to which different layers display similar topological properties and whether these are correlated, or causally linked. We also identify the main determinants of these correlations and we find that economic and demographic country size, as well as geographical distance, play a key role in explaining differences and similarities between the two sub-networks topologies.

Then we aim at better understand the structure of the GVC studying the interaction between trade and FDI with the H2S selection model to control for the large number of zeros in the data: we confirm the positive correlation found before and its dependence on economic and demographic countries characteristics. Next we study the interaction with distance and upstreamness, to check for possible non-linearity of this relation depending on countries distance and the industry position in the Global Supply Chain (GSC).
About distance we found a positive effect of the interaction with FDI on trade, i.e. trade and FDI tend to be more positively correlated as the geographical distance between the exporter and the importer countries increases; next, to account for the GSC, we introduce \textit{upstreamness} \citep{Antras2012}, that is a measure of how much of an industry's output goes for final use, finding that trade is more intense for products closer to the final use and that, the more the industry is upstream (far from final use), the more FDI and trade tend to be substitutes. 
A possible interpretation for this result being vertical investments: consistent parts of the production process are outsourced to other countries and then final goods are sold (traded) from few assembling facilities.

Moreover we ask what is the influence of Regional Trade Agreements (RTA) and if there is any special pattern for investments in Asian countries. Our findings are that, as expected, RTA are positively correlated with trade and that, when RTA are in place, trade and FDI tend to become substitutes, i.e. when some RTA is in place, by favoring trades, they tend to reduce investments. As for Asian countries, we discovered there are in general larger volumes of trade and a stronger complementarity with FDI when the exporter country is Asian.

Finally we quantify the correlation between trade and FDI by distinguishing for the three economic macro-sectors: a positive correlation is found for the secondary sector, while a negative one exist for the tertiary and we propose to justify this by thinking that firms make a discrete choice between the two channels.

The remainder of this chapter is structured as follow: Section \ref{chapt4_sec:data-and-definitions} describes our dataset as well as introducing our multiple-layer network; Section \ref{chapt4_sec:networks-comparison} discusses the correlation and topological properties of the trade and FDI sub-networks; in Section \ref{chapt4_sec:economic-and-econometric-approach} our econometric methodology is introduced, while Section \ref{chapt4_sec:results} presents and discusses the results. Section \ref{chapt4_sec:conclusions} concludes.

\section{Data and definitions}\label{chapt4_sec:data-and-definitions}
Trade data is taken from the BACI dataset \citep{CEPII:2010-23}, which originates from the data reported by over 150 countries to the United Nations Statistics Division (COMTRADE database) but also integrates new approaches to reconcile those reports, in order to have a single consistent figure of a bilateral flow.
The version we use (with HS96) covers more than $200$ countries and $5000$ products, between 1998 and 2012.

The Foreign Direct Investment dataset has been taken from a recent work \citep{Altomonte2013} where the authors have reconstructed the international business groups structure starting from worldwide proprietary linkages and firm-level financial accounts, where control was assumed if the parent exceeded the majority (50.01\%) of voting rights of the affiliate. In this way they ended up with data about $270,374$ headquarters of business groups, controlling a total of $1,519,588$ affiliates in 207 countries in the year 2010. The dataset thus indicates for each pair of countries the number of control links present between the two, i.e. counts the business groups branching between countries, and the industry (NAICS) sectors of both the parent and affiliate firms. For the sectors-level analysis in this work we always considered only the affiliate industry, as parents are usually big holding groups which official nature may not be meaningful for the effective sector they operate into.
In the dataset, two thirds of business groups are originated in OECD economies, with those headquarters controlling around 75\% of affiliates recorded in the data. Headquarters located in countries of the European Union, in particular, control 48\% of total affiliates, of which roughly one third ($259,278$) are located abroad. The situation is different in the US, where around 46\% of the affiliates controlled by American headquarters are located abroad. Developing countries, not surprisingly, have a larger share of domestic groups, with about 80\% of the $371,577$ affiliates controlled by non-OECD headquarters located domestically.
In the remainder of this chapter we will interpret this corporate control network as stock FDI.

Using the two datasets together and matching the industries (NAICS) and products (HS96) codes \citep{ramon} we retain only those countries that are present in both of them to enhance comparability, obtaining a set of 194 countries (nodes). The only year present in both dataset is 2010, thus it will be the only analyzed here.

We employ additional country-specific data as real Gross Domestic Product per-capita (\textit{rGDPpc}) and population (\textit{POP}) from the World Bank. We also use bilateral country geographic, political and socioeconomic data from the CEPII GeoDist dataset \citep{mayer2011notes}: this includes information about between-country geographical distance ($\Delta$);\footnote{We employ the great-circle definition of country distances. Our results do not change using alternative distance definitions.} geographical contiguity (\textit{contig}), i.e. whether two countries share a border; colony relationship (\textit{colony}), i.e. whether one of the two countries have ever been a colony of the other; the case where two countries have ever been unified (\textit{smctry}) and ethnical language commonality, as spoken by at least 9\% of population (\textit{comlang\_ethno}). We will use these variables to perform the gravity regressions of Section \ref{chapt4_sec:results}.

Both trade and FDI can be viewed as macroeconomic networks where we identify countries as nodes and interaction channels between them as edges, thus representing two possible interaction layers connecting world countries we will call \textit{World Trade Web} (WTW) and \textit{FDI Network} (FDIN). In fact we will have a network composed of $N=194$ nodes, with two kinds of weighted directed edges, i.e. two $N\times N$ adjacency matrices, one relative to bilateral trade flows ($T$) and the other to bilateral FDIs ($F$). The generic element of $T$ ($F$) represent the value of exports $t_{ij}$ (amount of FDI $f_{ij}$) from country $i$ to country $j$.

The analysis we have done are for the total flows and for the three main economic macro-sectors, namely the \textit{primary} (raw materials), the \textit{secondary} (manufacturing) and the \textit{tertiary} (services). In terms of NAICS codes we identify the raw materials with sectors 11, 21 and 22, respectively ``Agriculture, Forestry, Fishing and Hunting'', ``Mining, Quarrying, and Oil and Gas Extraction'' and ``Utilities''; the manufacturing with codes 31-33 that is exactly ``Manufacturing'' in the NAICS classification; services with sectors 51, 54 and 71, that correspond respectively to ``Information'', ``Professional, Scientific, and Technical Services'' and ``Arts, Entertainment, and Recreation''.\footnote{Trade datasets essentially contain data only about the first two macro-sectors, as sale of services are not registered by customs. However the sectors we identify as tertiary contain some products that can work as proxies for trade in services: just to give a few examples ``Information'' relate for example to trade of newspapers and journals, books, photographic plates and films; in the ``Professional, Scientific, and Technical Services'' we find, among others, trade of ``Plans And Drawings For Architectural, Engineering, [...] purposes'' and in the ``Arts, Entertainment, and Recreation'' we find clothes and costumes, toys and paintings. In particular the choice of the sectors to use has been based on the direct correspondences between CN and CPA, two classifications derived from HS and NACE respectively, found at \citet{ramon}.}

To give an idea of the intensity of these flows, in Figure \ref{chapt4_fig:maps} we plot the top 5\% of the direct weighted links in the four cases: the edges color identify whether the flow is predominantly in trade (blue), in FDI (red) or both (green); the edges thickness is proportional to the log of links weight;\footnote{In the cases where both trade and FDI have a significant flow intensity (green edges) the weight is calculated as the mean of the two after normalization.} while nodes size and color are proportional, respectively, to the log of POP and the log of rGDPpc.

These maps allows one to appreciate the differences among the four cases and their main characteristics, e.g. the predominance of long distance trade-only and FDI-only in the primary sector; the intense mixed edges between Asian countries and the United States in the secondary sector; or the intense and diverse relations between the United States and the United Kingdom in tertiary, FDI-only in one direction and mixed trade-FDI in the other. Also notice the central role of China as partner of the  United States and many European countries in all the four cases. These many differences justify the interest in analyzing the three macro-sectors separately.

\begin{landscape}
	\begin{figure}[tb]
		\caption[Map of the World Trade Web and International FDI Network flows, total and for the three macro-sectors]{The World Trade Web and International FDI Network for total flows (\ref{chapt4_fig:maptot}) and the three macro-sectors (\ref{chapt4_fig:mapprim}, \ref{chapt4_fig:mapsec}, \ref{chapt4_fig:mapter}). The figure plots the directed top 5\% links by weight. Blue links represent a trade-only relation and the red a FDI-only, while green indicate the presence of both a trade and FDI relation. Edges thickness is proportional to the log of link weights. Nodes size is proportional to the log of POP, while color (from blue to red) is proportional to the log of rGDPpc.}
		\label{chapt4_fig:maps}
		\centering
		\subfloat[Total\label{chapt4_fig:maptot}]{\includegraphics[width=\myFiguresSize]{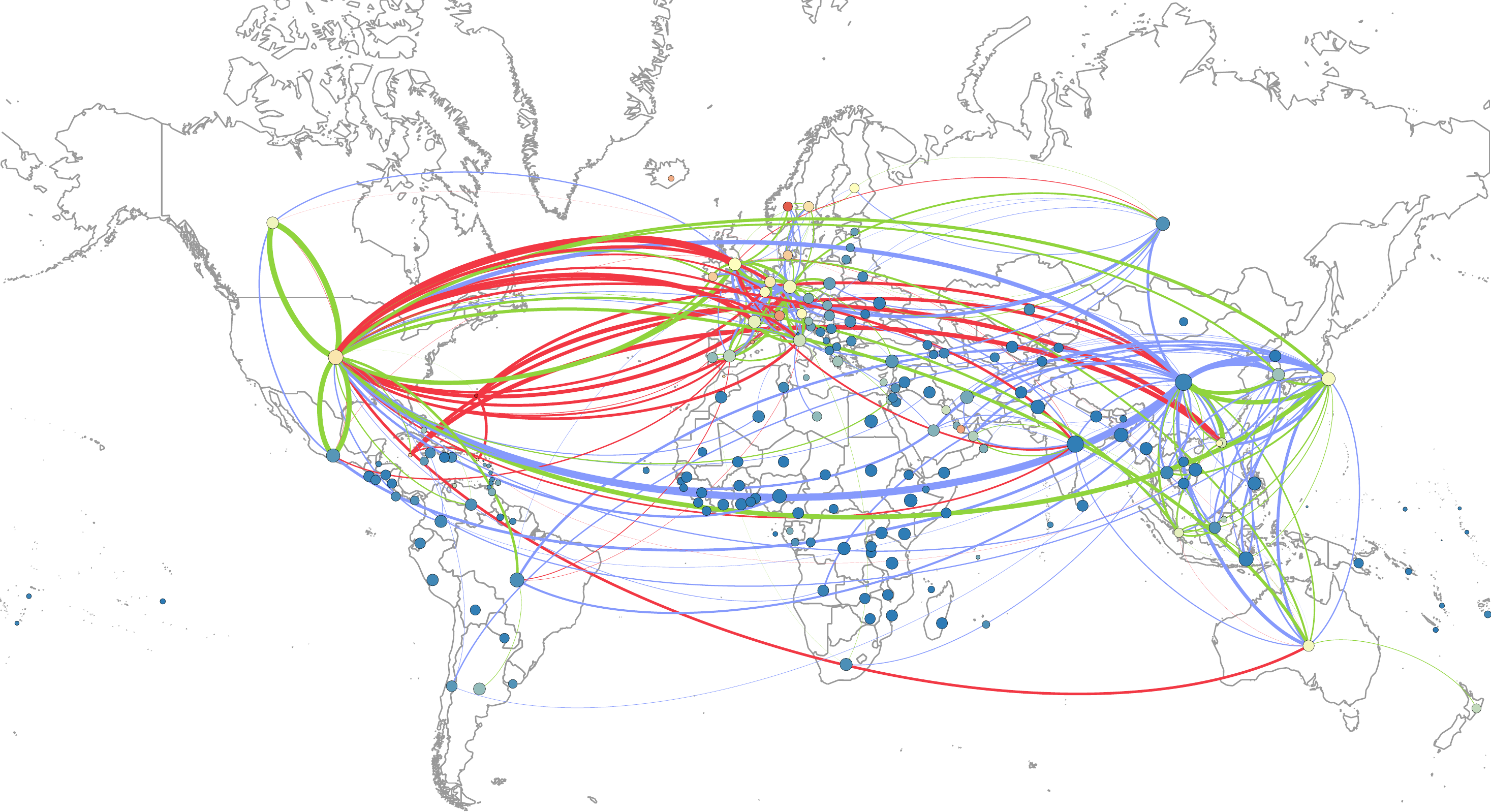}}	
	\end{figure}
\end{landscape}

\begin{figure}[tb]
	\ContinuedFloat
	\caption[]{See caption on previous page.}
	\label{chapt4_fig:maps2}
	\centering
	\subfloat[Primary\label{chapt4_fig:mapprim}]{\includegraphics[width=\myFiguresSize]{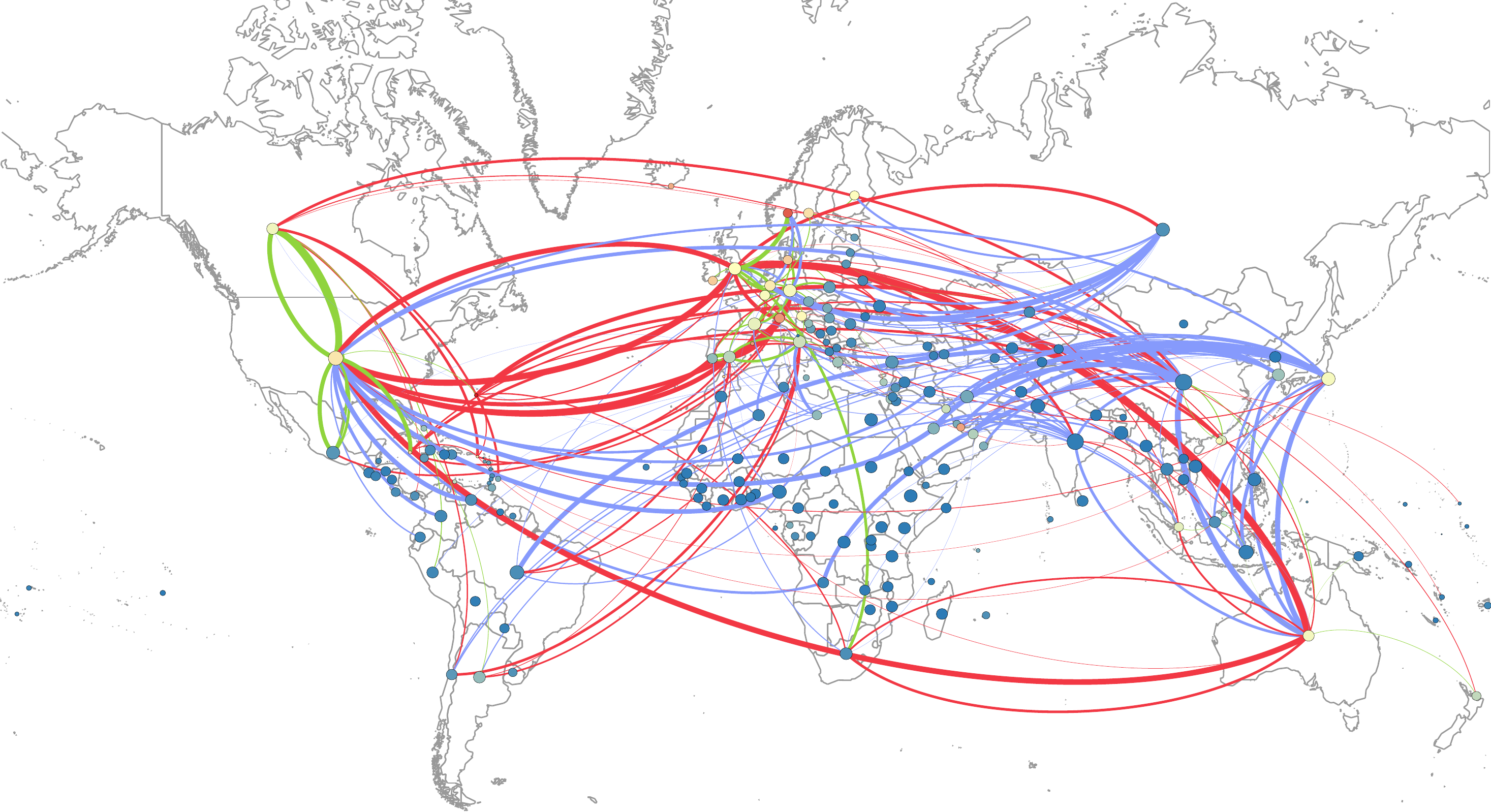}} \par
	\subfloat[Secondary\label{chapt4_fig:mapsec}]{\includegraphics[width=\myFiguresSize]{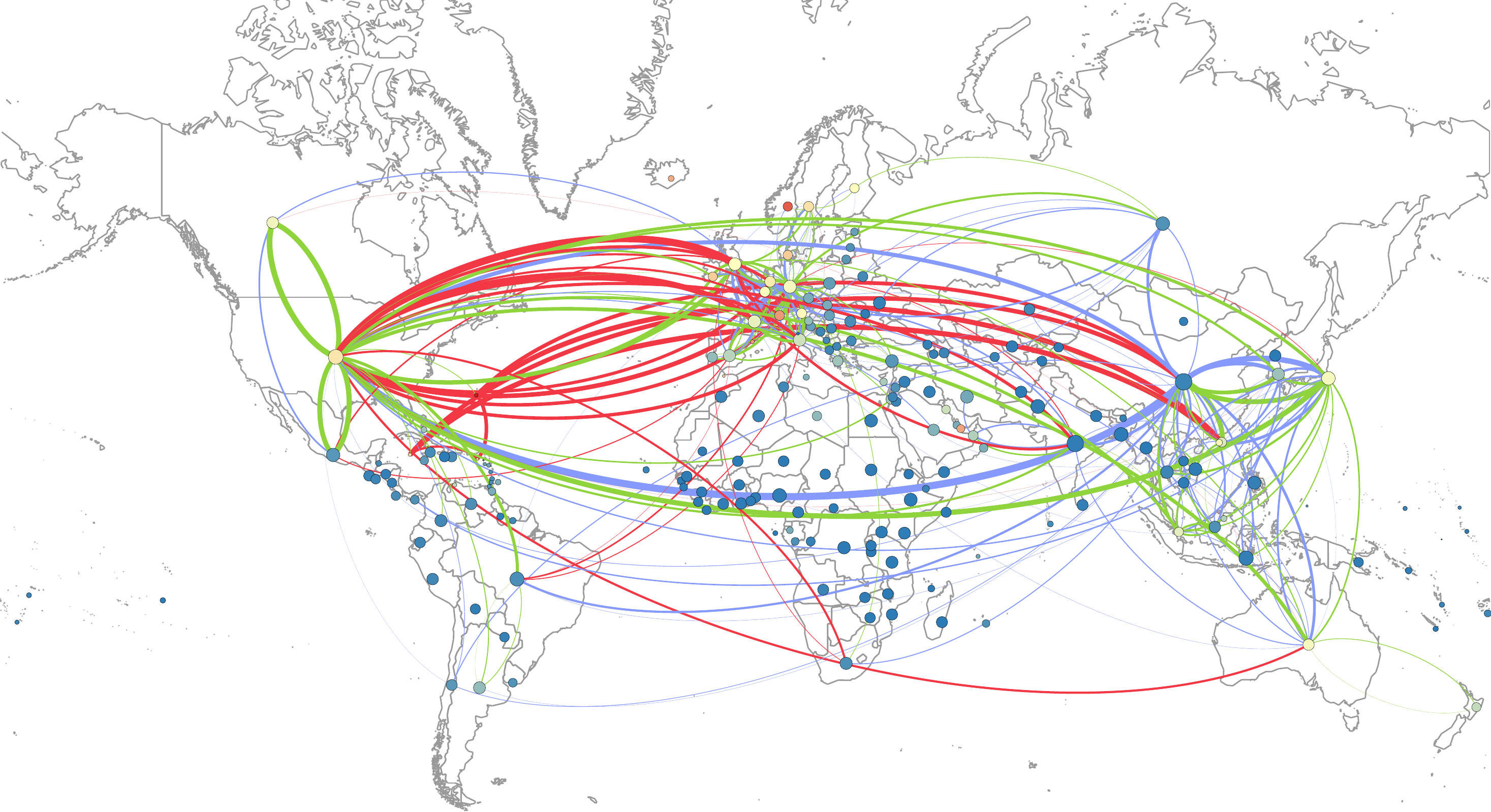}} \par
	\subfloat[Tertiary\label{chapt4_fig:mapter}]{\includegraphics[width=\myFiguresSize]{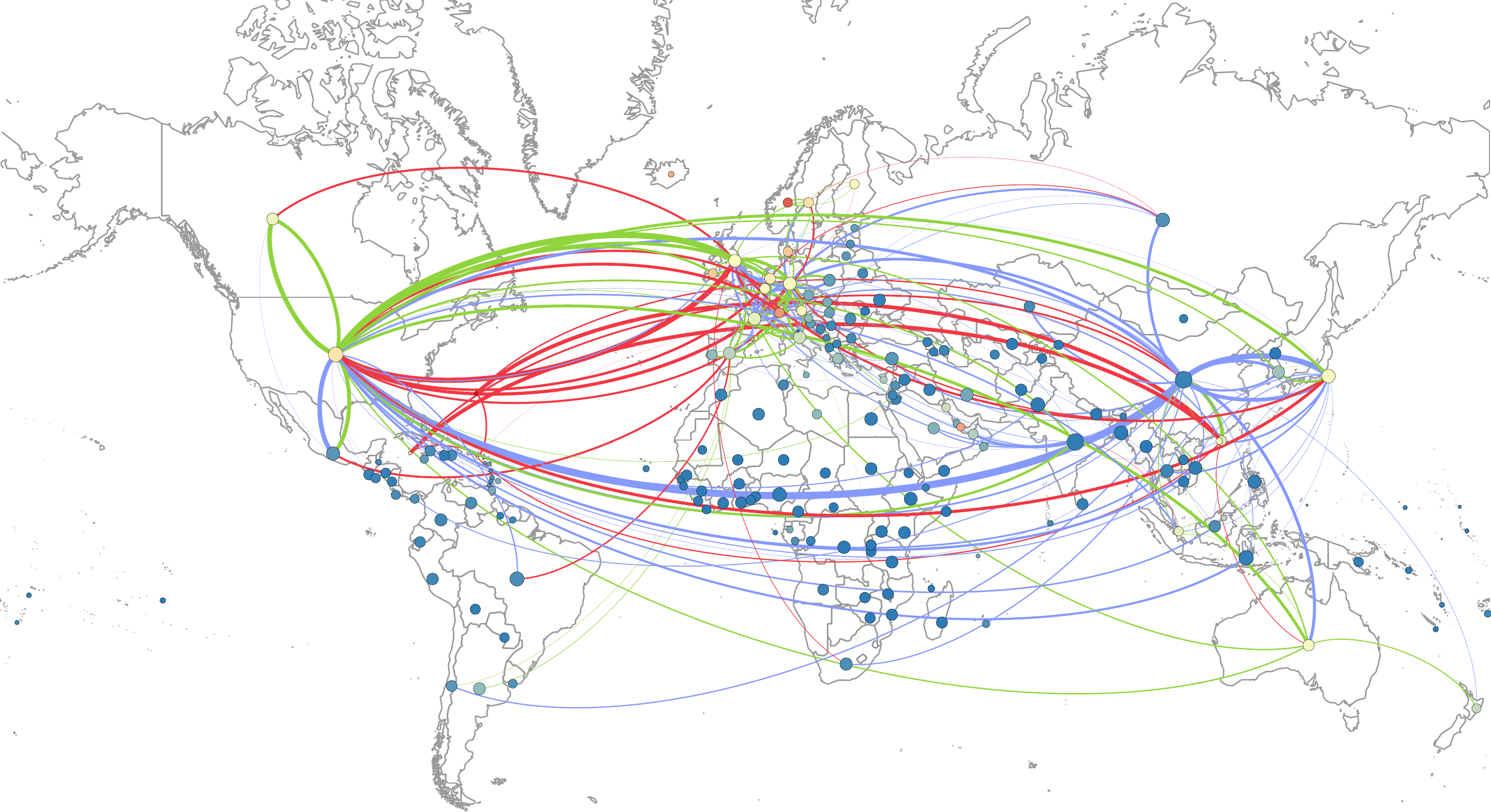}}
\end{figure}


\section{Networks comparison}\label{chapt4_sec:networks-comparison}
As a first step in the analysis of trade and FDI we present some descriptive statistics of the two networks, WTW and FDIN, separately, moving then to some comparative analysis of the two. In particular we present results for the total and for the three main economic macro-sectors networks as introduced before.\footnote{We consider the presence of a link when there is a non-zero flow between any two countries.}

\begin{table}[tb]
	\centering
	\caption[WTW and FDIN descriptive network statistics]{WTW and FDIN descriptive network statistics.}
	\label{chapt4_tab:zeros}
	\begin{adjustbox}{width=\textwidth,keepaspectratio}
		\begin{tabular}{lcccccccc}
			\hline \hline
			\rule{0pt}{2ex} & \multicolumn{2}{c}{Tot} & \multicolumn{2}{c}{Primary} & \multicolumn{2}{c}{Secondary} & \multicolumn{2}{c}{Tertiary} \\
			& {\footnotesize WTW} & {\footnotesize FDIN} & {\footnotesize WTW} & {\footnotesize FDIN} & {\footnotesize WTW} & {\footnotesize FDIN} & {\footnotesize WTW} & {\footnotesize FDIN} \\
			\hline
			\rule{0pt}{3ex}Average degree (\%) & 62.4 & 10.4 & 41.4 & 4.4 & 61.7 & 8.4 & 47.9 & 6.1 \\
			Density (\%) & 70.9 & 16.5 & 51.9 & 7.3 & 70.5 & 13.5 & 58.3 & 9.8 \\
			Reciprocity (\%) & 76.7 & 26.5 & 60.3 & 19.9 & 75.8 & 25.2 & 65.3 & 25.7 \\
			In-Out corr. coeff. & 0.92 & 0.73 & 0.87 & 0.72 & 0.91 & 0.70 & 0.90 & 0.76 \\
			Average cluster coeff. & 0.20 & 0.04 & 0.18 & 0.03 & 0.20 & 0.04 & 0.17 & 0.04 \\
			Assortativity & -0.327 & -0.412 & -0.325 & -0.404 & -0.328 & -0.398 & -0.350 & -0.380 \\
			\hline \hline
		\end{tabular}
	\end{adjustbox}
\end{table}


The first results of Table \ref{chapt4_tab:zeros} is that the FDIN is much more sparse than the WTW, with density and average degree 5 to 10 times larger and the presence of reciprocal links approximately threefold for trade with respect to FDI.

An important aspect of networks is their hierarchical structure, which is usually analyzed by means of the clustering coefficient and degree-degree correlation.
The \textit{clustering coefficient} of vertex $i$, with degree $k_i$, is defined as $c \equiv 2n_i/k_i(k_i-1)$, where $n_i$ is the number of neighbors of $i$ that are interconnected. Its values for our networks are $\simeq 0.2$ for WTW and $\simeq 0.05$ for FDIN, capturing the fact that WTW has a stronger clustered structure, i.e. the neighbors of a given vertex are interconnected with higher probability.

\begin{figure}[tb]
	\centering
	\caption[Disassortativity patterns of WTW and FDIN]{Disassortativity patterns of WTW and FDIN. Marker size is proportional to logs of $\text{POP}_i$. Colors scale (blue to red) is from lower to higher values logged $\text{rGDPpc}_i$.}
	\vspace{-10pt}
	\subfloat[WTW\label{chapt4_fig:s_vs_anns_trade}]{\includegraphics[width=\myFiguresSizeBis]{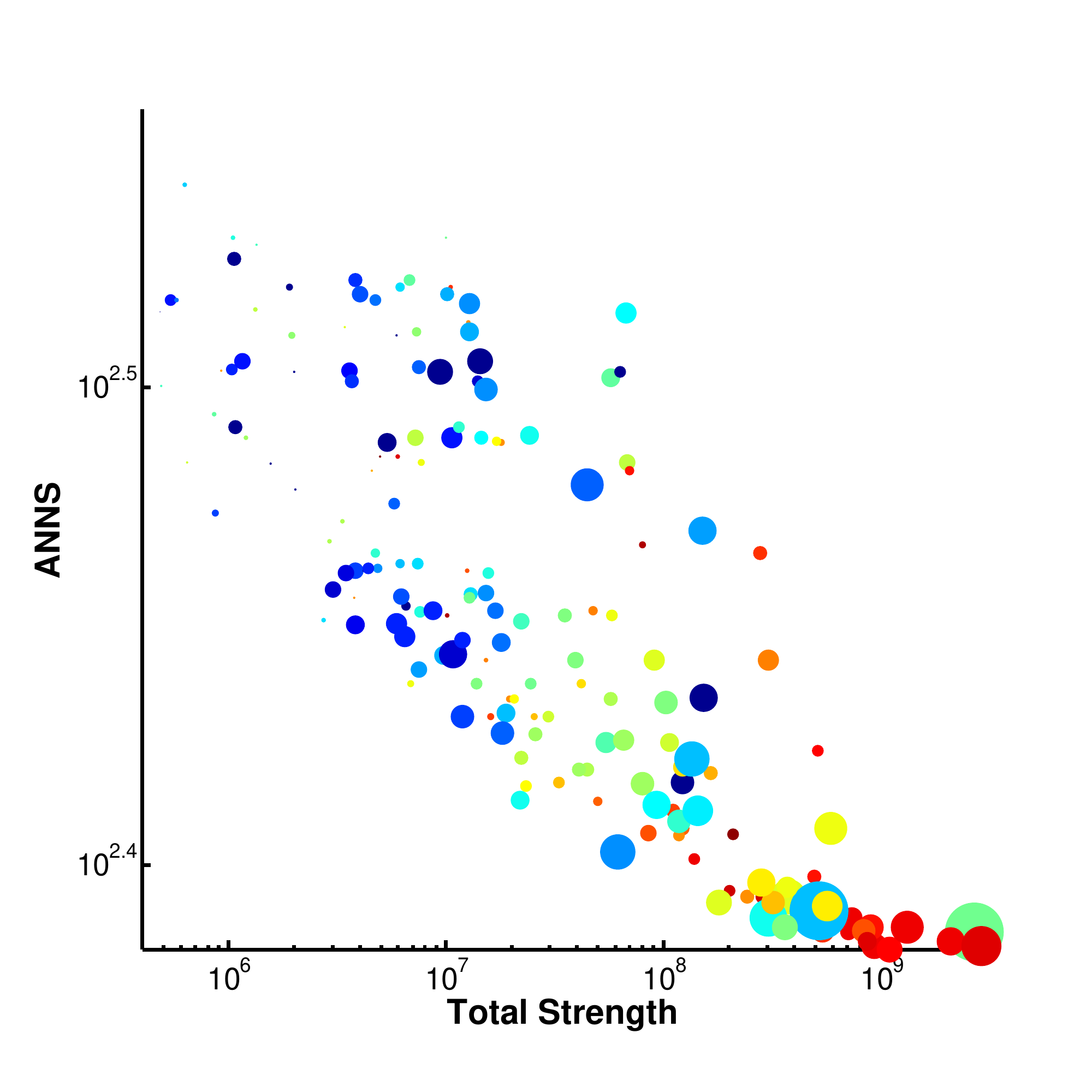}}
	\subfloat[FDIN\label{chapt4_fig:s_vs_anns_fdi}]{\includegraphics[width=\myFiguresSizeBis]{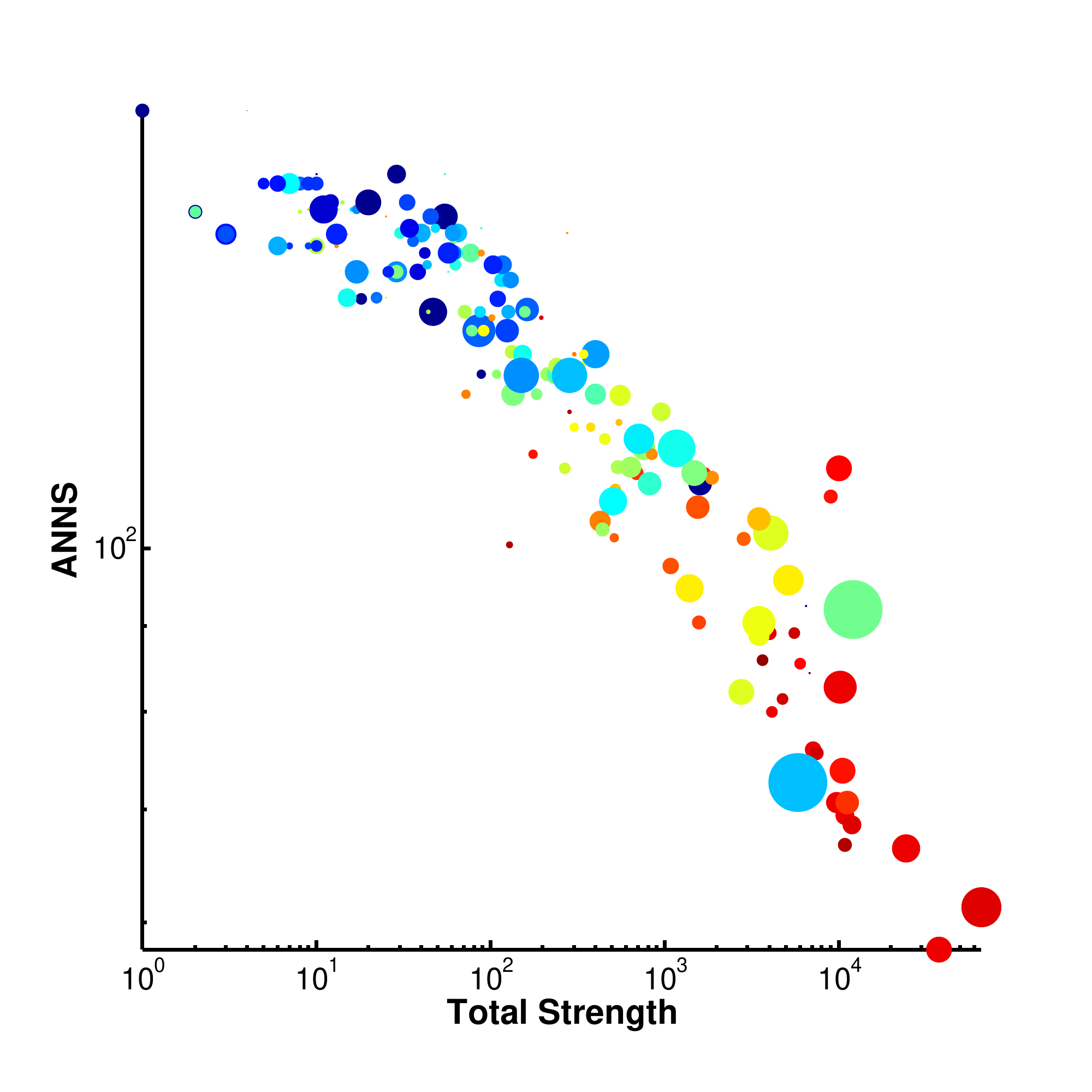}}
	\vspace{-10pt}
	\subfloat[WTW\label{chapt4_fig:degree_vs_annd_trade}]{\includegraphics[width=\myFiguresSizeBis]{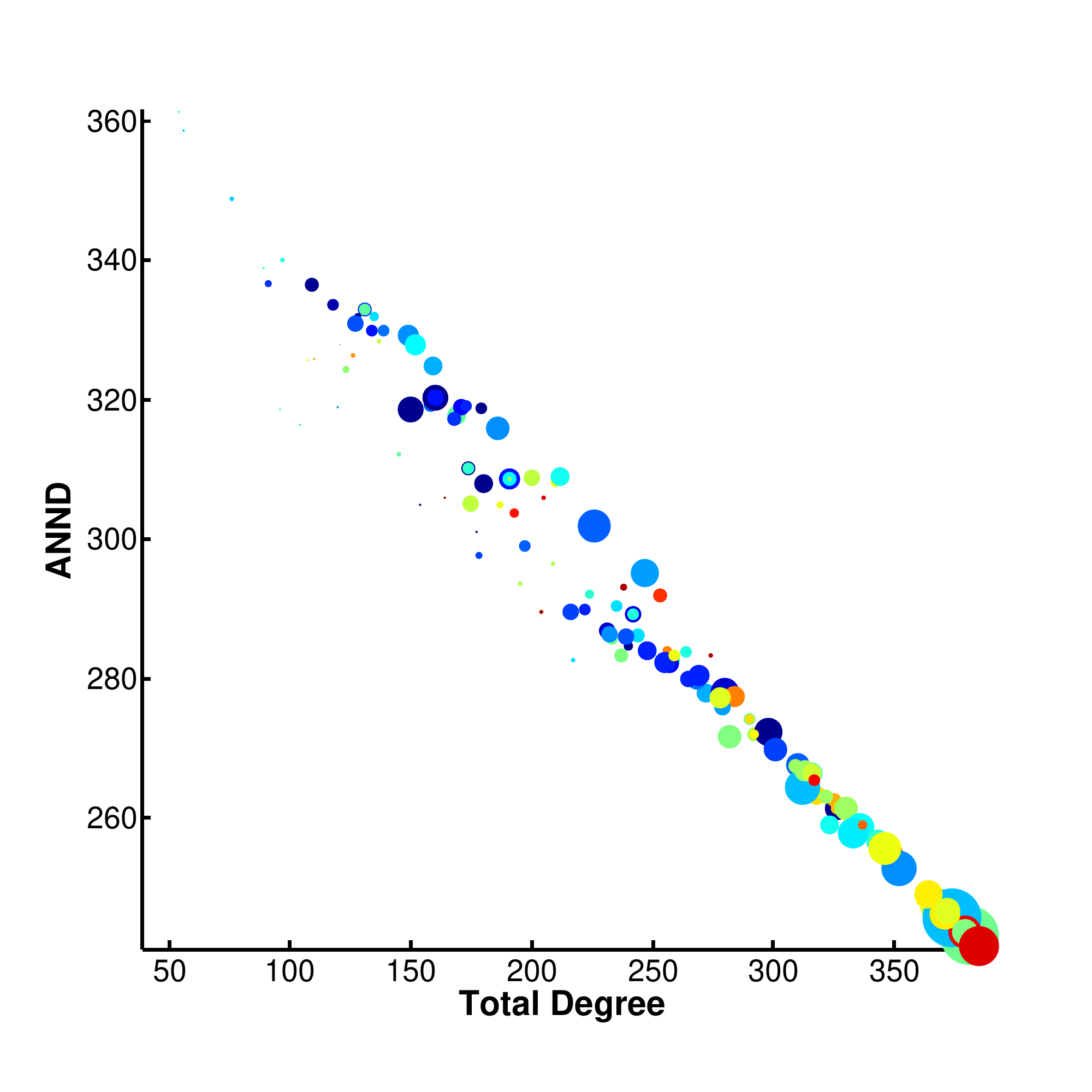}}
	\subfloat[FDIN\label{chapt4_fig:degree_vs_annd_fdi}]{\includegraphics[width=\myFiguresSizeBis]{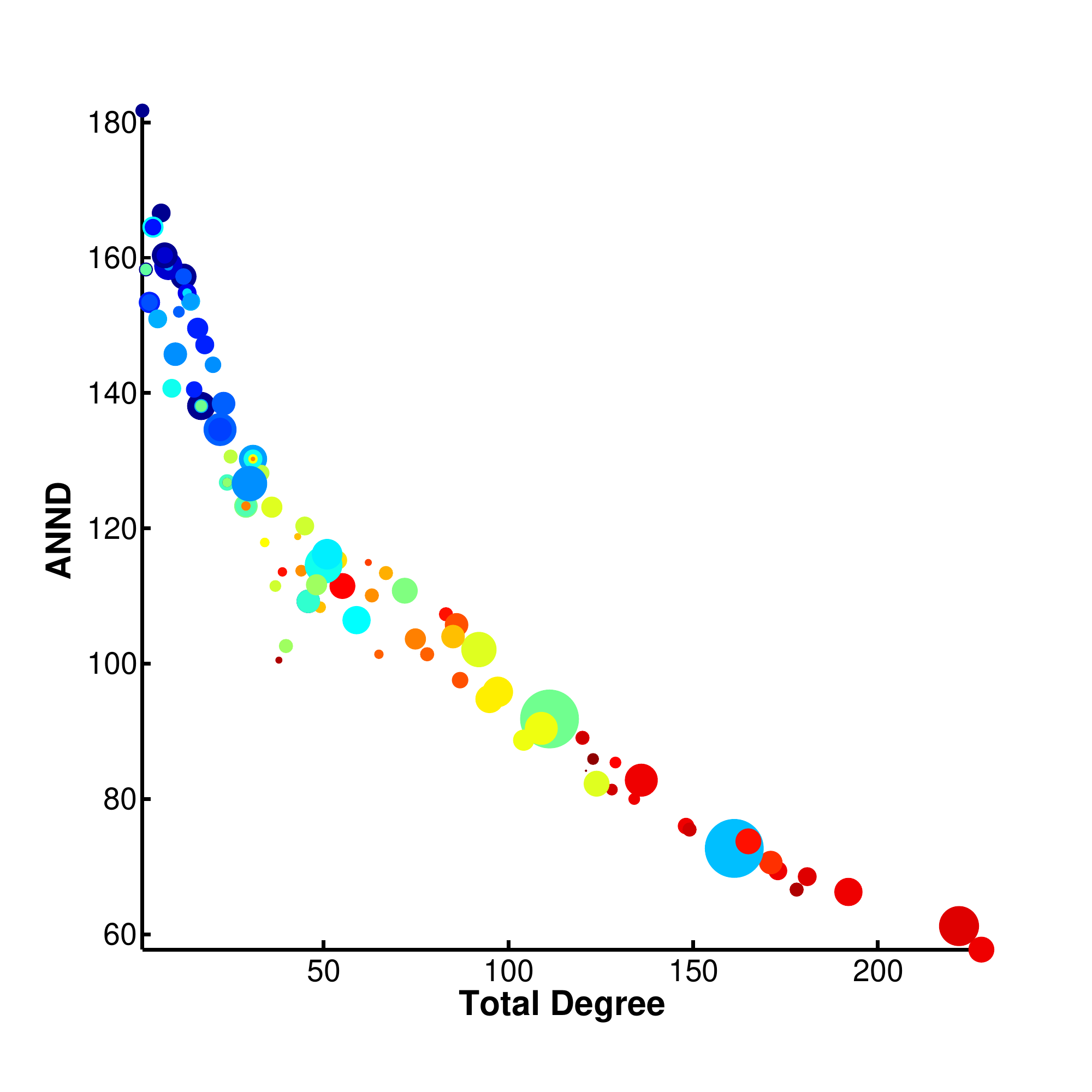}}
	\vspace{-10pt}
	\label{chapt4_fig:disassortativity}
\end{figure}

Moreover hierarchy is also reflected in the degree-degree correlation through the conditional probability $P(k|k')$, i.e. the probability that a vertex of degree $k'$ is linked to a vertex of degree $k$. This function is difficult to measure, due to statistical fluctuations, and it is usually substituted by the \textit{Average Nearest Neighbors Degree} (\textit{ANND}), defined as $\left\langle k_{nn}(k)\right\rangle =\sum_{k'}k'P(k'|k)$ \citep{pastor2001dynamical}. For independent networks this quantity would result independent of $k$. In analogy with ANND, if we use nodes' strength instead of degree, we obtain the \textit{Average Nearest Neighbor Strength} (\textit{ANNS}). Figure \ref{chapt4_fig:disassortativity} report ANND (ANNS) for the two networks, showing the dependency on the vertex's degree (strength) and indicating that, in both cases, highly connected vertexes tend to connect to poorly connected ones, i.e. they show a \textit{disassortative} behavior. In the scatter plots we also relate this behavior with countries POP end rGDPpc: bigger countries (by both economic and demographic indicators) have in general higher levels of degree (strength), thus connecting to smaller ones with higher ANND (ANNS). In Table \ref{chapt4_tab:zeros} we also report the general assortativity coefficient for the networks, confirming our finding.

\begin{figure}[tb]
	\centering
	\caption[Undirected total strength cumulative distribution for WTW and FDIN]{Undirected total strength cumulative distribution for WTW and FDIN. Red lines are power law fits with coefficients $\xi=-0.258\pm0.006$ for WTW and $\xi=-0.259\pm0.009$ for FDIN.}
	\vspace{-20pt}
	\includegraphics[width=\myFiguresSize]{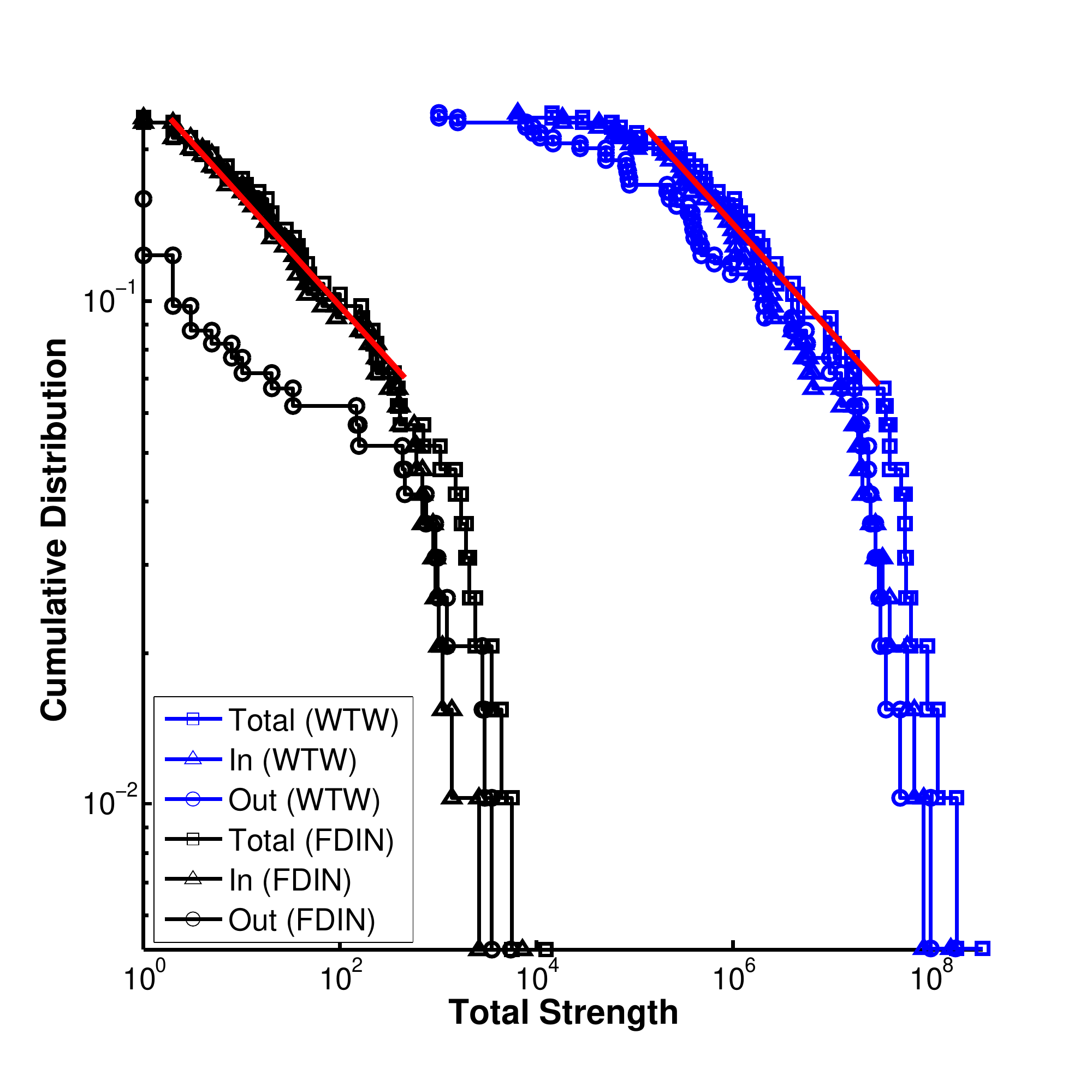}
	\vspace{-10pt}
	\label{chapt4_fig:cumdistr_totstrength}
\end{figure}

In Figure \ref{chapt4_fig:cumdistr_totstrength} we report the (undirected) cumulative distribution of the WTW and FDIN node strengths, defined as $P_c(s)\equiv \sum_{s*>s}P(s*)$. This quantity measures the probability of a randomly chosen vertex to have strength $s$ connecting to other vertexes. Red lines are power law fits for the total distributions, with coefficients $\xi=-0.258\pm0.006$ for WTW and $\xi=-0.259\pm0.009$ for FDIN, thus almost identical and anyhow equal within the error bars. Such power law behavior indicates an extremely high level of heterogeneity in the two networks.

Due to the directed nature of our networks, we also have to distinguish between in- and out-strength distributions: they are both shown in Figure \ref{chapt4_fig:cumdistr_totstrength} and present in general a very similar trend with respect to the total one, with the exception of the FDIN out-strength distribution, which markedly deviates for $s\lesssim500$. Such deviation indicates a lower probability of having a country doing many FDI, hence that a big share of world countries is engaged in very few investment connections.

\begin{figure}[t]
	\centering
	\caption[WTW vs FDIN link weights]{WTW vs FDIN link weights. Markers size is proportional to the log of $\text{POP}_i * \text{POP}_j / \Delta_{ij}$. Colors scale (blue to red) is from lower to higher values of logs of $\text{rGDPpc}_i$ $*$ $\text{rGDPpc}_j$ $/$ $\Delta_{ij}$.}
	\vspace{-10pt}
	\subfloat[Total\label{chapt4_fig:wtot}]{\includegraphics[width=\myFiguresSizeBis]{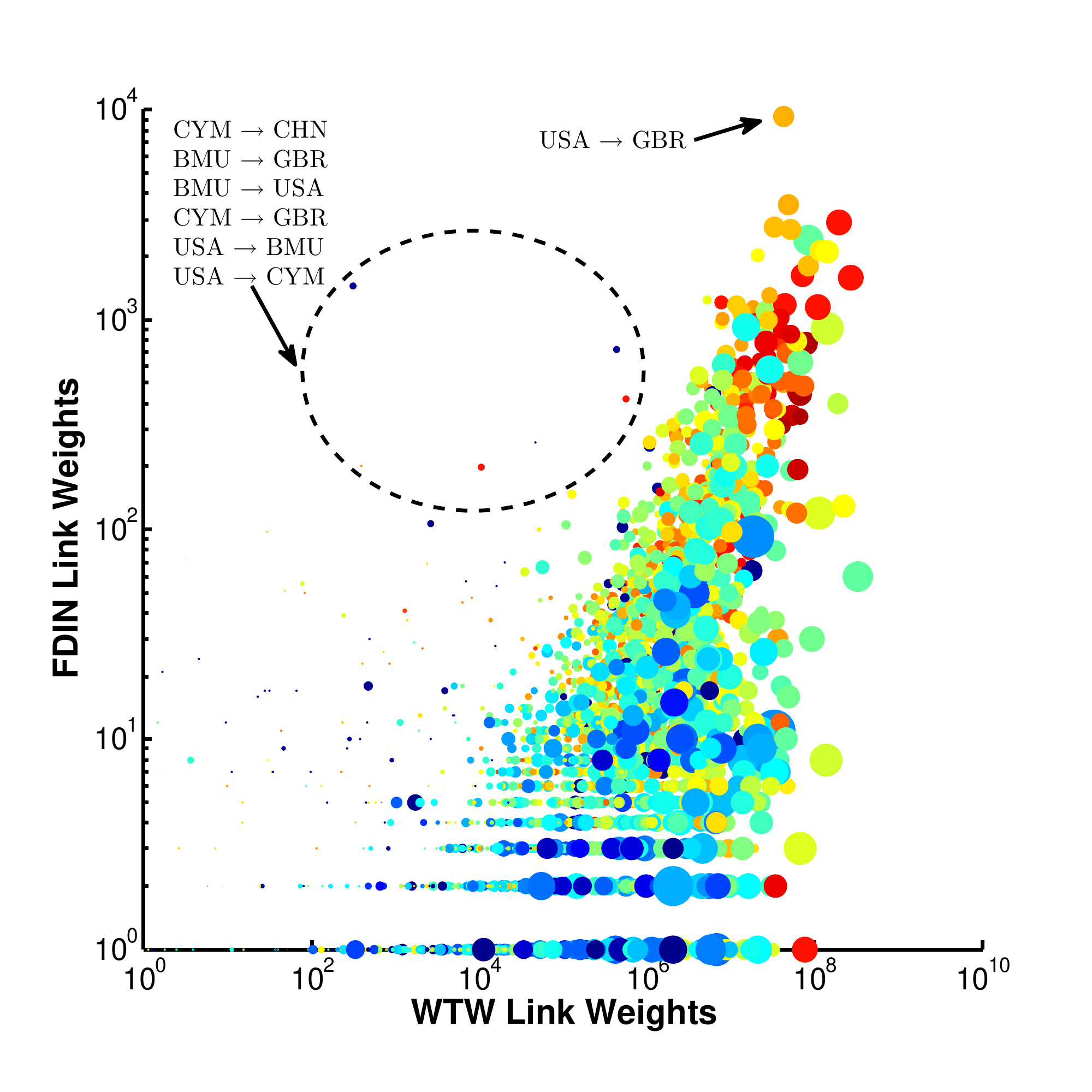}}
	\subfloat[Primary\label{chapt4_fig:wprim}]{\includegraphics[width=\myFiguresSizeBis]{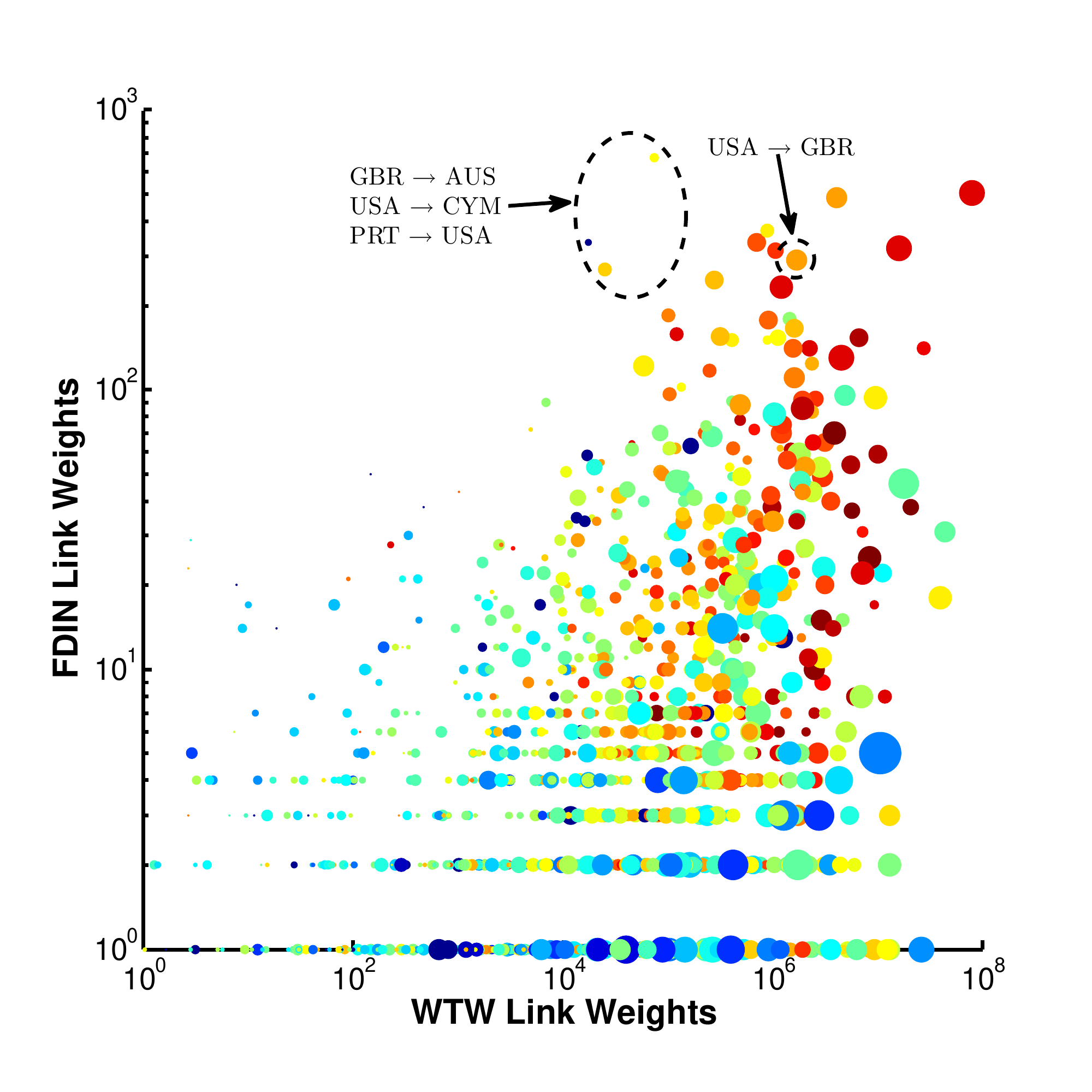}}
	\vspace{-10pt}
	\subfloat[Secondary\label{chapt4_fig:wsec}]{\includegraphics[width=\myFiguresSizeBis]{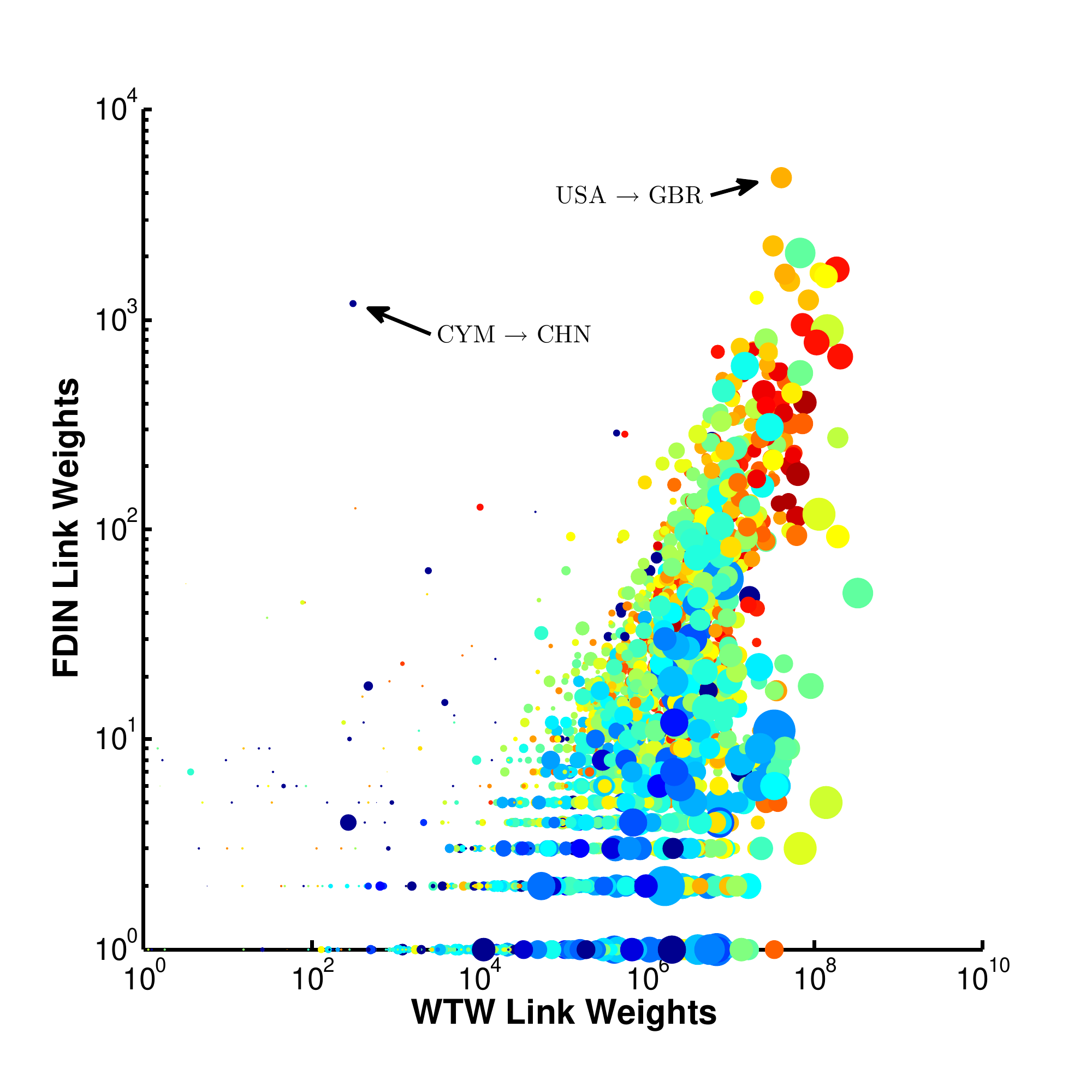}}
	\subfloat[Tertiary\label{chapt4_fig:wter}]{\includegraphics[width=\myFiguresSizeBis]{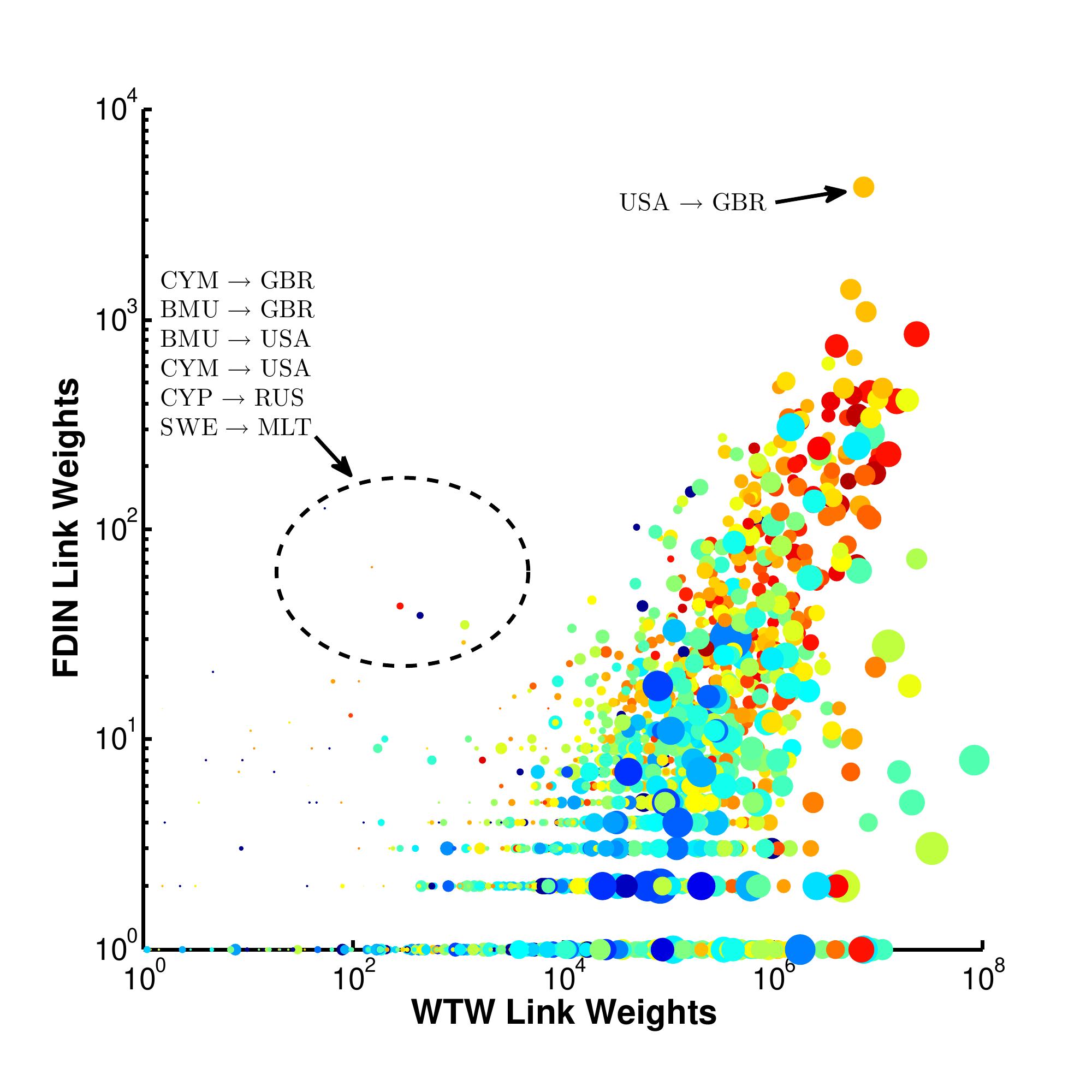}}
	\vspace{-10pt}
	\label{chapt4_fig:weights}
\end{figure}

We now want to analyze to which extent the two sub-networks display any correlated behavior. Figure \ref{chapt4_fig:weights} shows a log-log scatter plot of the WTW and FDIN link weights for the total and each macro-sector case: each dot is an element in the space $(t_{ij}, f_{ij})$, i.e. the space of the two networks link weights, which color is proportional to log of $\text{rGDPpc}_i * \text{rGDPpc}_j / \Delta_{ij}$ and size to log of $\text{POP}_i * \text{POP}_j / \Delta_{ij}$. The rationale behind this analysis resides in the well-known empirical success of the gravity model for FDI, but especially for trade: in both cases goods and investments flows are well explained by a gravity-like equation involving country sizes (rGDPpc and POP, respectively) and, inversely, geographical distance.

If this is the case, one should expect that most of the variation in the cloud of points can be explained by larger country sizes and smaller distances. In our case this is more evident with respect to gross domestic products (dots color) as richer pairs of countries tend to be located in the north-west portion of the plot, whereas it is less evident with respect to population (dots size).
However all the panels in Figure \ref{chapt4_fig:weights} suggest a direct relation between the link weights in the two networks, as a high level of exports is in general associated with a high level of FDI.\footnote{Although from Figure \ref{chapt4_fig:weights} we observe a proportional relation between link weights for all the sub-networks, in Section \ref{chapt4_sec:results} we will see how, using a gravity model with all the required regressors, this will not be the case: instead services will show an inverse relation and primary behavior will depend on the direction of FDIs one consider.}

It is also interesting to notice what are the outliers links (in Figure \ref{chapt4_fig:weights} we highlighted only a few): most of them are the so-called \textit{tax havens}. In fact these countries result to have very high levels of incoming and outgoing FDIs, with relatively low flows of goods.

Continuing to investigate the interplay between the two networks, an important characteristic is the number of common zero flows and the probability of having a link between the same two countries on both layers: we already know the two networks have different densities (from Table \ref{chapt4_tab:zeros}) and now we want to look at the percentage of common links. Our goal is to investigate characteristics that would suggest for possible \textit{substitutive} or \textit{complementary} interplay between the two phenomena.

\begin{table}[tb]
	\centering
	\caption[WTW and FDIN similarity statistics]{Statistics about the density, common links and networks similarity for the WTW and FDIN.}
	\label{chapt4_tab:matches}
	\begin{adjustbox}{width=\textwidth,keepaspectratio}
		\begin{tabular}{lcccccccc}
			\hline \hline
			\rule{0pt}{2ex} & \multicolumn{2}{c}{Tot} & \multicolumn{2}{c}{Primary} & \multicolumn{2}{c}{Secondary} & \multicolumn{2}{c}{Tertiary} \\
			& {\footnotesize WTW} & {\footnotesize FDIN} & {\footnotesize WTW} & {\footnotesize FDIN} & {\footnotesize WTW} & {\footnotesize FDIN} & {\footnotesize WTW} & {\footnotesize FDIN} \\
			\hline
			\rule{0pt}{3ex}\% matches & \multicolumn{2}{c}{47.3} & \multicolumn{2}{c}{62.3} & \multicolumn{2}{c}{46.3} & \multicolumn{2}{c}{57.6} \\
			\% WTW links in FDIN & \multicolumn{2}{c}{16.0} & \multicolumn{2}{c}{9.8} & \multicolumn{2}{c}{13.3} & \multicolumn{2}{c}{12.1} \\
			\% FDIN links in WTW & \multicolumn{2}{c}{96.3} & \multicolumn{2}{c}{92.4} & \multicolumn{2}{c}{97.3} & \multicolumn{2}{c}{95.1} \\
			Jaccard index (J)& \multicolumn{2}{c}{0.159} & \multicolumn{2}{c}{0.097} & \multicolumn{2}{c}{0.132} & \multicolumn{2}{c}{0.120} \\
			\hline \hline
		\end{tabular}
	\end{adjustbox}
\end{table}

From Table \ref{chapt4_tab:matches} we observe that the \% of matches varies roughly around half of the links, with just a few percentage of FDIN links without the correspondent trade counterpart.

The Jaccard index, also presented in Table \ref{chapt4_tab:matches}, is a measure of similarity of the two networks: given two sets of events $A$ and $B$, we define $J=\left| A\cap B\right| /\left| A\cup B\right|$, therefore representing the ratio between the number of events shared by the two sets, over the number of events in at least one of them \citep{jacc1901}. In our case the events are the presence of a link. This measure gives very low values of similarity, apparently in contrast with the previous results, but justifiable if considering that the Jaccard index gives much importance to zero flows, thus making the sparsity of our networks greatly influence its outcome.

\begin{figure}[tb]
	\centering
	\caption[Correlation of node-network statistics between WTW and FDIN]{Correlation of node-network statistics between WTW and FDIN. Marker size is proportional to the log of $\text{POP}_i$. Color scale (from blue to red) is proportional to the log of $\text{rGDPpc}_i$.}
	\vspace{-10pt}
	\subfloat[Total Degree\label{chapt4_fig:degree}]{\includegraphics[width=\myFiguresSizeBis]{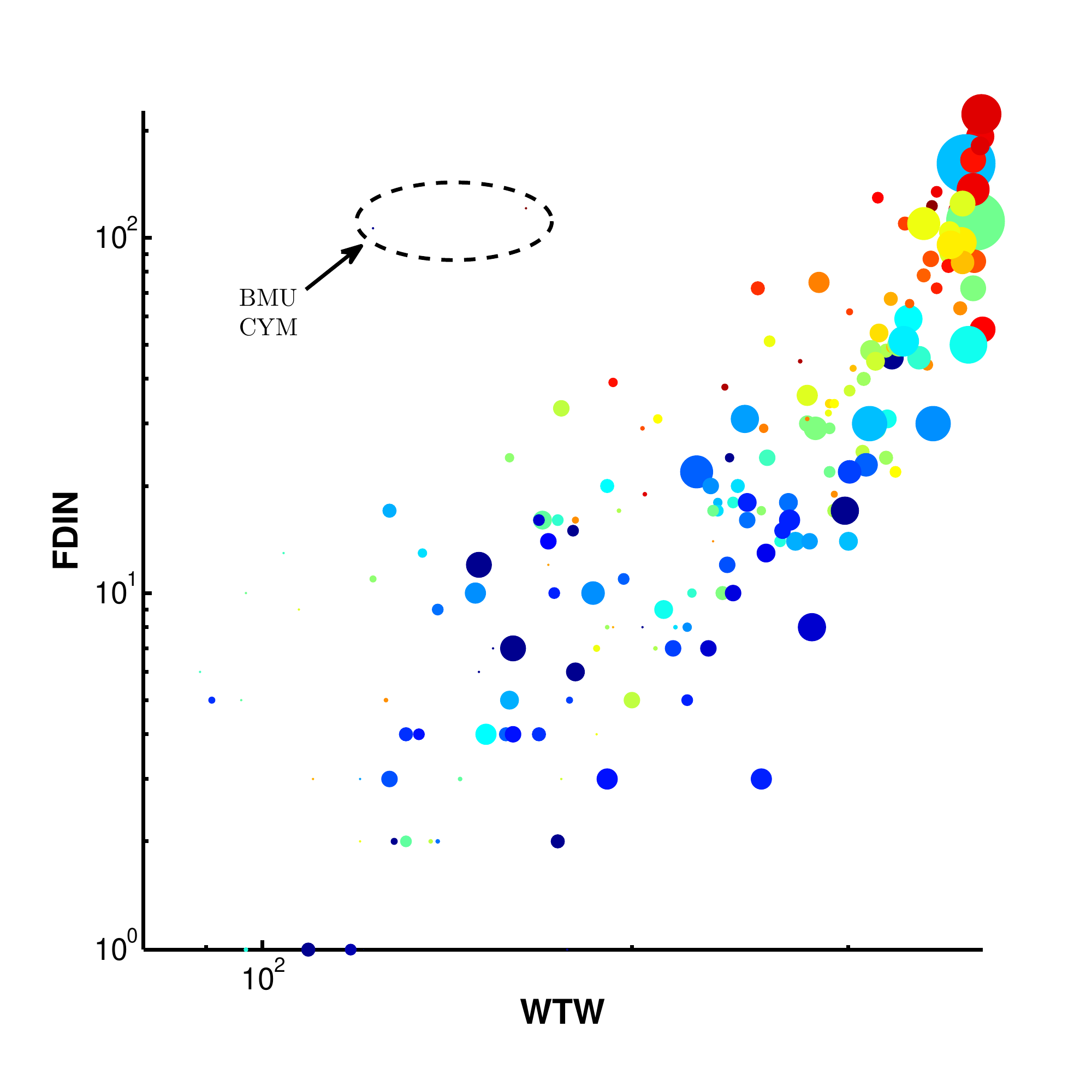}}
	\subfloat[Total Strength\label{chapt4_fig:strength}]{\includegraphics[width=\myFiguresSizeBis]{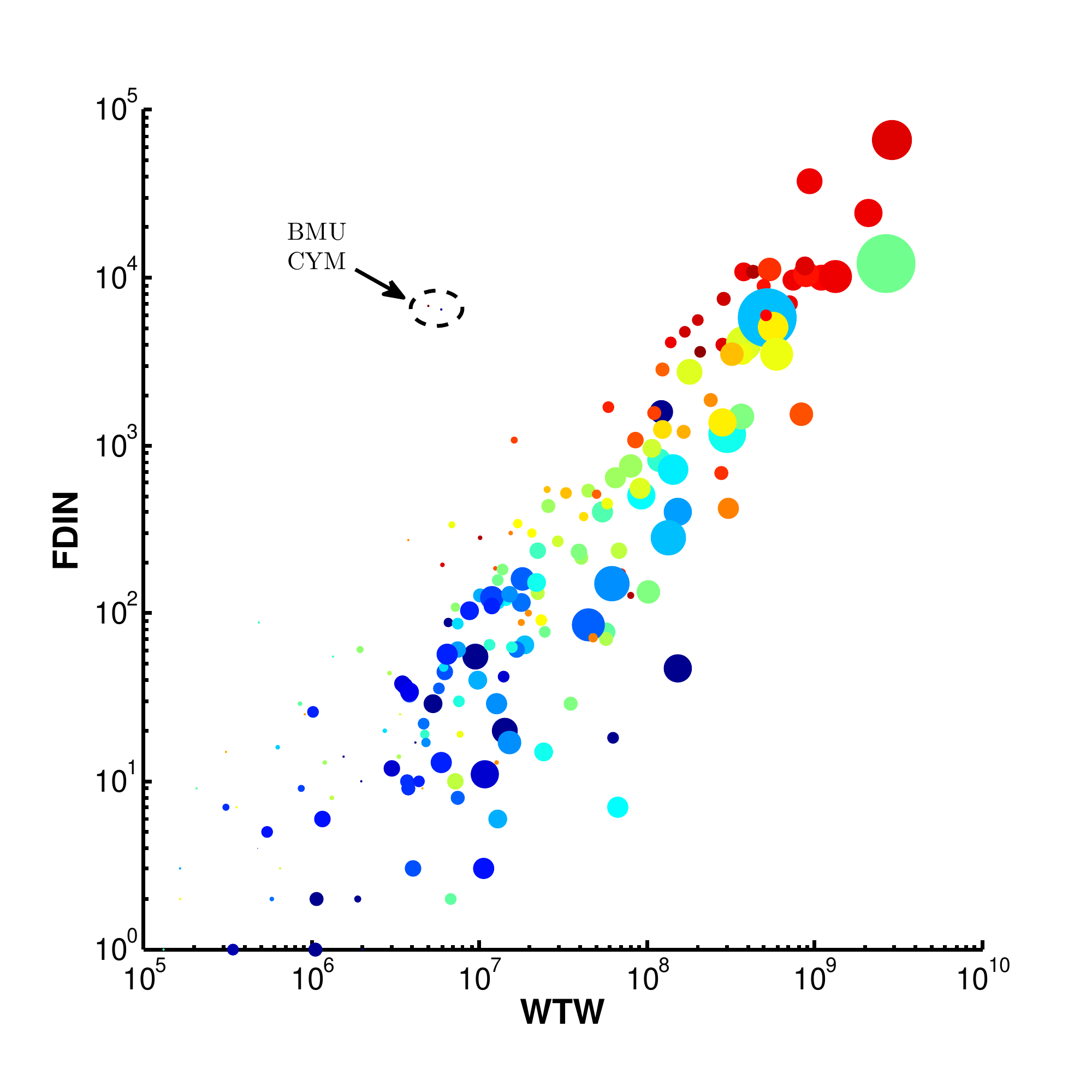}}
	\vspace{-10pt}
	\subfloat[Total ANND\label{chapt4_fig:annd}]{\includegraphics[width=\myFiguresSizeBis]{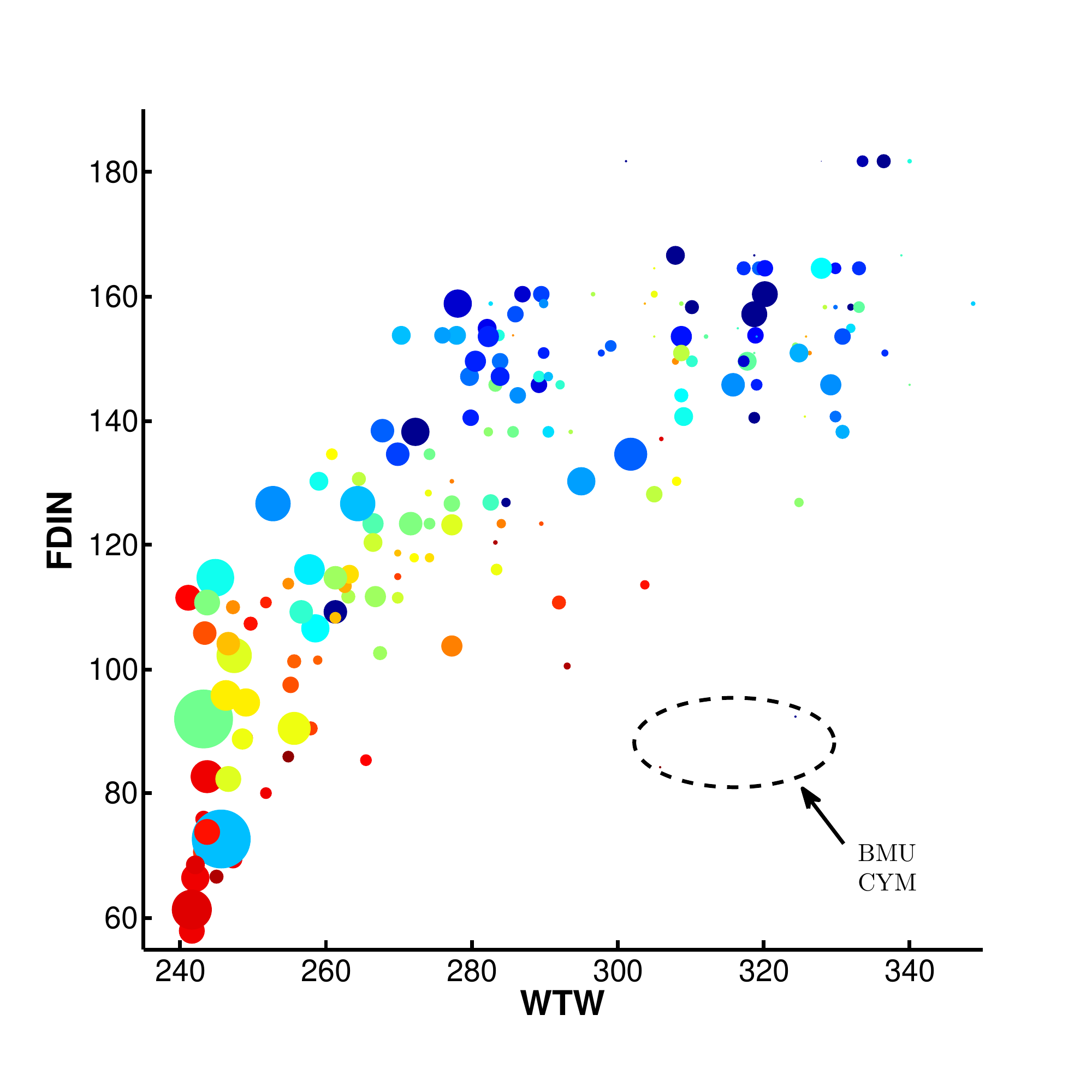}}
	\subfloat[Total ANNS\label{chapt4_fig:anns}]{\includegraphics[width=\myFiguresSizeBis]{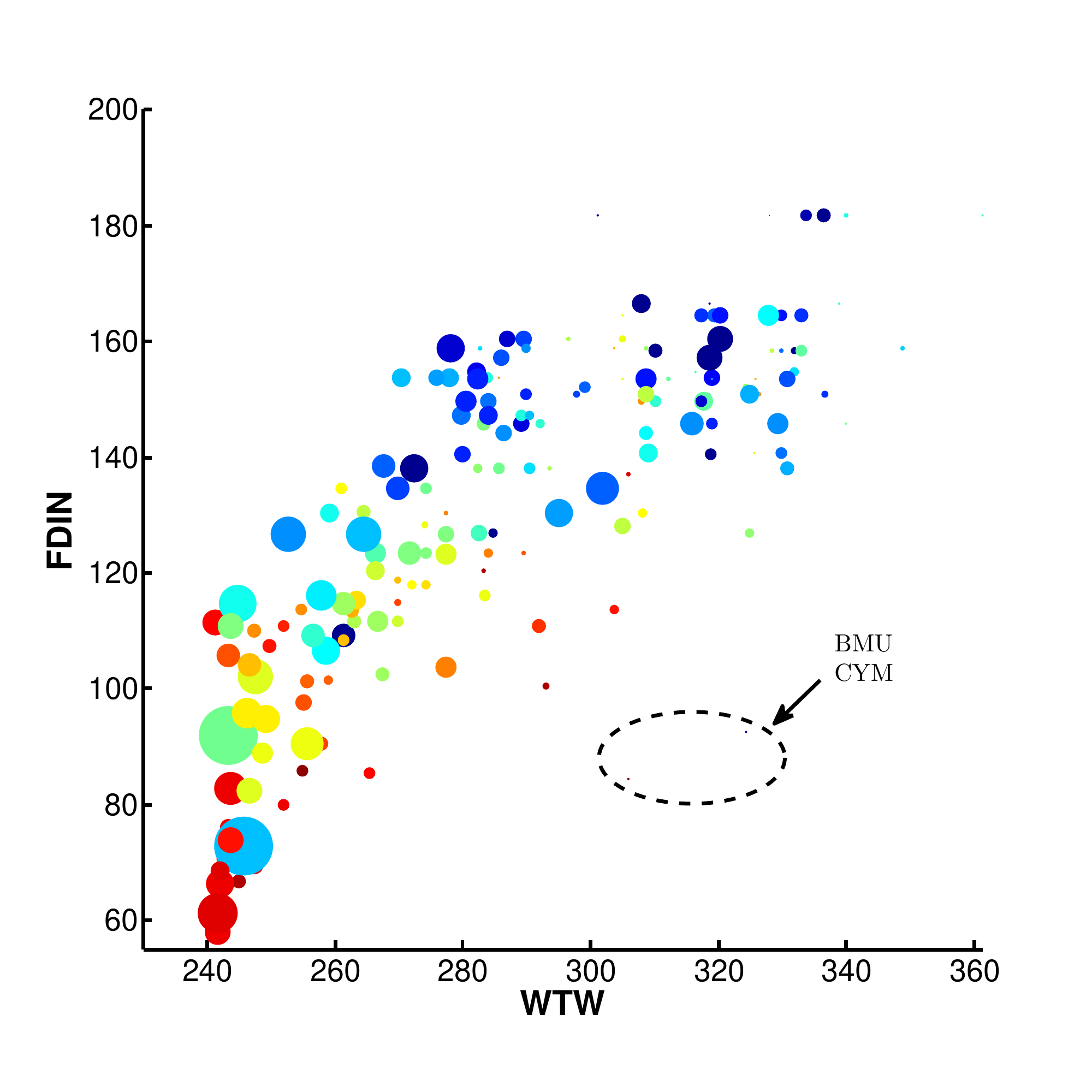}}
	\vspace{-10pt}
	\label{chapt4_fig:compare}
\end{figure}

To conclude this section we compare in Figure \ref{chapt4_fig:compare} some single node properties for the two layers of our network: total degree, total strength, ANND and ANNS. Again markers aspect is determined by countries size. We can observe how both the degree and strength (panels \ref{chapt4_fig:degree} and \ref{chapt4_fig:strength}) are positively correlated, i.e. countries with many trade channels (respectively, that trade more) also have many FDI partners (respectively, many investments), it is clear how this is related to their size: bigger countries (red and larger dots) lies in the north-east portion of the plot, indicating they have larger number (respectively, volume) of connections in both WTW and FDIN.
In panels \ref{chapt4_fig:annd} and \ref{chapt4_fig:anns} we show ANND and ANNS. In this case, although the two properties are still proportional between the two layers (i.e. countries linked with highly connected partners (partners with large exchange volumes) in WTW tend to do the same in FDIN), the relation with country size is inverse with respect to the previous cases: here bigger countries tend to have lower ANND and ANNS in both WTW and FDIN.

Again it is interesting to notice how \textit{tax havens} appear as outliers, in particular Bermuda (BMU) and Cayman Islands (CYM): in all the panels they stand out from the general trend of other countries, behaving similar to much richer ones with respect to FDI but not with respect to trade. This is clear in light of the results of Figure \ref{chapt4_fig:disassortativity}, where we saw that bigger countries (in both economic and demographic terms) have smaller ANND and ANNS, in both trade and FDI sub-networks.

\section{Economic and econometric approach}\label{chapt4_sec:economic-and-econometric-approach}
In the following analysis we make use of the gravity equation to model the relation between trade and FDI in an econometric perspective. The H2S selection model is also used to account for zero flows and analyze both the extensive and intensive margin effects.

Since the seminal works \citet{linnemann1966econometric, tinbergen1962shaping}, the gravity model has been widely used because of its excellent fit with empirical trade data \citep{frankel2002estimate, egger2002econometric}, as well as giving good results in modeling FDI \citep{blonigen2007fdi, baltagi2007estimating} and other phenomena like migration and tourism.
Moreover a lot of effort has been put in refining it and giving it a consistent economic foundation \citep{bergstrand1985gravity, anderson2003gravity}. The gravity model is based on a generalized form of Newton's law of universal gravitation: trade is directly proportional to countries size and inversely proportional to their geographical distance,\footnote{Newton's law of universal gravitation radically depends on the inverse of the distance \textit{squared}, although when used for international economics the important aspect is just the inverse relation.} i.e.

\begin{equation}\label{chapt4_eq:newton}
F_{ij} = \beta_0 \frac{s_i^{\beta_1} s_j^{\beta_2}}{d_{ij}^{\beta_3}}\ ,
\end{equation} 
where $F_{ij}$ represents the flow (being it trade, FDI, etc.) between countries $i$ and $j$, $s_i$ and $s_j$ their respective (economic) sizes and $d_{ij}$ the distance between the two.
Applying the logarithm on both sides of Eq. \ref{chapt4_eq:newton}, we obtain a linear multivariate equation like

\begin{equation}
\widetilde{F}_{ij} = \widetilde{\beta}_0 + \beta_1 \widetilde{s}_i + \beta_2 \widetilde{s}_j - \beta_3 \widetilde{d}_{ij} + \ldots + \epsilon_{ij}\ ,
\end{equation} 
where $\epsilon_{ij}$ are stochastic residual terms, usually assumed to be i.i.d. and $\sim N(0, \sigma^2)$, and we indicated with ``$\ldots$'' the possibility to add more regressors (countries- or link-specific characteristics) to the model specification.

Empirically, the proposed variables to be used as proxy for origin and destination size ($s_i$ and $s_j$) has been many, e.g. population, area size or Gross Domestic Product per-capita (\textit{GDPpc}). However in the following analysis we will use country-specific dummies instead of explicit factors: these have the advantage of yielding consistent estimates and account for any unobservable that contributes to shift the overall level of exports or imports of a country \citep{anderson2004trade,Harrigan1996,hema13}.

A variety of impedance factors have been incorporated in different gravity model specifications, aiming to explain potential barriers to trade flows. They typically are common language, contiguity, landlocked location, etc. Belonging to a customs union or trade agreements are also frequently found in gravity model specifications. Similarly, sharing the same currency, have been part of a same nation, past (or present) colonial relations or even have had a common colonizer, are also factors expected to affect trade flows between regions. All these are typically introduced as dummy variables \citep{baier2007free,glick2002does,bun2007euro}.

Transportation costs are the main resistance factors we include by considering geographical distance ($d_{ij}$) and they are typically approximated by considering the countries most important cities / agglomerations, i.e. the great-circle definition of distance. Nevertheless, more detailed method for calculating distances as a function of country size have been proposed in the literature, for example \citet{nitsch2000national,nowak2007impact}.

The first and most widespread technique used for estimating the baseline gravity model in its log-linear form is the Ordinary Least Squares (OLS). However, as pointed out by \citet{silva2006log}, using OLS in the presence of many zeros in the dependent variable (as is the case for trade) lead in general to inconsistent coefficient estimates in the log-linear form of the gravity equation.\footnote{Unless taking much stronger assumptions on the functional form of the error term than one have to do estimating a truly linear model with OLS.}

In recent years, as the theoretical and empirical studies in trade have evolved, it became clear that the large amount of zeros has to be treated with particular attention as it is a substantial fraction and an important feature of the data. In the recent literature \citet{Bikker1992} conclude that zero flows are the result of microeconomic decision-making based on the potential profitability of engaging in bilateral trade; in another work \citet{anderson2004trade} point at the importance of fixed costs associated with international, such as border costs \citep{hillberry2002aggregation}, search costs and other specific investments to enter foreign markets \citep{romer1994new}. Apart from the decision to trade or not, the size of expected potential trade is historically well determined by the conventional gravity model, as first introduced by \citet{tinbergen1962shaping} and discussed at the beginning of this section. In case of actual zero trade, potential trade is simply unobserved.

In this context, 2-step sample selection estimators have become a well-established approach to model bilateral trade in the presence of zero flows. These let us remove the effect of the extensive firm margin so as to correctly estimate the intensive effect, in contrast with other approaches which calculate coefficients that combine both the extensive and intensive margins \citep{hema13}. \citet{helpman2008estimating} also show that traditional estimates are biased and that most of the bias is due not to selection but rather to the omission of the extensive margin. Moreover \citet{linders2006estimation} conclude that censored or truncated regression, and replacement of zero flows with arbitrary numbers are not preferable as these approaches may yield misleading results and as they rely on ad-hoc assumptions and artificial censoring. Sample selection models, on the other hand, allows zero flows and the size of potential trade to be explained jointly and proved to be the best choice, both from an economic and econometric point of view.

In particular \citet{helpman2008estimating} provide a theoretical framework jointly determining both the set of trading partners and their trade volumes, that can be estimated using the H2S selection model \citep{Heckman1979}: they develop a model of international trade in which firms face fixed and variable costs of exporting and where productivity varies both firm- and destination-based. Trade channels depend on its profitability, therefore for any pair of countries there may be no firm productive enough to profitably export. As a result, the model it is consistent with zero trade flows in both directions between some countries, as well as positive --- though asymmetric --- trade flows in both directions for some country pairs. Finally, the model generates a gravity equation.

Following this literature we will carry out all the analysis in this chapter using the H2S, that involves first a probit to estimate the probability of a positive trade flow among any pair of countries and a second step that estimates the log-linear specification of the gravity equation on the positive-flow observations, including a selection correction.\footnote{Many other models have been introduced in the literature that employ 2-steps techniques to correctly take into account the excess of zero flows. Among them the most common are the Zero Inflated Poisson (ZIP), the Zero Inflated Negative Binomial (ZINB). To better justify our use of the H2S we should check our results against at least these two models. We leave this robustness check for the future development of this work.}

In the literature many studies focused on modeling both trade and FDI using the gravity specification, but much less is known about the interplay between the two, since the traditional gravity model for trade does not account for FDI and vice-versa. This is exactly the gap we want to close in this chapter, using the aforementioned H2S selection model.

In today's global economy, these two phenomena are intimately related through the processes of the GVC.
Among the many possible ways trade and FDI can relate and shape the GVC \citep{Baldwin2013a,Gereffi2005,Ntras2013}, we are interested in the so called \textit{vertical supply chain}: production outsourcing in another country, through FDI, fosters goods trade between the two countries involved. Therefore what we expect is to see a strong correlation between trade and FDI flows in the two possible ways this can happen: both when the exchanges are in the same direction (export-export, e.g. a foreign affiliate that have to import its production input) and when they happen in opposite ways (export-import, e.g. when moving a production stage and the parent firm needs to import back the output produced).\footnote{We introduced the concept of the vertical supply chain as it is the mechanism through one would expect trade and FDI are related the most, as opposed the horizontal one where investments are used to simply replicate the production chain and thus not giving rise to a significant increase in trade flows. However in general the GVC is a complex combination of vertical and horizontal strategies \citep{Baldwin2013a} and isolating the two in data is a non-trivial exercise. In this sense, while thinking of the vertical supply chain as the main source of the relation between trade and FDI we observe in the data, we will not be able to distinguish and quantify the contribution of other effects in our results. Moreover in this chapter we are looking only at direct effects, on the same dyad, therefore excluding all other mechanisms involving third countries. We leave this analysis for the future developments of this work.}

In particular production processes increasingly involve a sequential, vertical trading chain stretching across many countries, with each country specializing in particular stages of a good production sequence. This phenomenon is usually called \textit{vertical specialization} \citep{Hummels2001}. More recently, a measure for the position in the production line, \textit{upstreamness}, has been introduced by \citet{Antras2012}: it ranges from a minimum of $1$ (all output goes only to final use), to $4.65$ (Petrochemicals). Using it we investigate for possible non-linearities in the relation between FDI and trade as one moves along the GSC: adding an interaction term between FDI and upstreamness we study the composition of the trade-FDI relation with the position in the multi-stage production chain.

Another important feature of today's globalized and integrated economy, where production is dispersed and coordination costs have taken the place of transportation costs \citep{baldwin2012global}, is the proliferation of Regional Trade Agreements (RTA): according to WTO data, there are over 300 regional trade agreements, up from less than a hundred in the early 1990s and today more than half of world trade is governed by at least one RTA \citep{WTO2011,Damuri2012}. While one might be tempted to think that the trade have mainly a global span, it in fact mostly regional,\footnote{With ``regional'' we refer here to big world economic blocks as, for example, EU, NAFTA or Asean.} with for example intra-EU trades accounting for the 75\% in 2011 \citep{WTO2013}). In this scenario, and without a widespread harmonization of trade and investment agreement \citep{WTO2013}, it is plausible to expect a different interaction between trade and FDI when some RTA is in place or not: trade agreement should in this view, by favoring trades, reduce investments.

An aspect we already introduced in Section \ref{chapt4_sec:networks-comparison} is the difference between the merchandise and the services sectors. While the former has been widely analyzed and still represent the bulk of international trade, high-income countries nowadays are primarily service economies (e.g. in 2007, services accounted for nearly 75\% of GDP of high-income OECD countries) and foreign affiliate sales of services have grown faster in the last two decades than direct cross-border sales \citep{Francois2010}; however, very little is known about how this trade is being conducted. The General Agreement on Trade in Services (GATS) defines four modes of supply, though the bulk of service trade uses mainly two: cross-border trade and services rendered by a foreign affiliate, thus making our framework right to understand how these two possible choices relate. In particular we expect firms to make discrete choices, i.e. to use either one or the other channel separately \citep{kelle2013cross}.

\section{Results}\label{chapt4_sec:results}
We start by estimating a traditional log-in-log gravity model with the H2S procedure in the same fashion proposed by \citet{helpman2008estimating}, in order to account for the zero flows in the dependent variable (i.e. trade flows).

In our baseline model we consider the logarithm\footnote{In all the analysis shown here we always used the natural logarithm. Although regressions were also run using $\log_{10}$: the only difference was that dummy variables coefficients were reduced approximately to half the value, but all significance levels were unchanged.} of unilateral trade flows ($\ln tr$) as the dependent variable and the logarithm of the number of FDIs ($\ln FDI$) among the explanatory variables. In addition to these the model includes origin and destination country-specific dummies to control for importer/exporter characteristics, such as GDP and POP;\footnote{Not shown in tables for brevity.} the logarithm of the geographical distance (great-circle definition, $\ln \Delta$); and the traditional gravity dummy variables that accounts for common borders (\textit{contig}), colony relations (\textit{colony}), whether they have ever been unified (\textit{smcrtry}), common language (as spoken by at least 9\% of the population, \textit{comlang\_ethno}). Note that all our results are robust to additional controls such as common religion, common colonial ties, and landlocking effects.

\begin{table}[!t]
	\centering
	\caption[Regression results for the plain model, upstreamness and distance]{Regressions for the base model and for the specifications with interactions of FDI with \textit{upstreamness} and \textit{distance}. Country dummy variables for importer/exporter fixed effects included. The right part refers to the case where trade and FDI were considered in opposite directions.}
	\label{chapt4_tab:res_ups_dist}
	\begin{adjustbox}{width=\textwidth,totalheight=\textheight,keepaspectratio}
		\begin{tabular}{lccc|ccc}
			\hline\hline
			\rule{0pt}{2ex} & \multicolumn{1}{c}{Plain} & \multicolumn{1}{c}{Upstreamness} & \multicolumn{1}{c}{Distance} & \multicolumn{1}{c}{Plain (inv)} & \multicolumn{1}{c}{Upstremness (inv)} & \multicolumn{1}{c}{Distance (inv)}\\
			\hline
			\rule{0pt}{3ex}$\ln trade$ & & & & & & \\
			$\ln FDI$ & 0.120\sym{***}& 1.496\sym{***}& -0.842\sym{***}& & & \\
			& (0.0237) & (0.0703) & (0.101) & & & \\
			$\ln FDI$\_inv& & & & 0.305\sym{***}& 1.114\sym{***}& -0.672\sym{***}\\
			& & & & (0.0168) & (0.0673) & (0.100) \\
			$\ln ups$ & & -0.860\sym{***}& & & -0.940\sym{***}& \\
			& & (0.0405) & & & (0.0405) & \\
			$\ln \Delta$ & -0.796\sym{***}& -0.800\sym{***}& -0.872\sym{***}& -0.767\sym{***}& -0.773\sym{***}& -0.803\sym{***}\\
			& (0.0192) & (0.0192) & (0.0205) & (0.0170) & (0.0169) & (0.0173) \\
			$\ln ups$ \# $\ln FDI$& & -1.746\sym{***}& & & & \\
			& & (0.0837) & & & & \\
			$\ln \Delta$ \# $\ln FDI$& & & 0.126\sym{***}& & & \\
			& & & (0.0129) & & & \\
			$\ln ups$ \# $\ln FDI$\_inv& & & & & -1.027\sym{***}& \\
			& & & & & (0.0827) & \\
			$\ln \Delta$ \# $\ln FDI$\_inv& & & & & & 0.121\sym{***}\\
			& & & & & & (0.0122) \\
			contig & 0.835\sym{***}& 0.831\sym{***}& 0.884\sym{***}& 0.805\sym{***}& 0.802\sym{***}& 0.862\sym{***}\\
			& (0.0537) & (0.0535) & (0.0536) & (0.0536) & (0.0534) & (0.0538) \\
			colony & 0.636\sym{***}& 0.639\sym{***}& 0.606\sym{***}& 0.613\sym{***}& 0.619\sym{***}& 0.583\sym{***}\\
			& (0.0582) & (0.0580) & (0.0581) & (0.0582) & (0.0580) & (0.0582) \\
			smctry & 0.460\sym{***}& 0.474\sym{***}& 0.437\sym{***}& 0.470\sym{***}& 0.484\sym{***}& 0.448\sym{***}\\
			& (0.0751) & (0.0748) & (0.0747) & (0.0750) & (0.0748) & (0.0750) \\
			comlang\_ethno& 0.455\sym{***}& 0.455\sym{***}& 0.450\sym{***}& 0.444\sym{***}& 0.445\sym{***}& 0.437\sym{***}\\
			& (0.0279) & (0.0277) & (0.0279) & (0.0278) & (0.0277) & (0.0278) \\
			\_cons & 10.71\sym{***}& 11.39\sym{***}& 11.11\sym{***}& 10.57\sym{***}& 11.31\sym{***}& 10.81\sym{***}\\
			& (0.228) & (0.229) & (0.231) & (0.226) & (0.227) & (0.227) \\
			\hline
			trade dummy & & & & & & \\
			FDI dummy & 1.465\sym{***}& 1.465\sym{***}& 1.465\sym{***}& 1.465\sym{***}& 1.465\sym{***}& 1.465\sym{***}\\
			& (0.0153) & (0.0153) & (0.0153) & (0.0153) & (0.0153) & (0.0153) \\
			$\ln \Delta$ & -0.330\sym{***}& -0.330\sym{***}& -0.330\sym{***}& -0.330\sym{***}& -0.330\sym{***}& -0.330\sym{***}\\
			&(0.00291) &(0.00291) &(0.00291) &(0.00291) &(0.00291) &(0.00291) \\
			\_cons & 2.477\sym{***}& 2.477\sym{***}& 2.477\sym{***}& 2.477\sym{***}& 2.477\sym{***}& 2.477\sym{***}\\
			& (0.0255) & (0.0255) & (0.0255) & (0.0255) & (0.0255) & (0.0255) \\
			\hline
			mills & & & & & & \\
			lambda & -1.179\sym{***}& -1.206\sym{***}& -0.948\sym{***}& -1.223\sym{***}& -1.243\sym{***}& -1.173\sym{***}\\
			& (0.0662) & (0.0660) & (0.0694) & (0.0473) & (0.0472) & (0.0474) \\
			\hline
			\(N\) & 336978 & 336978 & 336978 & 336978 & 336978 & 336978 \\
			\hline\hline
			\multicolumn{7}{l}{\footnotesize Standard errors in parentheses; \sym{*} \(p<0.1\), \sym{**} \(p<0.05\), \sym{***} \(p<0.01\)}\\
		\end{tabular}
	\end{adjustbox}
\end{table}

The results of this first specification (we called \textit{Plain}) are presented in the first column of Table \ref{chapt4_tab:res_ups_dist}.

To estimate the first step of the H2S selection model, i.e. the probability that a dyad will trade or not (\textit{extensive margins}), we use $\ln \Delta$ and a dummy variable for the presence of FDI (\textit{fdi\_dummy}), finding that such probability is positively correlated with the number of outgoing FDI and negatively correlated with the geographical distance between the two countries.\footnote{We show the results for the first-step estimation only in Table \ref{chapt4_tab:res_ups_dist}: in the analysis of Tables \ref{chapt4_tab:res_region} and \ref{chapt4_tab:res_sectors} the exact same procedure has been employed but we omit them there as they are identical in all cases.} Moreover the second step's explanatory variables have the expected coefficients: distance is negative, while all the dummies have positive and significant estimated coefficients. Most importantly, we find that the number of FDIs have a positive effect in determining the volume of trade (\textit{intensive margins}), i.e. the second step estimated coefficient for $\ln FDI$ is positive and significant.

We consider also the reverse relation between FDI and trade, i.e. the case where a country invests in another one and then imports goods from it, or vice-versa. The right part of Table \ref{chapt4_tab:res_ups_dist} (and in particular column 4) shows the results of the regressions done considering as covariate the number of FDIs from the importer to the exporter country: using this model specification the estimated coefficient of FDI increases, meaning that, on general, inverse FDIs affect more the volume of trade between two countries.

In the second and fourth columns of Table \ref{chapt4_tab:res_ups_dist} we replicate our baseline specification of the gravity model, introducing now an interaction term between FDI and \textit{upstreamness}.\footnote{As introduced in \ref{chapt4_sec:economic-and-econometric-approach}, upstreamness inversely measures how much of an industry output goes to the final use, i.e. the average position of a certain product in the production line: it ranges from 1, when all output goes to final use, to 4.65 (Petrochemicals)} Its coefficient alone ($\ln ups$) is negative, meaning that international trade is more developed for products that are relatively downstream (closer to final use) in the production chain; but what is more interesting is its interaction coefficient with FDI: this indicates that the relation is non-linear and that there is a composition of the effects of FDI and the position in the GSC on trade. Its negative sign indicates that the more the industry is upstream, the more FDI and trade tend to be substitutes.
In particular with an easy exercise it is possible to identify the value of upstreamness for which the observed negative correction make trade and FDI substitutes: lets rewrite the model equation as
\begin{equation}\label{chapt4_eq:easymodel}
T = \beta_{F} F + \beta_{U} U + \beta_{FU} F U + \ldots\ ,
\end{equation}
where clearly $T$ stands for trade, $F$ for FDI, $U$ per upstreamness and where we indicated with ``\ldots'' all the other terms that are now irrelevant. Remembering that in our model we used all variables in logarithm, the value of upstreamness we are looking for is the one that solve
\begin{equation}\label{chapt4_eq:dT_dF}
0 = \frac{dT}{dF} = \beta_{F} + \beta_{FU} \ln \widehat{U}\ ,
\end{equation}
from which $\widehat{U}=2.356$.\footnote{This value of upstreamness correspond approximately to the sectors of \textit{wood} and \textit{metal} product manufacturing.} Therefore, for $U<\widehat{U}$ trade and FDI are complements, while for $U>\widehat{U}$ they are substitutes.
Similar results hold, even if with different magnitudes, for the inverse FDIs.

In the third column of Table \ref{chapt4_tab:res_ups_dist} we instead investigate a different interaction, the one with distance. With this new regressor we can analyze the (non-linear) relation between trade and FDI, conditionally to the geographical distance between countries. The negative (and significant) coefficients of FDI in column 3 and 6 tell us that, at zero distance, the relation between trade and FDI is negative (the two are substitutive), for both the possible directions of investment; however, distance gives an additional positive term to be composed with the general coefficient, i.e. trade and FDI tend to become complements (positively correlated) as the geographical distance between the exporter and the importer countries increases. Using the same reasoning of Equations \ref{chapt4_eq:easymodel} and \ref{chapt4_eq:dT_dF}, we find $\widehat{\Delta}=798.34 \mbox{Km}$. Hence we conclude that (after controlling for the extensive margin and the other explanatory variables) country couples have both the FDI and trade channels open if $\Delta>\widehat{\Delta}$, while they prefer one way or the other if $\Delta<\widehat{\Delta}$. In other words, since almost all country pairs in our dataset are separate by distances $\Delta\gg\widehat{\Delta}$,\footnote{In fact just the $2.2\%$ of countries are at distances $\Delta<\widehat{\Delta}$, e.g. Austria and Italy, Jordan and Israel or Argentina and Uruguay. Remember here we used the great-circle definition of country distances.} this result means that trade and FDI are complements and that this relation become stronger as the geographical distance increases.\\

In Table \ref{chapt4_tab:res_region} (columns 1 and 2) we present an analysis about the influence of \textit{Regional Trade Agreements (RTA)}: we are interest in studying if the belonging to a common RTA fosters trade and what influence this has on the its relation with FDI. As expected the presence of RTA (variable \textit{rta}) is positively correlated with trade as its coefficient is positive and highly significant. However the interaction term between RTA and FDI is negative: this means that the presence of RTA, i.e. some agreement that ease trade between two countries, increment trade flows but reduce the amount of FDI. One can think of this effect as countries that already have a facilitated channel of interaction, as given by RTA, do not bother also investing and prefer the first, easier way. In other words when RTA are in place, trade and FDI tend to become substitutes. This result hold again for both the directions of FDI.

\begin{table}[!tb]
	\centering
	\caption[Regression results for regional trade agreements and the Asian countries]{Regressions with \textit{Regional Trade Agreements} and the special case of Asian countries. Country dummy variables for importer/exporter fixed effects included.}
	\label{chapt4_tab:res_region}
	\begin{adjustbox}{width=\textwidth,totalheight=\textheight,keepaspectratio}
		\begin{tabular}{lcc|c}
			\hline\hline
			\rule{0pt}{2ex} &\multicolumn{1}{c}{RTA}&\multicolumn{1}{c}{RTA (inv)}&\multicolumn{1}{c}{Asean + China}\\
			\hline
			\rule{0pt}{3ex}$\ln trade$ & & & \\
			$\ln FDI$ & 0.245\sym{***}& & \\
			& (0.0293) & & \\
			$\ln FDI$\_inv& & 0.408\sym{***}& 0.293\sym{***}\\
			& & (0.0210) & (0.0176) \\
			rta & 0.510\sym{***}& 0.495\sym{***}& \\
			& (0.0322) & (0.0322) & \\
			rta \# $\ln FDI$ & -0.239\sym{***}& & \\
			& (0.0305) & & \\
			rta \# $\ln FDI$\_inv& & -0.263\sym{***}& \\
			& & (0.0296) & \\
			Asean+China dummy & & & 5.436\sym{***}\\
			& & & (0.163) \\
			Asean+China dummy \# $\ln FDI$\_inv& & & 0.102\sym{**} \\
			& & & (0.0443) \\
			$\ln \Delta$ & -0.739\sym{***}& -0.694\sym{***}& -0.767\sym{***}\\
			& (0.0205) & (0.0179) & (0.0170) \\
			contig & 0.842\sym{***}& 0.821\sym{***}& 0.806\sym{***}\\
			& (0.0536) & (0.0536) & (0.0536) \\
			colony & 0.630\sym{***}& 0.603\sym{***}& 0.614\sym{***}\\
			& (0.0581) & (0.0582) & (0.0582) \\
			smctry & 0.353\sym{***}& 0.360\sym{***}& 0.467\sym{***}\\
			& (0.0752) & (0.0753) & (0.0750) \\
			comlang\_ethno& 0.450\sym{***}& 0.439\sym{***}& 0.443\sym{***}\\
			& (0.0278) & (0.0278) & (0.0278) \\
			\_cons & 10.18\sym{***}& 10.00\sym{***}& 10.57\sym{***}\\
			& (0.233) & (0.230) & (0.226) \\
			\hline
			mills & & & \\
			lambda & -1.038\sym{***}& -1.177\sym{***}& -1.221\sym{***}\\
			& (0.0680) & (0.0472) & (0.0473) \\
			\hline
			\(N\) & 336960 & 336960 & 336978 \\
			\hline\hline
			\multicolumn{4}{l}{\footnotesize Standard errors in parentheses; \sym{*} \(p<0.1\), \sym{**} \(p<0.05\), \sym{***} \(p<0.01\)}\\
		\end{tabular}
	\end{adjustbox}
\end{table}

Next we want to study the special case of Asia: we isolate countries belonging to the Asean\footnote{These are Brunei, Cambodia, Indonesia, Laos, Malaysia, Myanmar, Philippines, Singapore, Thailand, Vietnam.} trade agreement plus China and add a dummy identifying them as trade exporters, plus an interaction term of such dummy with FDI. Notice that in this case we are just considering the inverse channel of FDI. In other words we want to determine if there is a premium for countries to invest in Asia and them import goods from there. Column 3 of Table \ref{chapt4_tab:res_region} shows the results for this regression: one can observe how both the \textit{Asean+China dummy} as well as its interaction with FDI have positive (and significant) coefficients, meaning that there are in general larger volumes of trade when the exporter country is one of those considered and that investments in such countries are positively correlated with larger volumes of imports, that is trade and (inverse) FDI are more strongly complements, if the country which exports the goods (alternatively, the one that receive the investments) is China or belongs to the Asean trade agreement.

Now to analyze differences among economic macro-sectors,\footnote{For the definition of macro-sectors refer to Section \ref{chapt4_sec:data-and-definitions}.} we now add to the gravity model specification a set of fixed effects: in the first column of Table \ref{chapt4_tab:res_sectors} we can see the correspondent dummies, that have to be added to the constant term of the model (\textit{\_cons}), that is relative to the primary sector, our baseline. Hence all the three \textit{fixed effects} are positive but slightly different, meaning that, without any other regressor in place, goods in the secondary sector are traded slightly more than those in the primary and that goods in the tertiary are traded slightly less than products in the primary. Baseline coefficients do not change significantly, though the coefficient of distance increases in absolute value after macro-sector fixed effects inclusion.

\begin{figure}[tb]
	\centering
	\caption[Normalized fixed effects and interactions between FDI and the trade macro-sectors]{Normalized fixed effects and interactions between FDI and the trade macro-sectors. Figure \ref{chapt4_fig:bars_1} refers to exports in both trade and FDI; figure \ref{chapt4_fig:bars_2} to exports in trade and imports in FDI. Data is taken from Table \ref{chapt4_tab:res_sectors}.}
	\vspace{-10pt}
	\subfloat[Same direction\label{chapt4_fig:bars_1}]{\includegraphics[width=\myFiguresSizeBis]{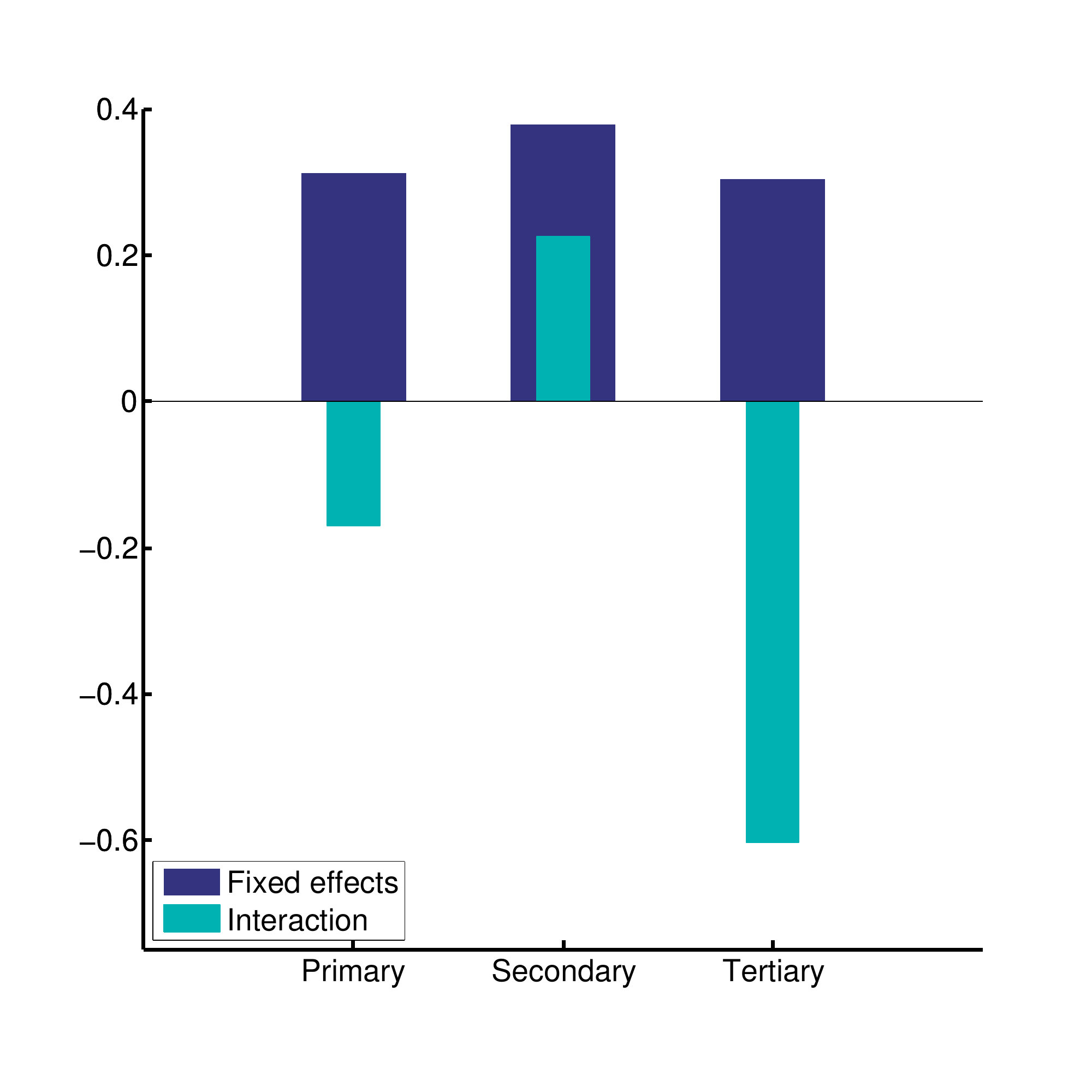}}
	\subfloat[Opposite direction\label{chapt4_fig:bars_2}]{\includegraphics[width=\myFiguresSizeBis]{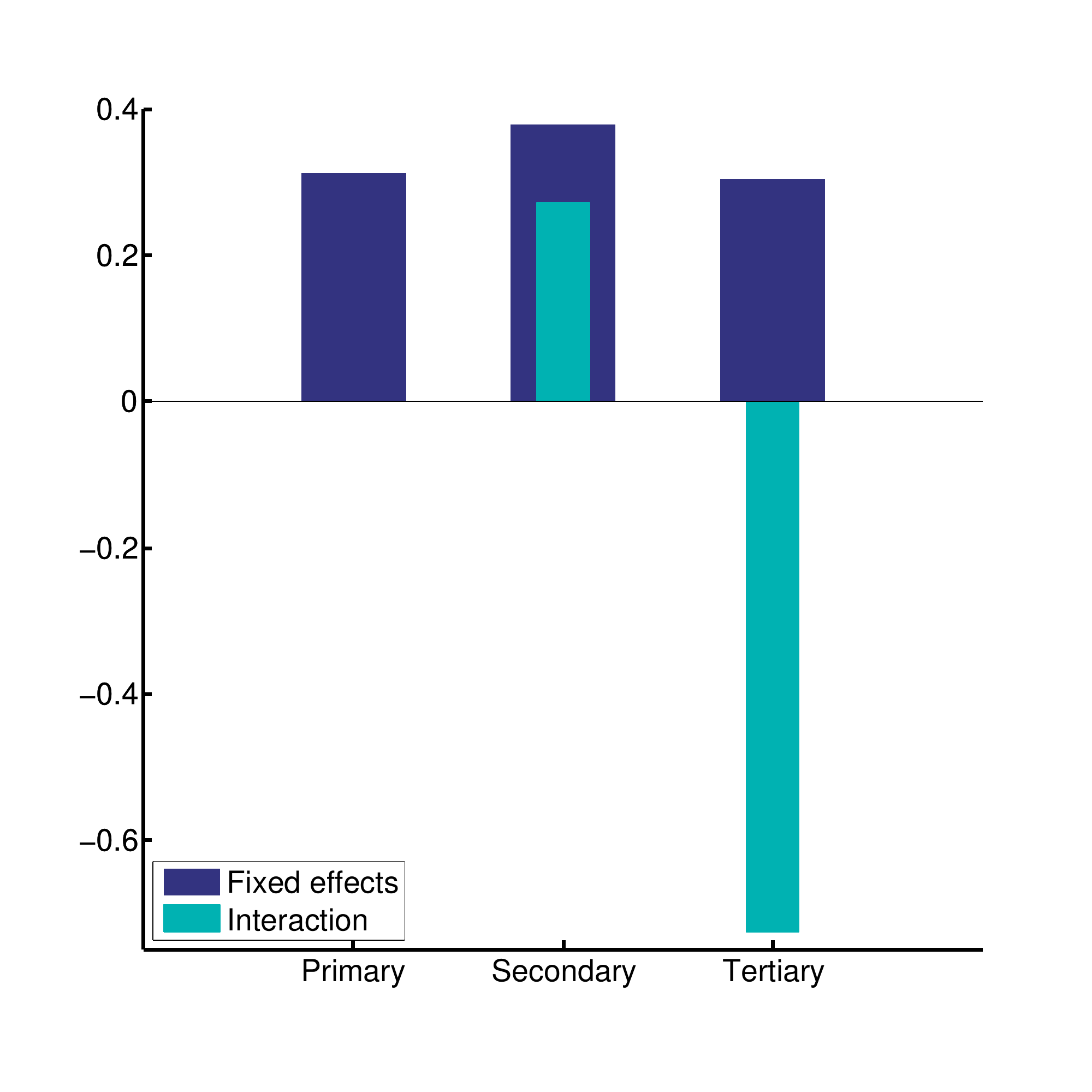}}
	\vspace{-10pt}
	\label{chapt4_fig:bars}
\end{figure}

We introduce now in our analysis an interaction coefficient between FDI and macro-sectors, i.e. we let the coefficient of FDI to change for different sectors. We find (in column two of Table \ref{chapt4_tab:res_sectors}) that the sign for the secondary sector is positive (and significant), while those for the primary and tertiary are negative\footnote{Again the primary sector is our baseline, hence to obtain the correct coefficients one has to sum the base coefficient for $\ln FDI$ with the interaction ones.} (and significant). This is a strong evidence of a \textit{complementarity} effect of FDI on trade for the secondary sector and of a substitutive effect for the primary and tertiary. The interaction and fixed effects coefficients are shown (normalized) in Figure \ref{chapt4_fig:bars_1}.

\begin{table}[!tb]
	\centering
	\caption[Regression results for the analysis of macr-sectors]{Regressions with \textit{fixed effects} and \textit{interactions} for the industry macro-sectors. Country dummy variables for importer/exporter fixed effects included. The right part refers to the case where trade and FDI were considered in opposite directions.}
	\label{chapt4_tab:res_sectors}
	\begin{adjustbox}{width=\textwidth,totalheight=\textheight,keepaspectratio}
		\begin{tabular}{lcc|cc}
			\hline\hline
			\rule{0pt}{2ex} & \multicolumn{1}{c}{Fixed Effects} & \multicolumn{1}{c}{Interaction} & \multicolumn{1}{c}{Fixed Effects (inv)} & \multicolumn{1}{c}{Interaction (inv)}\\
			\hline
			\rule{0pt}{3ex}$\ln trade$ & & & & \\
			$\ln FDI$ & -0.123\sym{***}& -0.351\sym{***}& & \\
			& (0.0206) & (0.0359) & & \\
			$\ln FDI$\_inv& & & -0.0170 & 0.0212 \\
			& & & (0.0149) & (0.0322) \\
			primary & 0 & 0 & 0 & 0 \\
			& (.) & (.) & (.) & (.) \\
			secondary & 2.456\sym{***}& 2.333\sym{***}& 2.450\sym{***}& 2.367\sym{***}\\
			& (0.0179) & (0.0181) & (0.0180) & (0.0182) \\
			tertiary & -0.441\sym{***}& -0.280\sym{***}& -0.444\sym{***}& -0.268\sym{***}\\
			& (0.0203) & (0.0205) & (0.0203) & (0.0205) \\
			primary \# $\ln FDI$& & 0 & & \\
			& & (.) & & \\
			secondary \# $\ln FDI$& & 0.821\sym{***}& & \\
			& & (0.0345) & & \\
			tertiary \# $\ln FDI$& & -0.899\sym{***}& & \\
			& & (0.0382) & & \\
			primary \# $\ln FDI$\_inv& & & & 0 \\
			& & & & (.) \\
			secondary \# $\ln FDI$\_inv& & & & 0.453\sym{***}\\
			& & & & (0.0346) \\
			tertiary \# $\ln FDI$\_inv& & & & -1.203\sym{***}\\
			& & & & (0.0387) \\
			$\ln \Delta$ & -1.123\sym{***}& -1.129\sym{***}& -1.168\sym{***}& -1.152\sym{***}\\
			& (0.0168) & (0.0166) & (0.0149) & (0.0147) \\
			contig & 0.856\sym{***}& 0.880\sym{***}& 0.837\sym{***}& 0.864\sym{***}\\
			& (0.0472) & (0.0464) & (0.0469) & (0.0463) \\
			colony & 0.801\sym{***}& 0.841\sym{***}& 0.800\sym{***}& 0.827\sym{***}\\
			& (0.0513) & (0.0505) & (0.0513) & (0.0506) \\
			smctry & 0.552\sym{***}& 0.521\sym{***}& 0.554\sym{***}& 0.529\sym{***}\\
			& (0.0659) & (0.0648) & (0.0657) & (0.0648) \\
			comlang\_ethno& 0.588\sym{***}& 0.586\sym{***}& 0.588\sym{***}& 0.585\sym{***}\\
			& (0.0247) & (0.0243) & (0.0247) & (0.0243) \\
			\_cons & 11.13\sym{***}& 11.21\sym{***}& 11.29\sym{***}& 11.26\sym{***}\\
			& (0.201) & (0.198) & (0.200) & (0.197) \\
			\hline
			mills & & & & \\
			lambda & -0.640\sym{***}& -0.616\sym{***}& -0.406\sym{***}& -0.478\sym{***}\\
			& (0.0575) & (0.0567) & (0.0409) & (0.0405) \\
			\hline
			\(N\) & 336978 & 336978 & 336978 & 336978 \\
			\hline\hline
			\multicolumn{5}{l}{\footnotesize Standard errors in parentheses; \sym{*} \(p<0.1\), \sym{**} \(p<0.05\), \sym{***} \(p<0.01\)}\\
		\end{tabular}
	\end{adjustbox}
\end{table}

Considering the inverse relation, fixed effects, as well as interaction coefficients, are consistent with the former case for the secondary and tertiary sectors. Interestingly the relation between trade and FDI for the raw materials sector is no more statistically significant: this means that if FDI ``exports'' are substitutes with trade exports, they have no relation with trade imports, i.e. in the primary sector, investing in another country does not lead to an increase imports in that same sector, from that same country. As before, we find a complementarity relation in the secondary and a substitutive one in the tertiary. A graphical representation of the estimated fixed effects and coefficients is shown in Figure \ref{chapt4_fig:bars_2}.

These findings are also consistent with the results about product upstreamness. This is particularly so in the direct case (export-export) where for raw materials trade and FDI are substitutes, while for manufacturing (thus generally closer to the final use, downstream) they are complements.

\section{Conclusions}\label{chapt4_sec:conclusions}
In this chapter we investigated the relation and interplay between international trade and Foreign Direct Investments\footnote{As proxied by the number of transnational corporations (TNC) affiliates} (FDI). Initially looking at them as two networks, with countries as nodes and trade or FDI channels as directed weighted edges, we study their topological properties: considering single network characteristics they have very similar features (they are both markedly disassortative and have nodes strength power law distribution with the same exponent), as different ones (the World Trade Web (WTW) is more dense than the FDI network (FDIN) and have a stronger cluster structure); then we look for correlation between the two and find that their edges weight, as well as nodes strength and degree and Average Nearest Neighbor Degree (ANND) and Strength (ANNS), are positively correlated and that such correlation can be mostly explained by country economic/demographic size and geographical distance.

Next we employ the Heckman 2-step (H2S) selection model to quantify the correlations already found and study other properties of the interplay between the two phenomena, in particular seeking for causal relations. We find that in general trade and FDI channels are positively correlated, both when they are considered in the same and opposite directions.
Moreover we study the interaction of FDI with distance and upstreamness, finding a positive effect the interaction of distance and FDI has on trade, i.e. trade and FDI tend to be more positively correlated as the geographical distance between the exporter and the importer country increases; as for upstreamness, we highlight how trade is more intense for products closer to the final use and how, the more the industry is upstream, the more FDI and trade tend to be substitutes.
Then we investigated the influence of Regional Trade Agreements (RTA) and the case of Asian countries. Our findings are that, as expected, RTA are positively correlated with trade and, when these are in place, trade and FDI tend to become substitutes. As for Asian countries, we discover there are in general larger volumes of trade and a stronger complementarity with FDI when the exporter country is Asian. These findings confirm the prominent role of Asia in the WTW and its role of, as it is sometimes called, ``world factory''.
Finally we quantify the correlation between trade and FDI by distinguishing for the three economic macro-sectors: in fact we find a positive correlation for the secondary, but negative one for the tertiary. For the primary sector we find that trade and FDI are substitutes if we consider the two flows in the same direction, while no statistically significant relation is observed if we consider them in opposite ways. Consistently with the results about product upstreamness.

Our work can be extended in at least two directions. First, one can further investigate endogeneity issues arising in gravity-model exercises, due to the reverse-causation link possibly existing from trade to FDI. A possible way out might involve instrumenting investments stocks (e.g., using a simple FDI gravity model) and replace FDI-related regressors with the predictions from the instrumental-variable estimation. Second, some robustness checks are required to test the goodness of our gravity specification and of the H2S selection model against the family of Poisson regressors.